\documentclass{iopjournal}

\usepackage[english]{babel}
\usepackage{amssymb}
\usepackage{amsmath}
\usepackage[style=numeric, sorting=none, firstinits=true, maxcitenames=3, backend=biber]{biblatex}
\usepackage{bm}
\usepackage{booktabs}
\usepackage{multirow}
\usepackage{ragged2e}
\usepackage{siunitx}
\usepackage{subcaption}
\usepackage{tabularx}

\newcommand{\opensc}{OPENSC$^2$}
\newcommand{\dd}{\,\mathrm{d}}
\newcommand{\ddt}[1]{\frac{\dd #1}{\dd t}}

\begin{document}

\articletype{Paper} 

\title{Winding-Pack Quench Simulation of Stellarator Coils: Hydraulic Field-Network Coupling and Transverse Heat Conduction in \opensc}

\author{Laura A.~M.~D'Angelo$^{1,*}$\orcid{0000-0003-1009-1324} and Felix Warmer$^{1,2}$\orcid{0000-0001-9585-5201}}

\affil{$^1$Max Planck Institute for Plasma Physics, Wendelsteinstr.~1, 17491 Greifswald, Germany}\\
\affil{$^2$Eindhoven University of Technology, PO Box 513, 5600 MB, Eindhoven, Netherlands}

\affil{$^*$Author to whom any correspondence should be addressed.}

\email{laura-anna-maria.dangelo@ipp.mpg.de}

\keywords{quench simulation, cable-in-conduit-conductors, stellarator, thermo-hydraulics, Wendelstein 7-X}

\begin{abstract}
Quench propagation in cable-in-conduit conductors is thermo-hydraulically driven, yet a standalone 1D model treats a conductor as one straight adiabatic cable with prescribed pressure and mass-flow boundaries, so secondary quenches and flow redistribution between windings cannot be resolved without coupling to further, so far proprietary, codes. We extend the open-source code \opensc\ with transverse heat conduction between turns, a 3D field resolving the current-sharing temperature, induced losses, and a lumped hydraulic network coupled to the conductors by a mass-flow formulation. Compared to the 2006 W7-X coil tests, it predicts the quench site and cuts the outlet-temperature overestimation from 55\,K to 25\,K. The identified mechanisms indicate that the position of the helium inlets and outlets relative to the field, the hydraulic interconnection of the windings, the turn insulation and the current dump should be considered together in the quench design of future stellarator coils.
\end{abstract}

\justifying 

\section{Introduction}
Stellarator-type fusion reactors are currently regarded as a very promising route to practically inexhaustible carbon-free energy~\cite{warmer_overview_2024}. Within such a reactor, the magnet system, consisting of inherently three-dimensional superconducting coils, plays a central role: physically, as the source of the magnetic confinement of the plasma, and economically, as a major cost driver of the reactor~\cite{warmer_system_2016, prost_economically_2024}. Quench protection is therefore a crucial aspect of the design of these coils. Future fusion power-plant magnets will store magnetic energy on the order of gigajoules, so an uncontrolled quench would have disastrous consequences. Stellarators in particular cannot afford damaged coils, since replacing their geometrically complex three-dimensional windings would be practically impossible without dismantling major parts of the machine, shutting down the entire power plant for several months, or even longer if the affected coil had to be re-manufactured.

\begin{figure}[b]
    \centering
    \includegraphics[height=4cm]{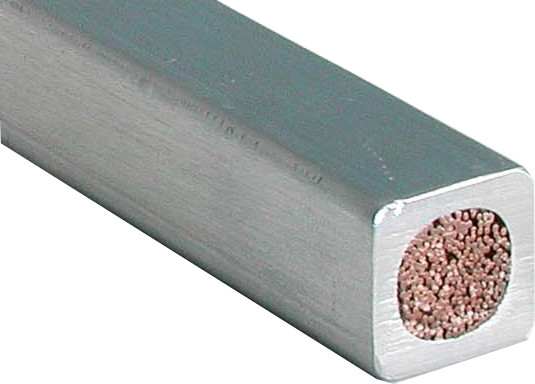} \hspace{1cm}
    \includegraphics[height=4cm]{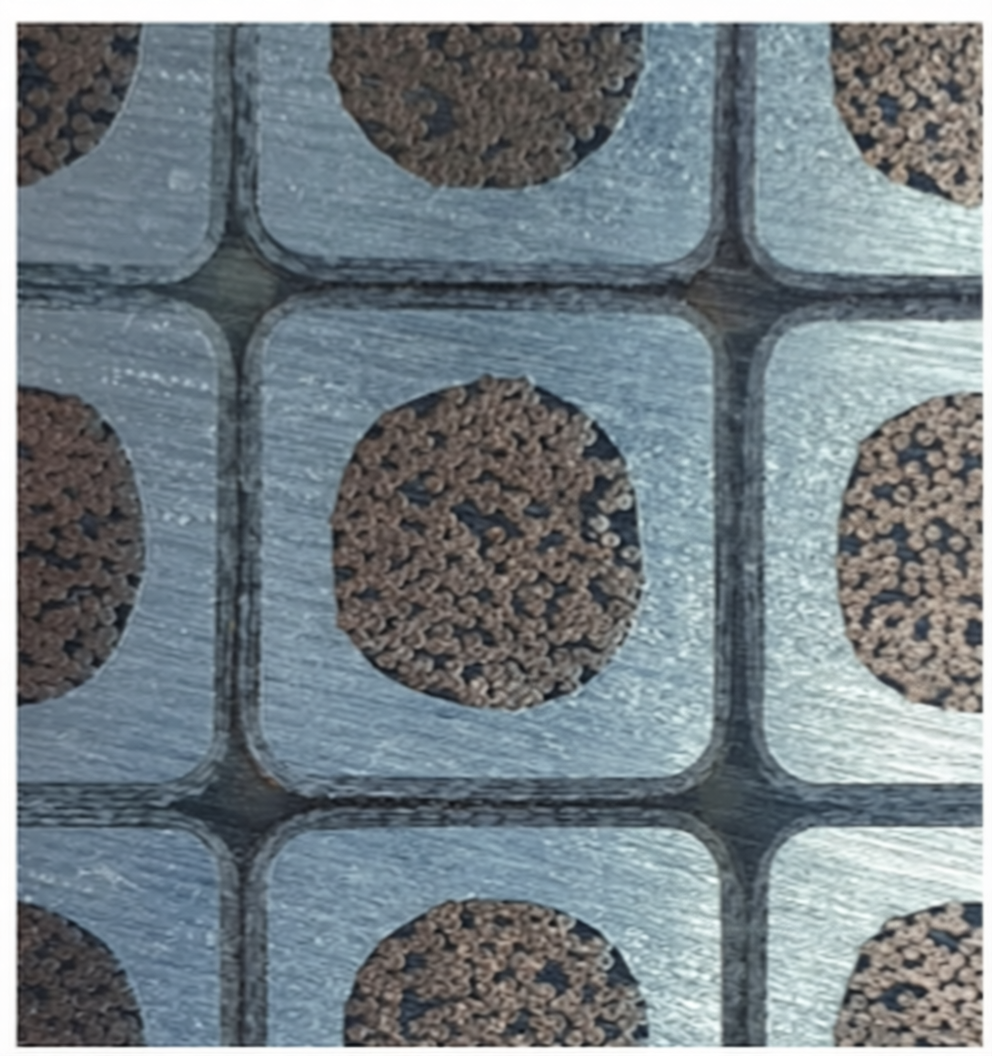}
    \caption{The W7-X conductor is a CICC made of superconducting NbTi strands enclosed by an aluminum jacket. The coolant (supercritical helium) is forced through the void between the strands. Figures adapted from~\cite{rummel_superconducting_2012, risse_thermohydraulic_2024}.}
    \label{fig:w7x-conductor}
\end{figure}

Cable-in-conduit conductors (CICCs) are a class of superconducting cables that are particularly attractive for high-field and high-current applications such as fusion reactor magnets~\cite{rummel_superconducting_2012, bagni_analysis_2017, muzzi_cableinconduit_2015}.
CICCs are characterized by a conduit enclosing either superconducting strands, in the case of low-temperature superconductors (LTS), or a number of high-temperature superconductor (HTS) stacks, together with the coolant, which is confined within the cable. The coolant is forced through the CICC along one or more flow channels, wetting and cooling the superconductor either directly or indirectly through a former. Figure~\ref{fig:w7x-conductor} shows, as an example, the CICC used in the Wendelstein 7-X (W7-X) fusion experiment at the Max Planck Institute for Plasma Physics in Greifswald, Germany. In this conductor, superconducting NbTi strands are enclosed by an aluminum jacket, and supercritical helium is forced through the void between the strands, cooling them directly. The conduit gives CICCs a high robustness against thermal and structural loads and simplifies the insulation of adjacent turns, permitting large potential differences between them. This makes CICCs intrinsically better protected against quench than other superconducting cable types such as Rutherford cables.

\begin{figure}[tbp]
    \centering
    a) \includegraphics[height=4cm]{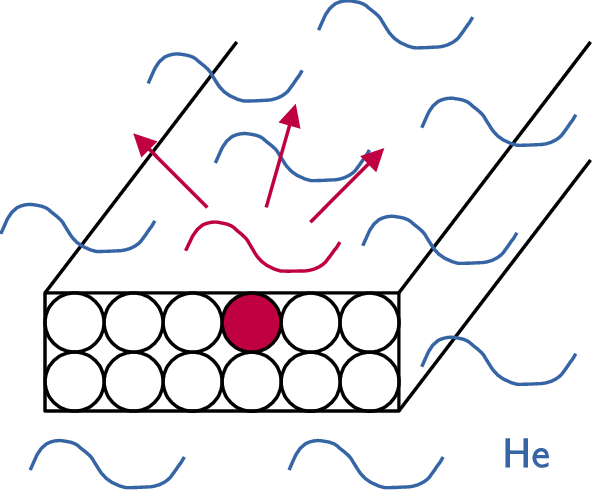} \hspace{1cm}
    b) \includegraphics[height=4cm]{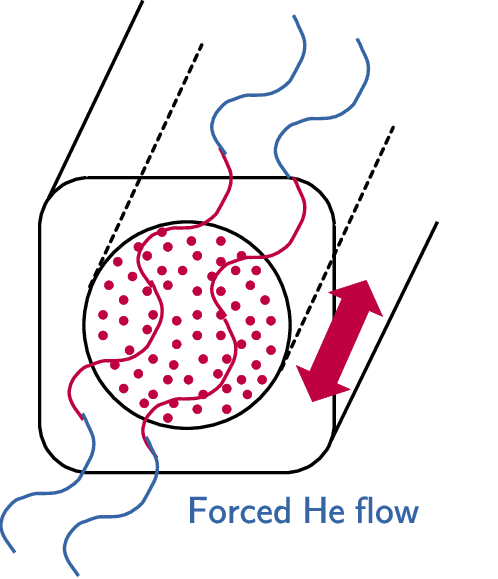}
    \caption{Bath-cooled superconductors like Rutherford cables (a) compared to forced-flow cooled superconductors like CICCs (b) in the event of a quench: While helium can freely escape in the bath-cooled scenario, the helium is confined to its channel in the CICC case leading to a pressure rise and helium expulsion.}
    \label{fig:bath-vs-forced-flow}
\end{figure}

Numerical simulation tools have become indispensable for studying quench behavior and for designing quench protection systems. Most of these tools originated in the accelerator magnet community. LEDET, for example, employs a lumped-element electro-thermal model to simulate the superconducting coil in the event of a quench and to analyze the performance of quench protection systems, representing the physical phenomena in terms of electric circuits and thermal networks~\cite{ravaioli_lumpedelement_2016}.
ROXIE, the workhorse of the magnet design for the Large Hadron Collider at CERN, instead offers high-fidelity magneto-thermal quench simulation based on a reduced magnetic vector potential approach combined with a finite-element-boundary-element coupling~\cite{schwerg_quench_2008}.
FiQuS is a more recent tool that supports both 2D and 3D quench simulation, employs advanced numerical techniques and is built on the open-source meshing and finite-element (FE) tools Gmsh and GetDP~\cite{vitrano_opensource_2023, dular_general_1998}.
What most accelerator magnet simulation tools such as LEDET, ROXIE and FiQuS have in common is that they solve a coupled electro- or magneto-thermal problem in order to simulate quench propagation in a superconducting coil. This is appropriate because accelerator magnets typically use Rutherford cables cooled by a static helium bath. In the event of a quench, the helium evaporating at the quenching conductor can escape freely and is immediately replaced by fresh helium, as sketched in figure~\ref{fig:bath-vs-forced-flow}a. The fluid behavior of the helium is therefore irrelevant for the quench, whose propagation is driven purely by heat conduction within the cable. The cooling can be reduced to a Robin boundary condition representing the heat transfer between the cable and the surrounding helium at a fixed temperature.

In a CICC, by contrast, the heated helium is confined inside the conduit and causes a hydraulic pressure rise (see figure~\ref{fig:bath-vs-forced-flow}b). The warm helium is expelled along the winding, carrying its heat into coil sections that have not (yet) quenched. Quench propagation in a CICC is therefore thermo-hydraulically driven, so the helium hydraulics cannot be neglected. Depending on the topology of the hydraulic network, warm helium from the initially quenched winding may even redistribute into adjacent windings and trigger secondary quenches. Dedicated software tools for CICCs that account for the hydraulics do exist. THEA, for example, is a commercial code written in Fortran that solves the full set of electric, thermal and hydraulic equations~\cite{bottura_general_2000}. 4C~\cite{savoldirichard_4c_2010} and VINCENTA~\cite{bessette_simulations_2006} are both proprietary codes that perform thermo-hydraulic simulations of CICCs and have been used for the simulation of the toroidal field coils in ITER. 

A more recent addition to this list is \opensc, developed by the MAHTEP research group at Politecnico di Torino, Italy~\cite{savoldi_thermalhydraulic_2022}. It is particularly attractive for the current fusion research community because it is open-source, written in a modern programming language (Python) and capable of accurately simulating HTS CICCs, a task with which other simulation tools usually struggle~\cite{lewandowska_modelling_2024}. As in THEA, \opensc\ models the CICC as a set of straight, mutually coupled 1D components, on which a coupled electro-thermo-hydraulic problem is solved in space and time.

For stellarator coils, this modeling approach reaches its limits. A 1D model of this kind is intrinsically adiabatic in the transverse direction: it accounts only for longitudinal heat conduction along the conductor, but not for heat conduction from turn to turn, so that the resulting hot-spot temperature is conservative. Although most of the heat is indeed conducted longitudinally, owing to the insulation around the turns, such an approach still overestimates the actual hot-spot temperature, since transverse heat diffusion is small but not zero. Consequently, \opensc\ cannot capture secondary quenches in adjacent windings caused by transverse heat conduction or other effects. Quench measurements from the W7-X coil tests at CEA in 2006, however, indicate that such secondary quenches do occur~\cite{baldzuhn_coil_2006}. A further limitation is the finite extent of the 1D domain, which requires prescribed boundary conditions for pressure and mass flow. In reality, these are determined by the external hydraulic network that supplies the coil with helium. To date, it has therefore not been possible to reproduce the measurement data of the W7-X coil tests using 1D quench simulation tools~\cite{baldzuhn_coil_2006}.

To overcome these limitations, this work extends \opensc\ with several new features, with the aim of reproducing the coil test conditions as accurately as possible. We add transverse heat conduction between turns, space- and time-dependent magnetic flux densities affecting the current-sharing temperature of the superconductor, and induced currents. Finally, we derive a hydraulic network model based on the transmission-line model for pipelines~\cite{wylie_fluid_1978} and couple it to the coil field model by means of a hydraulic field-network coupling approach adapted from the electric field-circuit coupling method well established in the computational electromagnetics community~\cite{schops_winding_2013}. Similar approaches, coupling lumped-element hydraulic models to field models, already exist in the hemodynamics community for computing human blood flow~\cite{quarteroni_coupling_2001} and, in the form of the Flower model, for cryogenic networks and the calculation of quench behavior~\cite{bottura_flower_2003}. Such models differ in how the lumped elements, i.e.\ the hydraulic resistances, inductances and capacitances, are defined for the problem at hand, and in the choice of the primary variables. Whereas the Flower model uses pressure, fluid velocity and temperature as primary variables~\cite{bottura_flower_2003}, this work selects the mass flow rate instead of the velocity which comes with multiple advantages: it is continuous across the coupling interfaces while the velocity is not, it avoids a variable transformation at the coupling interfaces, it is flux conservative by construction, it resolves the steady-state exactly (no spurious mass accumulation), imposing the mass flow becomes a pure Dirichlet boundary condition, and during a quench event the mass flow stays comparatively smooth while the velocity varies violently. 

This work is structured as follows. First, the governing equations of the 1D electro-thermo-hydraulic quench model in \opensc\ are introduced in Sec.~\ref{sec:model}, and \opensc\ is benchmarked against a W7-X simulation in THEA. Then, the new features are introduced and discussed next: Sec.~\ref{sec:2d} deals with the transverse thermal coupling of turns and layers, the inclusion of space- and time-dependent external magnetic fields and the modeling of induced effects such as hysteresis, coupling and eddy-current losses, while Sec.~\ref{sec:hydraulic-network} derives the hydraulic network equations and couples them to the 1D field models. Finally, \opensc\ with all new features is applied to the W7-X quench test measurement scenarios from 2006 at CEA~\cite{baldzuhn_coil_2006} (Sec.~\ref{sec:validation}). The purpose of this comparison is to assess whether the winding-pack features improve agreement with the measurements compared to the established adiabatic 1D model, and to identify the physical mechanisms that can explain the observed phenomena. Section~\ref{sec:conclusion} concludes and gives an outlook on further work.

\section{1D Electro-Thermo-Hydraulic Model in \opensc}\label{sec:model}
This section briefly recapitulates how a superconducting cable is represented and which electro-thermo-hydraulic model is originally constructed and solved in \opensc. More details can be found in the original work of Savoldi et al.~\cite{savoldi_thermalhydraulic_2022}.

\subsection{Cable Model}
In the \opensc\ framework, a conductor consists of $N_\mathrm{f}$ fluid components (channels) and $N_\mathrm{s}$ solid components such as strands and jackets. Each conductor (and with it, its components) is represented by a 1D straight domain $\Omega = [0, L]$ with $L$ being the length of the conductor in \si{\meter}. These 1D domains can be thermally and electrically linked with each other. To visualize the components and their links, \opensc\ provides a custom scheme: each component type and each link type have their own symbol, e.g.\ red pentagons for strand components and zigzag lines (US-style resistors) for a thermal conductive link. A complete list of symbols is found in~\cite{savoldi_thermalhydraulic_2022}. The cable model is discretized spatially using the FE method and temporally using a user-defined time integrator.

\begin{figure}[tbp]
    \centering
    \includegraphics[width=0.7\linewidth]{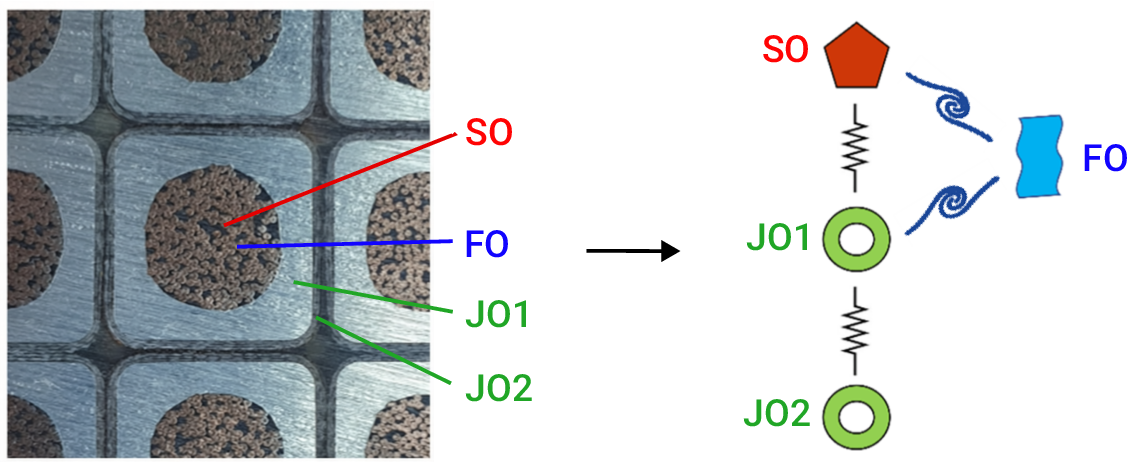}
    \caption{The W7-X conductor model visualized with the \opensc\ custom scheme~\cite{savoldi_thermalhydraulic_2022}: The conductor is resolved by four components: Superconducting strands as a strand object (SO, red), the aluminum jacket and the insulation as jacket objects (JO, green) and the helium flowing through the void inside the jacket as fluid object (FO, blue). The thermal links between the objects are visualized by zigzags in the case of thermal conduction and by swirls in the case of thermal convection.}
    \label{fig:w7x-opensc-model}
\end{figure}

Figure~\ref{fig:w7x-opensc-model} shows the scheme applied to the W7-X conductor: The superconducting strands are modeled as a strand object (SO), the aluminum jacket and insulation as jacket objects (JO), while the coolant flowing through the void between the superconducting strands is modeled as a fluid object (FO). The strands and the aluminum as well as the aluminum and the insulation are linked to each other by thermal conduction, and the coolant is linked to both the strands and the aluminum by thermal convection. The scheme only shows the thermal links, not the electrical ones, so the electrical link between strands and aluminum jacket is not visualized herein.

\subsection{Thermal Model}
Each solid component $i$ obeys a homogenized 1D heat conduction equation
\begin{equation}
    \frac{A_i \varrho_i c_{\mathrm{p},i}}{\cos(\theta_i)} \, \partial_t T_i - \partial_x \left( \frac{A_i \lambda_i}{\cos(\theta_i)} \partial_x T_i \right) + \sum\limits_k \alpha_{ik} P_{ik} \left( T_i - T_k \right) = q_i,  
\end{equation}
with $x$ the 1D longitudinal coordinate in \si{\meter}, $T_i$ the temperature in \si{\kelvin}, $A_i$ the cross-section in \si{\square\meter}, $\varrho_i$ the homogenized density in \si{\kilo\gram\per\cubic\meter}, $c_{\mathrm{p},i}$ the homogenized isobaric specific heat in \si{\joule\per\kilo\gram\per\kelvin}, $\lambda_i$ the homogenized thermal conductivity in \si{\watt\per\meter\per\kelvin}, $\cos(\theta_i)$ the dimensionless twist pitch factor, $P_{ik}$ the contact perimeter in \si{\meter} and $\alpha_{ik}$ the heat transfer coefficient in \si{\watt\per\square\meter\per\kelvin} towards the contacting component $k$~\cite{savoldi_thermalhydraulic_2022}. The heat source term $q_i$ in \si{\watt\per\meter} collects the various heat sources, 
\begin{equation}
    q_i = q_\mathrm{ext} + q_\mathrm{Joule}^{||} + q_\mathrm{Joule}^\perp + q_\mathrm{rad}, 
    \label{eq:heat-source-term}
\end{equation}
i.e.\ external heating, longitudinal and transverse Joule power from the electric problem (see Sec.~\ref{subsec:em-model}), and radiative heat exchange in case of a jacket component.

The temperature in a fluid component $j$ is resolved by the energy balance equation
\begin{equation}
    \partial_t T_j + v_j \partial_x T_j + \phi_j T_j \partial_x v_j - \frac{2 f_j |v_j|}{D_{\mathrm{h},j}} \frac{v_j^2}{c_{v,j}} = \sum\limits_i \frac{\alpha_{ij} P_{ij}}{\varrho_j c_{v,j} A_j} (T_i - T_j), \label{eq:thermal-fluid}
\end{equation}
where $v$ is the fluid velocity in \si{\meter\per\second}, $\phi$ is the dimensionless Grüneisen parameter, $c_v$ the isochoric specific heat in \si{\joule\per\kilo\gram\per\kelvin}, $f$ the dimensionless Fanning friction factor and $D_\mathrm{h}$ the hydraulic diameter in \si{\meter}. The right-hand side describes the heat exchange term with a solid component $i$. 

\subsection{Hydraulic Model}\label{subsec:opensc-hydraulic-model}
\opensc\ implements the hyperbolic compressible Euler system in terms of the velocity $v$ in \si{\meter\per\second}, pressure $p$ in \si{\pascal} and temperature $T$, reading
\begin{align}
    \partial_t v + v \partial_x v + \frac{1}{\varrho} \partial_x p + \frac{2f|v|}{D_\mathrm{h}}v &= 0, \label{eq:velocity-system-velocity} \\
    \partial_t p + v \partial_x p + \varrho c^2 \partial_x v - \frac{2f|v|}{D_\mathrm{h}} \phi \varrho v^2 &= \dot{q}_{p}, \label{eq:velocity-system-pressure} \\
    \partial_t T + v \partial_x T + \phi T \partial_x v - \frac{2f|v|}{D_\mathrm{h}} \frac{v^2}{c_v} &= \dot{q}_T, \label{eq:velocity-system-temperature}
\end{align}
with $c$ the speed of sound in \si{\meter\per\second}, $\dot{q}_T$ in \si{\kelvin\per\second} representing the heating of the fluid by the heat exchange with the solid components, i.e.\ the right-hand side of~\eqref{eq:thermal-fluid}, and $\dot{q}_p = \phi \varrho c_v \dot{q}_T$ in \si{\pascal\per\second} representing the resulting pressurization. In matrix form, these equations form the quasi-linear system 
\begin{equation}
    \partial_t \mathbf{U} + \mathbf{K}(\mathbf{U}) \partial_x \mathbf{U} + \mathbf{S}(\mathbf{U}) \mathbf{U} = \mathbf{q} \label{eq:original-quasilinear-system}
\end{equation}
with $\mathbf{U} = [v, p, T]^\top$ gathering the primary variables, $\mathbf{q} = [0, \dot{q}_p, \dot{q}_T]^\top$, and the flux Jacobian $\mathbf{K}$ and the friction matrix $\mathbf{S}$ reading, respectively, 
\begin{equation}
    \mathbf{K} = \begin{bmatrix}
        v & \varrho^{-1} & 0 \\
        \varrho c^2 & v & 0 \\
        \phi T & 0 & v
    \end{bmatrix}, \quad \mathbf{S} = \frac{2f|v|}{D_\mathrm{h}} \begin{bmatrix}
        1 & 0 & 0 \\
        -\phi \varrho v & 0 & 0 \\
        -v c_v^{-1} & 0 & 0 
    \end{bmatrix}.
\end{equation}

\subsection{Helium Model}
\opensc\ employs the open-source library CoolProp to evaluate the thermophysical properties of helium~\cite{bell_pure_2014}. A major limitation of CoolProp is that it does not provide a two-phase flow model~\cite{bell_pure_2014}, i.e.\ it cannot model helium as a liquid-vapor mixture. So when the fluid state crosses the saturation curve, it does not boil or condense in the model; instead, its properties jump discontinuously from the liquid to the vapor state and vice versa, possibly causing numerical instabilities and solution divergence in doing so. As long as the pressure stays above the critical pressure $p_\mathrm{c} \approx 0.227\,$\si{\mega\pascal}, helium is supercritical, i.e.\ it avoids a liquid-vapor separation and this limitation does not matter. For superconducting magnet cooling, it is anyway imperative to use supercritical helium, since the heat transfer properties of two-phase helium are considerably more uncertain, and so the pressure of the forced flow is chosen accordingly to ensure the supercritical state~\cite{bald_supercritical_1971}. Nevertheless, this is an important limitation to keep in mind, especially when considering an external hydraulic network which might have relief valves to release helium in order to depressurize the network, thus possibly introducing a two-phase state which cannot be modeled accurately.

\subsection{Electric Model}\label{subsec:em-model}
The cable's current distribution in \opensc\ is modeled as a distributed electric circuit: the longitudinal edges (one per strand per mesh element) are coupled node-to-node by transverse contact conductances. Unknowns are the edge currents $i_e$ in \si{\ampere} and the nodal electric potentials $\varphi_n$ in \si{\volt} of all current-carrying strands. The governing differential-algebraic system is 
\begin{equation}
    \begin{bmatrix}
        \mathbf{L} & \mathbf{0} \\ 
        \mathbf{0} & \mathbf{0}
    \end{bmatrix}
    \frac{\mathrm{d}}{\mathrm{d}t} 
    \begin{bmatrix}
        \mathbf{i} \\ \bm{\varphi}
    \end{bmatrix}
    +
    \begin{bmatrix}
        \mathbf{R} & \mathbf{A} \\ 
        -\mathbf{A}^\top & \mathbf{G}
    \end{bmatrix}
    \begin{bmatrix}
        \mathbf{i} \\ \bm{\varphi}
    \end{bmatrix}
    = 
    \begin{bmatrix}
        \mathbf{0} \\ \mathbf{i}_\mathrm{src}
    \end{bmatrix},
\end{equation}
i.e.\ a lumped transmission-line model with per-strand inductive-resistive longitudinal edges (first row) and Kirchhoff's current conservation (second row). Here, $\mathbf{A}$ is the edge-to-node incidence matrix, $\mathbf{R}$ the diagonal longitudinal resistance matrix in \si{\ohm}, $\mathbf{G}$ the nodal transverse conductance matrix in \si{\siemens}, and $\mathbf{L}$ the full (self and mutual) inductance matrix in \si{\henry}.

\subsection{Coupling}\label{subsec:coupled-model}
\begin{figure}[tbp]
    \centering
    \includegraphics[width=0.85\linewidth]{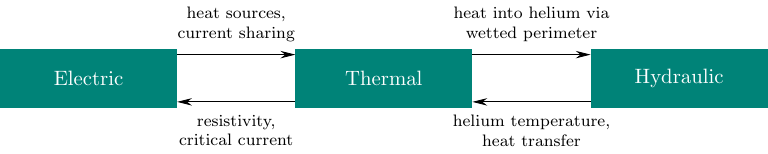}
    \caption{Multi-physical coupling scheme of the electro-thermo-hydraulic model.}
    \label{fig:coupling-diagram}
\end{figure}

The thermal, hydraulic and electric model are strongly coupled with each other as shown in figure~\ref{fig:coupling-diagram}: The electric model determines whether strand segments are superconducting, current-sharing or normal-conducting and, thus, whether they generate Joule losses, which are handed over to the thermal model as heat sources. The thermal model solves for the temperature, which determines the electric resistivity and the critical current for the electric model. The heat exchange between solids and fluids is what couples the thermal and hydraulic models.

\subsection{Benchmark against W7-X THEA Simulation}\label{subsec:validation-against-thea}
After refactoring the \opensc\ code and thereby improving its computational efficiency drastically, \opensc\ was able to simulate a full W7-X double layer winding ($L=144.5\,\si{\meter}$) in 19\,\si{\minute} instead of 22\,\si{\hour}. The highlights of this refactoring work are given in Appendix~\ref{app:refactoring}.

To validate \opensc\ and the refactoring work, the W7-X conductor is modeled in \opensc\ as a four-component conductor as shown in figure~\ref{fig:w7x-opensc-model} and benchmarked against a simulation carried out in THEA~\cite{risse_thermohydraulic_2024}. To make the comparison as fair as possible, both models use the same material characteristics, conductor geometry, number of mesh elements, time stepper and current transient. To that end, an option to scale the applied magnetic flux density proportionally to the source current dump has been added to \opensc\ to match THEA's model in this respect (see Sec.~\ref{subsec:3d-bfield}).

\begin{figure}[tbp]
    \centering
    \includegraphics[width=1\linewidth]{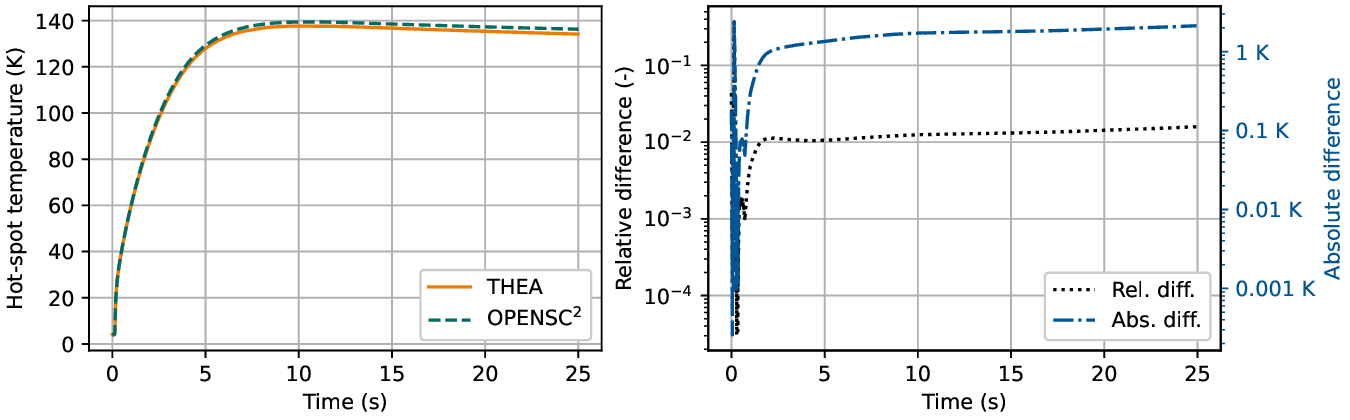}
    \caption{Hot-spot temperature over time in the W7-X strand, computed by THEA (orange, solid) and \opensc\ (green, dashed) (left), and the relative (black, dotted) and absolute (blue, dash-dotted) difference between the two results (right). After the first two seconds, the difference stays between 1.0 and 1.6\,\%, i.e.\ between 1.0 and \SI{2.1}{\kelvin}.}
    \label{fig:thea-verification-hot-spot}
\end{figure}

\begin{figure}
    \centering
    \includegraphics[width=1\linewidth]{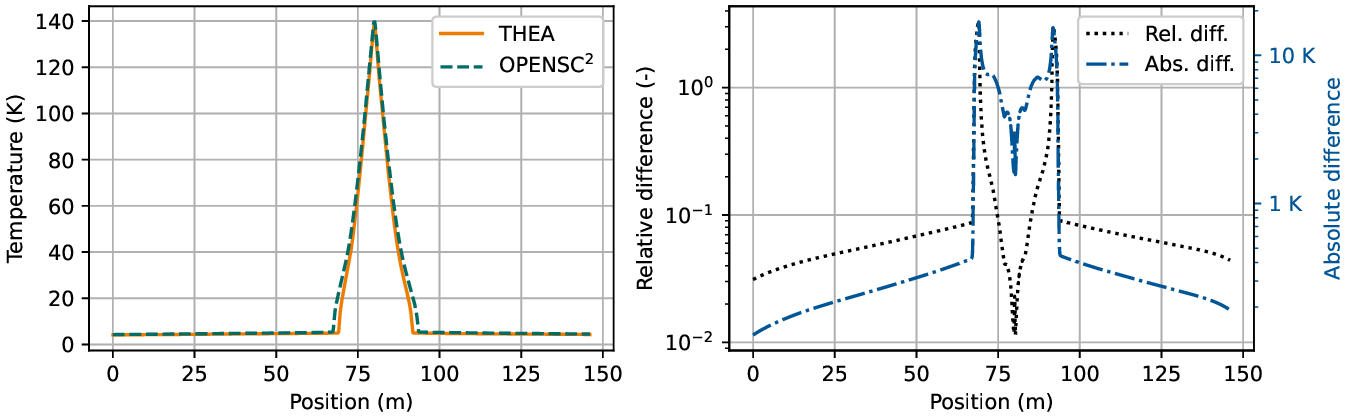}
    \caption{Temperature along the W7-X strand at $t=9\,$s, computed by THEA (orange, solid) and \opensc\ (green, dashed) (left), and the relative (black, dotted) and absolute (blue, dash-dotted) difference between the two results (right). Both differences are largest at the quench fronts, where the temperature gradient is very steep.}
    \label{fig:thea-verification-spatial-T}
\end{figure}

Figure~\ref{fig:thea-verification-hot-spot} shows the temporal evolution of the hot-spot temperature in the W7-X strand. The two models agree closely over the entire transient: with exception of the steep initial rise during the first two seconds, the relative difference stays between 1.0 and 1.6\,\%, which corresponds to an absolute difference of 1.0 to \SI{2.1}{\kelvin}.
Figure~\ref{fig:thea-verification-spatial-T} shows the corresponding spatial temperature distributions along the conductor at $t=9\,$s. Outside the normal zone, the two results differ by 0.1 to \SI{0.4}{\kelvin}, and at the hot spot by \SI{1.7}{\kelvin} (1.2\,\%). The largest differences, up to \SI{16.8}{\kelvin}, occur at the quench fronts: the normal zone computed by \opensc\ extends about \SI{1.5}{\meter} further at each end than that of THEA, and the steep temperature gradient translates this offset of the front position into a large local difference.

All in all, \opensc\ has been successfully benchmarked against THEA. The comparison also demonstrates the effect of the computational efficiency improvement of Sec.~\ref{subsec:efficiency}: on a standard working machine, the 25\,s transient runs in \opensc\ in 19\,min, compared with 22\,h before the refactoring and 15\,min for THEA, bringing the open-source \opensc\ implementation into the same range as the established commercial THEA software.

\section{Thermal and Electromagnetic Effects in the Winding Pack}\label{sec:2d}
\begin{figure}
    \centering
    \includegraphics[width=.7\textwidth]{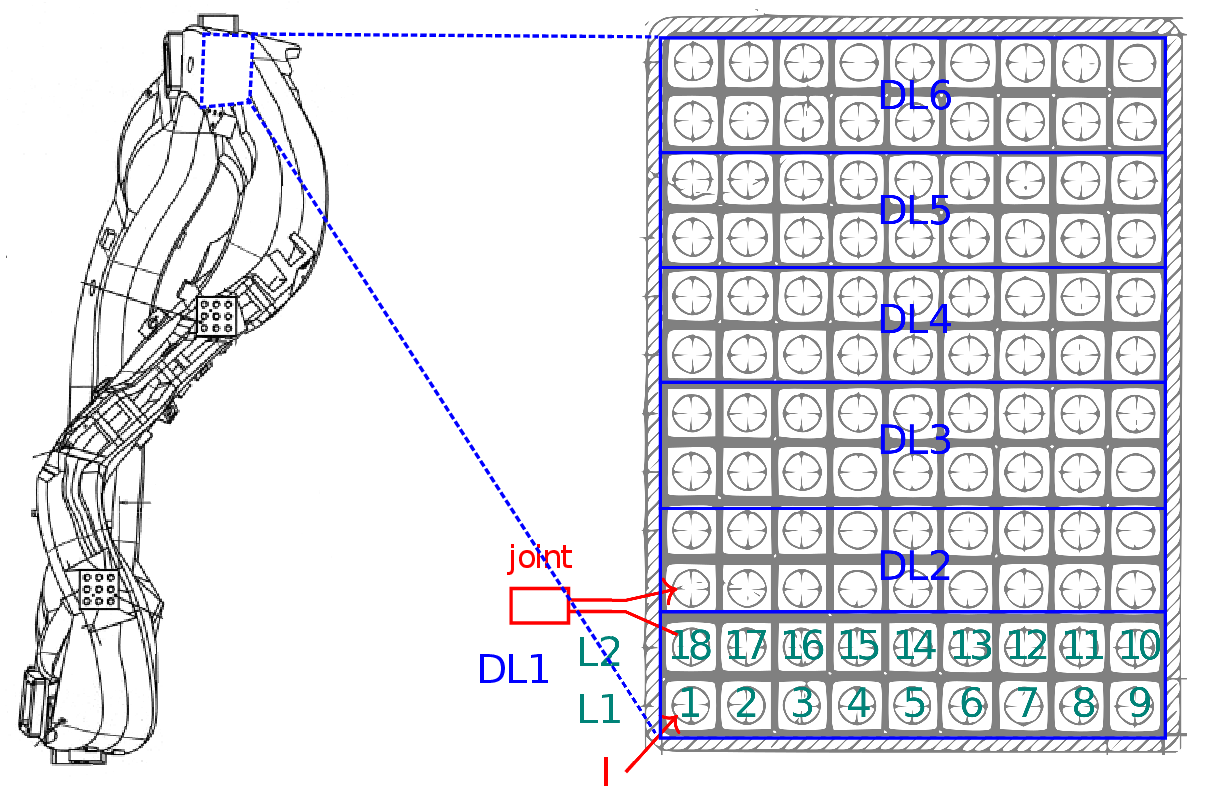}
    \caption{3D sketch of the W7-X non-planar coil of type 1 (left) and its winding pack cross-section (right): the coil consists of six double layer (DL) windings with nine turns per layer (L), so 18 turns per DL. The DLs are electrically connected in series via joints.}
    \label{fig:w7x-winding-pack}
\end{figure}

Stellarator coils consist of multiple CICC windings which form a winding pack (WP). Figure~\ref{fig:w7x-winding-pack} shows as an example the WP cross-section of a non-planar W7-X coil: it consists of six double layer (DLs) windings with nine turns per layer (L), so 18 turns per DL. The DLs are electrically connected in series via joints. A purely 1D quench model as implemented in \opensc\ is usually considering only one DL as done in Sec.~\ref{subsec:validation-against-thea}, which is then modeled as one unraveled straight piece of cable. However, this approach is missing several physical effects that play a role in the context of a WP: transverse heat conduction (Sec.~\ref{subsec:transverse-thermal}), 3D space- and time-dependent magnetic flux density (Sec.~\ref{subsec:3d-bfield}) and induced electromagnetic effects (Sec.~\ref{subsec:induced-effects}), which this work adds to the original \opensc\ implementation as discussed in the following.

\subsection{Transverse Heat Conduction Coupling}\label{subsec:transverse-thermal}
\begin{figure}[tbp]
    \centering
    \includegraphics[width=0.85\linewidth]{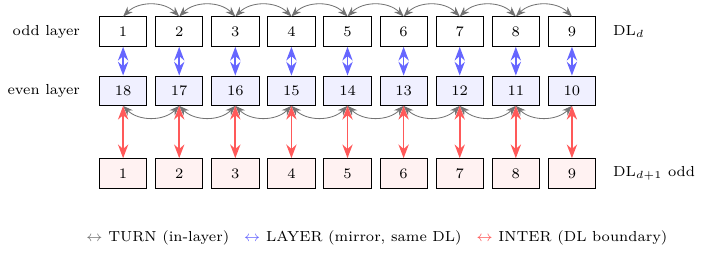}
    \caption{Transverse heat coupling topology of one W7-X DL (numbers are turn indices $k$, even-layer turns labeled by their $k=19-\text{vertical segment}$). Nine LAYER mirrors couple odd-layer turn $k$ to even-layer turn $k$ of the same DL; sixteen TURN pairs connect adjacent turns within each layer; nine INTER pairs face the even layer of DL$_d$ to the odd layer of DL$_{d+1}$.}
    \label{fig:transv_thermal}
\end{figure}

The 1D model of Sec.~\ref{sec:model} conducts heat only along the conductor axis. Within the WP, however, every turn is in contact with its neighbors in the same layer and the adjacent layer, separated only by the turn insulation, so that a hot spot heats the surrounding turns through the insulation as well, as shown in figure~\ref{fig:transv_thermal}. The resulting transverse heat flux is small compared to the longitudinal one, but it is the mechanism by which a quench can be transferred to a turn that carries no heat along its own axis, and it is therefore indispensable for resolving the secondary quenches observed in the W7-X coil tests~\cite{baldzuhn_coil_2006}.

\begin{figure}[tp]
    \centering
    \includegraphics[width=1\linewidth]{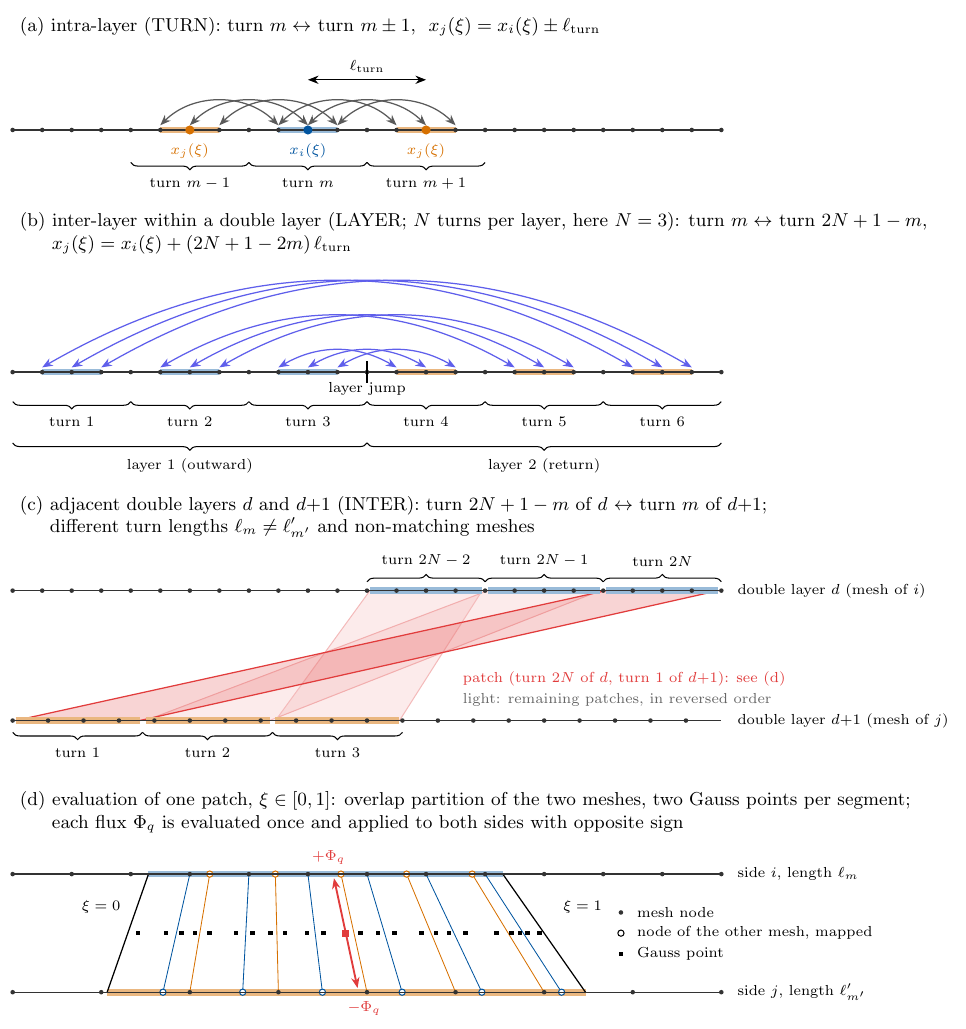}
    \caption{Transverse conduction patches on the 1D conductor meshes. A patch pairs one turn of a solid component $i$ (blue) with the adjacent turn of its partner $j$ (orange); arrows and bands connect positions of equal patch coordinate $\xi$, colored by contact type as in figure~\ref{fig:transv_thermal}. (a) Intra-layer contact, (b) inter-layer contact within a DL of $N$ turns per layer (here $N=3$), (c) contact between adjacent DLs $d$ and $d+1$, i.e.\ partners on the meshes of different conductors with different turn lengths. (d) Evaluation of one patch: the flux is computed once per Gauss point of the overlap partition of the two meshes and applied to both sides with opposite sign.}
    \label{fig:patch-mapping}
\end{figure}

In the 1D setting, however, two points that are neighbors in the 3D WP are far apart along the axial coordinate as shown in figure~\ref{fig:patch-mapping}: the neighboring turn in the same layer lies a turn length $\ell_\mathrm{turn}$ further along the same conductor, whereas the neighboring turn of the adjacent layer, whose turns progress in the opposite radial direction within a DL, lies at a mirrored position, either on the same conductor or on the conductor of the adjacent DL. Transverse heat conduction is thus a nonlocal coupling within and between the 1D conductor models of the coil. The coupling is described by patches: turn $m$ of a solid component $i$, starting at the axial position $s_m$ and of length $\ell_m$, is paired with a turn $m'$ of a partner component $j$, starting at $s'_{m'}$ and of length $\ell'_{m'}$. The partner turn follows from the winding geometry,
\begin{equation}
    m' = \left\lbrace \begin{array}{ll}
        m \pm 1    & \text{turn-to-turn contact within a layer } (j = i), \\
        2N + 1 - m & \text{inter-layer contact within a DL } (j = i), \\
        2N + 1 - m & \text{contact between adjacent DLs } (j \neq i),
    \end{array} \right.
    \label{eq:partner-turn}
\end{equation}
where $N$ is the number of turns per layer and turns are counted from the inlet of each conductor. A patch is parametrized by $\xi \in [0,1]$, with the positions on the two sides 
\begin{equation}
    x_i(\xi) = s_m + \xi \ell_m, \quad x_j(\xi) = s'_{m'} + \xi \ell'_{m'}.
\end{equation}
Since all turns run around the coil in the same circumferential direction, both sides are traversed in the same direction. For turns of equal length, $x_j = x_i + (m'-m)\ell_\mathrm{turn}$. The turns of adjacent DLs differ in length so that the two sides of such a patch have different lengths and, in general, non-matching meshes. 

The patch has the geometric factor
\begin{equation}
    g = \frac{w_\mathrm{ins}}{t_\mathrm{ins}} \frac{\ell_m + \ell'_{m'}}{2}
\end{equation}
in \si{\meter}, where $w_\mathrm{ins}$ is the contact width and $t_\mathrm{ins}$ the insulation thickness between two turns. 
For steady conduction through the insulation layer, the heat flux density is constant, and integrating Fourier's law across the thickness gives 
\begin{equation}
    q_{\mathrm{tr}} = \frac{1}{t_\mathrm{ins}} \int\limits_{T_i}^{T_j} \lambda(T) \, \dd T = \frac{\Lambda(T_j) - \Lambda(T_i)}{t_\mathrm{ins}}, \qquad \Lambda(T)=\int\limits_{T_0}^{T} \lambda(T') \, \dd T',
\end{equation}
with the thermal conductivity $\lambda$ in \si{\watt\per\meter\per\kelvin}, the conductivity integral $\Lambda$ in \si{\watt\per\meter} and the temperature $T_k$ on side $k \in \{i, j\}$ of the patch. This is exact for any temperature dependence of $\lambda$; no mean temperature has to be chosen. The lower limit $T_0$ drops out of the difference. The power that side $i$ receives from side $j$ is then obtained by
\begin{equation}
    \phi(\xi) = g [\Lambda(T_j(\xi))-\Lambda(T_i(\xi))]  = g \bar{\lambda}(\xi)[T_j(\xi) - T_i(\xi)], \qquad \bar{\lambda} = \frac{\Lambda(T_j)-\Lambda(T_i)}{T_j-T_i}
    \label{eq:secant-conductivity}
\end{equation}
with $T_k(\xi) = T_k(x_k(\xi))$ and $\bar{\lambda} \rightarrow \lambda(T)$ for $T_j \rightarrow T_i$. 

In the discrete setting, the temperature of a solid component is piecewise linear on its mesh, with the nodal shape functions $N_k^{(i)}$ and $N_r^{(j)}$ of the two sides. The weak form of the source gives the nodal powers in \si{\watt},
\begin{equation}
	F^{(i)}_k = \int_0^1 N^{(i)}_k\big(x_i(\xi)\big)\,\phi(\xi)\,\mathrm{d}\xi,
	\qquad
	F^{(j)}_r = -\int_0^1 N^{(j)}_r \big(x_j(\xi)\big)\,\phi(\xi)\,\mathrm{d}\xi.
	\label{eq:tr_weak}
\end{equation}
Both integrals are evaluated with the same quadrature on the overlap partition of the patch, i.e.\ the union of the mesh nodes of $i$ and $j$ inside the patch and the two patch ends, all expressed in $\xi$ (figure~\ref{fig:patch-mapping}d). On each segment of this partition both temperature fields and all shape functions are linear, and a two-point Gauss-Legendre rule with points $\xi_q$ and weights $w_q$, $\sum_q w_q = 1$, is applied. With the interpolation matrices $(\mathbf{P}_i)_{qk} = N^{(i)}_k(x_i(\xi_q))$ and $(\mathbf{P}_j)_{qr} = N^{(j)}_r(x_j(\xi_q))$, the temperatures of the two sides at the quadrature points are $\mathbf{P}_i\mathbf{T}_i$ and $\mathbf{P}_j\mathbf{T}_j$. The flux is evaluated once per quadrature point, $\Phi_q = w_q\,\phi(\xi_q)$, and distributed to the nodes of both sides,
\begin{equation}
    \mathbf{F}^{(i)} = \mathbf{P}_i^\top \bm{\Phi}, \qquad \mathbf{F}^{(j)} = -\mathbf{P}^\top_j \bm{\Phi}, \qquad \bm{\Phi} = \mathbf{D}\,(\mathbf{P}_j \mathbf{T}_j - \mathbf{P}_i \mathbf{T}_i ),
    \label{eq:tr_discrete}
\end{equation}
with $\mathbf{D} = \mathrm{diag}(w_q \, g \, \bar{\lambda}_q)$, where $\bar{\lambda}_q$ is the secant conductivity~\eqref{eq:secant-conductivity} evaluated with the two temperatures at $\xi_q$. Due to the partition-of-unity property of the nodal shape functions, it follows that
\begin{equation}
    \sum_k F^{(i)}_k = \sum_q \Phi_q = -\sum_r F^{(j)}_r,
    \label{eq:tr_conservation}
\end{equation}
i.e.\ the integrated powers of the two sides cancel exactly. This holds for any pair of turn lengths, any two meshes and any position of the patch ends within an element, and it does not depend on the accuracy of the quadrature, because both sides use the same values $\Phi_q$. With $\mathbf{B} = [\,-\mathbf{P}_i \;\; \mathbf{P}_j\,]$, the nodal powers of the patch can be written as
\begin{equation}
    \begin{bmatrix} \mathbf{F}^{(i)} \\ \mathbf{F}^{(j)} \end{bmatrix}
    = -\,\mathbf{B}^\top \mathbf{D}\,\mathbf{B}
    \begin{bmatrix} \mathbf{T}_i \\ \mathbf{T}_j \end{bmatrix}.
    \label{eq:tr_operator}
\end{equation}
Since $\bar{\lambda}_q > 0$, the matrix $\mathbf{B}^\top \mathbf{D}\,\mathbf{B}$ is symmetric and positive semi-definite: at every quadrature point the heat flows from the warmer to the colder side, so that the exchange is dissipative. For a patch within one component ($j = i$), both sets of nodal powers are added to the same component and their sum vanishes. Finally, the nodal powers are converted into the nodal linear power $q_{\mathrm{tr},k} = F_k/h_k$ in \si{\watt\per\meter}, with $h_k$ the lumped length of node $k$, i.e.\ half the length of each adjacent element. It is added to the heat source term~\eqref{eq:heat-source-term}, whose consistent load then deposits exactly $\sum_k F_k$ in the component.

An implicit treatment of the patches would couple mesh nodes that are a turn length apart in every row of the system of equations, destroying the banded structure that the solver in Sec.~\ref{subsec:field-network-coupling} relies on. The exchange is therefore evaluated explicitly: all patches are evaluated once per time step from the temperatures that resulted from the previous time step and enter the right-hand side. The sensitivity of the W7-X results to the time step is quantified in Appendix~\ref{subsec:time-step-sensitivity}.
All conductors linked by patches advance with a common time step. For conductors sharing a hydraulic network this is already required by the monolithic step discussed in Sec.~\ref{subsec:field-network-coupling}.

\subsection{3D Magnetic Flux Density}\label{subsec:3d-bfield}
\begin{figure}[tbp]
    \centering
    \includegraphics[width=.9\linewidth]{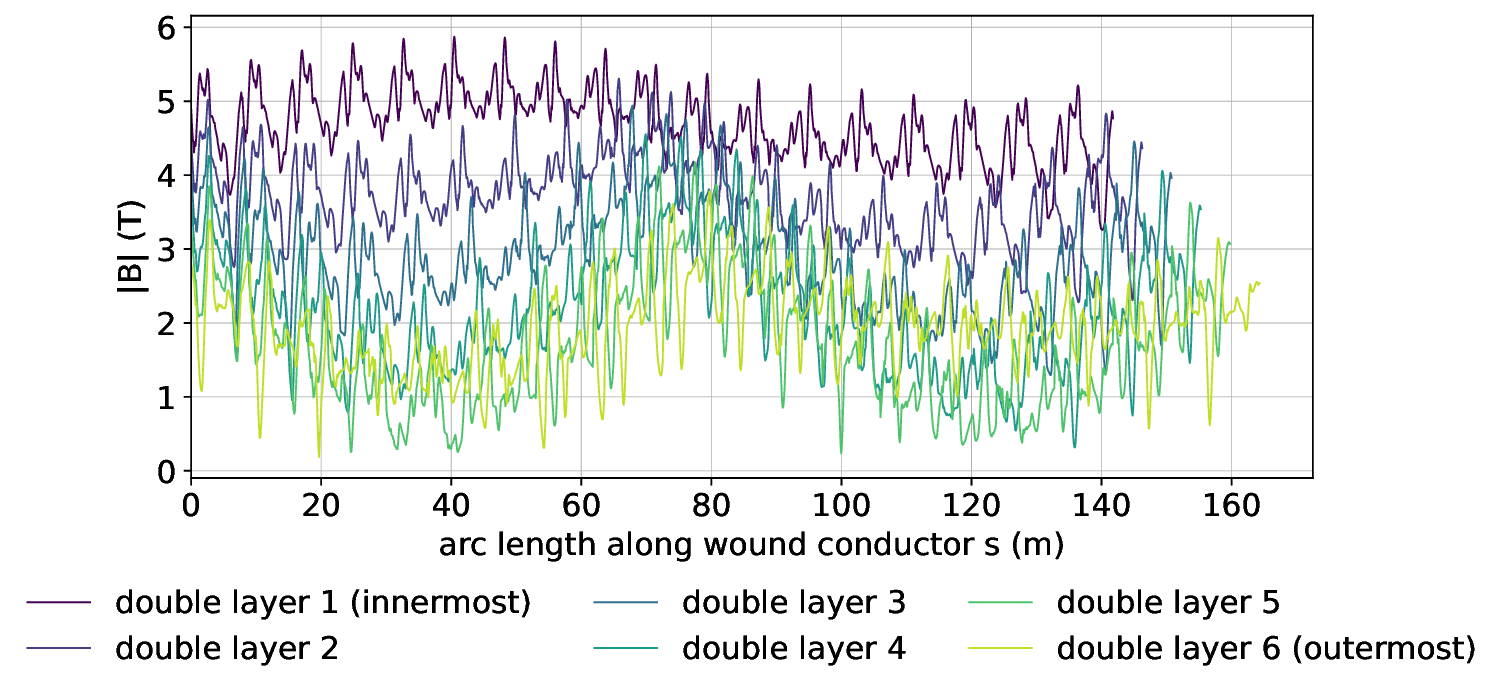}
    \caption{Magnitude of the magnetic flux density along the DLs of a non-planar W7-X coil.}
    \label{fig:bfield-along-dls}
\end{figure}

The critical current density of the superconductor, and with it the current-sharing temperature $T_\mathrm{cs}$ at which Ohmic heating sets in, depend on the local magnetic flux density $B$. In \opensc, $B$ along a conductor was originally prescribed as constant or as a linear profile between inlet and outlet, which is adequate for a straight cable sample but not for a stellarator coil: Figure~\ref{fig:bfield-along-dls} shows that $B$ strongly varies along the winding of a non-planar W7-X coil. Thus, the temperature margin $T_\mathrm{cs}-T$ varies just as strongly in space, and the location at which a quench initiates, the position of the hot spot and the susceptibility of neighboring turns to a secondary quench are all set by the local $B$ rather than by the heating location alone. Moreover, during a current dump, $B$ collapses together with the current, which raises $T_\mathrm{cs}$ everywhere along the conductor, thus benefiting recovery.

\begin{figure}[tbp]
    \centering
    \includegraphics[width=.95\linewidth]{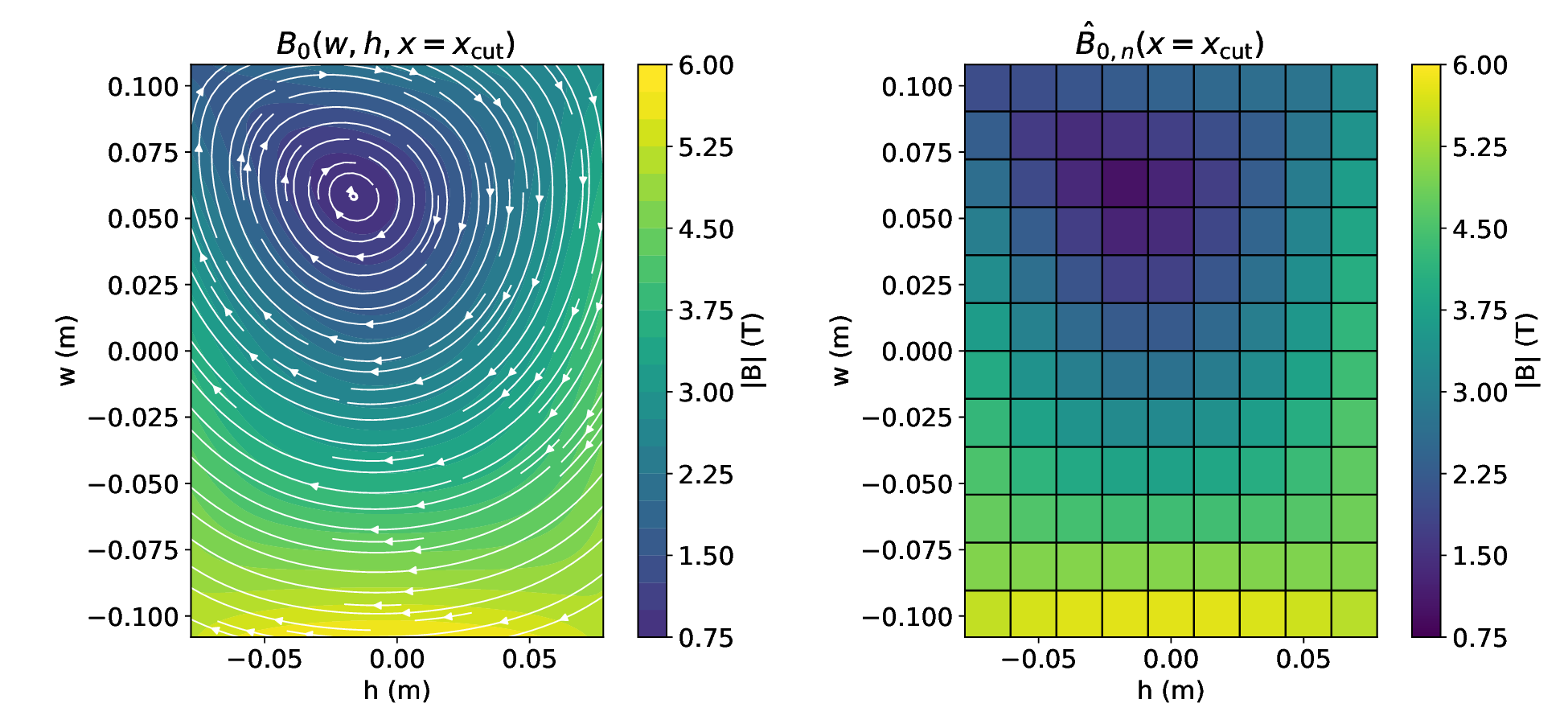}
    \caption{Magnetic flux density in the winding pack of a non-planar W7-X coil at a particular cut-plane at $x=x_\mathrm{cut}$. Left: continuous field $B_0$; right: peak field $\hat{B}_{0,n}$ per turn $n$.}
    \label{fig:w7x-field-in-wp}
\end{figure}

$B$ is therefore imposed along each conductor as space- and time-dependent data. To that end, the magnitude of $B$ at steady state, $B_0(w,h,x) = |\vec{B}_0(w,h,x)|$ at the nominal current $I_0$, is calculated within the 3D coil structure by employing the Landreman-Hurwitz method based on Biot-Savart's law~\cite{landreman_efficient_2025}. Herein, $(w, h, x)$ forms a Bishop frame~\cite{bishop_there_1975}, with $w$ and $h$ the transverse coordinates across the radial and the toroidal width of the WP, respectively, and $x$ the arc length of the coil. Being a rotation-minimizing frame, it follows the WP without twist, so that the turn cross-sections are fixed in $(w,h)$ along the coil.
Next, the WP cross-section in the normal plane of the Bishop frame is subdivided into the cross-sections $\Omega_n = [w_n^-, w_n^+] \times [h_n^-, h_n^+]$ of the individual turns $n$, and the peak magnetic flux density over each turn cross-section is taken along the arc length,
\begin{equation}
     \hat{B}_{0,n}(x) = \max\limits_{(w,h) \in \Omega_n} B_0(w,h,x).
    \label{eq:turn-peak-field}
\end{equation}
Figure~\ref{fig:w7x-field-in-wp} shows $B_0$ (left) and $\hat{B}_{0,n}$ (right) per turn $n$ at a particular cut-plane at $x=x_\mathrm{cut}$ for a non-planar W7-X coil. The innermost layer is located at the bottom.

Temporally, the applied $B$ in each turn is scaled with the source current,
\begin{equation}
    B_n(x,t) = \hat{B}_{0,n}(x) \frac{I(t)}{I_0},
\end{equation}
and then enters the model through the critical surface by determining the critical current density and current-sharing temperature and, thus, whether a strand element is superconducting, current-sharing or normal-conducting (see Sec.~\ref{subsec:coupled-model}). A consequence of the high spatial variation of $B$ and therefore $T_\mathrm{cs}$ is that not every point along the DLs is susceptible to quench initiation by imposing a heat disturbance; so one has to carefully choose the initiation point according to the location with a high $B$ $\leftrightarrow$ low $T_\mathrm{cs}$.

\subsection{Induced Electromagnetic Effects}\label{subsec:induced-effects}
A current dump collapses $B$ over the entire coil within the quench dump time constant $\tau$, so that $\partial_t B$ is large everywhere in the coil at once. The time-varying field induces currents in every conducting element of the cable: magnetization currents within the superconducting filaments; coupling currents between filaments and strands flowing through the resistive matrix and the strand contacts; and eddy currents in the resistive matrix and jacket~\cite{wilson_superconducting_1987}. All three lead to heat losses that are not confined to the quench, but exist distributed in the whole WP for the duration of the transient.

Following the classical literature on transient losses in multi-filamentary conductors~\cite{wilson_superconducting_1987, campbell_general_1982} and electromagnetic field theory~\cite{simonyi_foundations_1963, smythe_static_1968}, the hysteresis, coupling and eddy losses per unit length are 
\begin{align}
    q_{\mathrm{hyst}} &= \frac{2}{3\pi}\,J_c(T,B)\,d_{\mathrm f}\,|\partial_t B|\,A_{\mathrm{sc}}, \\
    q_{\mathrm{coup}} &= \frac{n\tau_{\mathrm{eff}}}{\mu_0}\,|\partial_t B|^2A_{\mathrm{str}}, \\
    q_{\mathrm{eddy}} &= \sigma(T,B)\,|\partial_t B|^2\,C_{\mathrm{geo}},
\end{align}
with $d_\mathrm{f}$ the filament diameter, $A_{\mathrm{sc}}$ and $A_{\mathrm{str}}$ the superconductor and strand cross-sections, $n\tau_{\mathrm{eff}}$ the effective coupling time constant of the cable, which combines the inter-filament and inter-strand contributions, $\mu_0$ the permeability of vacuum, $\sigma(T,B)$ the electrical conductivity of the eddy-current path including its magnetoresistance, from the material library in \opensc, and $C_{\mathrm{geo}}=\int_{A}\xi^{2}\,\mathrm dA$ the second moment of the conducting cross-section about its centroidal axis parallel to the field, $\xi$ being the transverse coordinate perpendicular to $\vec B$.

The hysteresis and coupling losses are deposited in the strand component, the eddy current loss in the jacket and mixed strand components, each entering the heat source term~\eqref{eq:heat-source-term} of the respective solid component. All three are evaluated once per thermal time step with the temperature of the previous time step.

\section{Hydraulic Field-Network Coupling}\label{sec:hydraulic-network}
The coupling of a field model to a lumped-parameter network is an established technique in computational electromagnetics~\cite{schops_winding_2013}: the field model, discretized by finite elements, is embedded as a device into an electric circuit whose branches carry resistors, inductors, capacitors and sources, and the resulting differential-algebraic system is assembled by modified nodal analysis (MNA)~\cite{ho_modified_1975}. The device exchanges nodal potentials and branch currents with the circuit through its terminals, either with the voltage impressed and the current returned (voltage-driven) or vice versa (current-driven)~\cite{schops_winding_2013}.

This section transfers that construction to hydraulics: the network equations are derived from the transmission-line model for pipelines~\cite{wylie_fluid_1978} in the form of an MNA (Sec.~\ref{subsec:hydraulic-network-equations} and~\ref{subsec:comparison-to-circuits}), the channel model of Sec.~\ref{subsec:opensc-hydraulic-model} is reformulated in the variables $(\dot{m},p,T)$ so that its port unknowns coincide with those of the network (Sec.~\ref{subsec:mass-flow-formulation}), and channels and network are coupled through pressure continuity and mass conservation at the ports into a single monolithic system (Sec.~\ref{subsec:field-network-coupling}). 

To the best of our knowledge, our addition of this feature into \opensc\ represents the first open-source implementation of such a hydraulic field-network coupling for quench simulation.

\subsection{Hydraulic Network Equations}\label{subsec:hydraulic-network-equations}
A hydraulic network is represented by a graph consisting of $\mathcal{N}$ nodes and $\mathcal{E}$ branches, whose connections are described by an incidence matrix $\mathbf{A} \in \mathbb{R}^{\mathcal{N}\times\mathcal{E}}$ with entries 
\begin{equation}
    a_{ij} = \left\lbrace \begin{array}{ll}
        -1 & \text{if $i$ is the starting node of branch $j$}, \\
        +1 & \text{if $i$ is the ending node of branch $j$}, \\
        0 & \text{otherwise}.
    \end{array} \right. 
    \label{eq:incidence-matrix}
\end{equation}
The primary variables in a hydraulic network are the nodal pressures $\mathbf{p} \in \mathbb{R}^\mathcal{N}$ in \si{\pascal} located on the graph nodes and the branch mass flows $\dot{\mathbf{m}} \in \mathbb{R}^\mathcal{E}$ in \si{\kilo\gram\per\second} located on the graph branches.

The network model follows most of the assumptions made in~\cite{matko_pipeline_2000}:
\begin{enumerate}
    \item the fluid is homogeneous, i.e.\ the density $\varrho$ is constant along the branches;
    \item the flow is viscous, i.e.\ shear stresses in a moving fluid are taken into account; 
    \item  resistive network branches are isenthalpic, i.e.\ pressure is dropped by friction without exchanging work, so the fluid leaves the branch with the same enthalpy it entered with, and its temperature shifts by the Joule-Thomson coefficient $(\beta T-1)/(\varrho c_p)$;
    \item  the flow is one-dimensional, i.e.\ all fluid characteristics depend on the axial coordinate along the pipes.
\end{enumerate}
The hydraulic network equations can be derived similarly to the electric circuit equations. For the first equation, the starting point is the continuity equation, 
\begin{equation}
    \partial_t \varrho + \nabla \cdot (\varrho \vec{v}) = 0.
    \label{eq:hydraulic-continuity-equation}
\end{equation}
Note that it is structurally identical to the electric current continuity equation, which is a starting point for the electric circuit equations. From~\eqref{eq:hydraulic-continuity-equation}, the nodal continuity equation in the hydraulic regime is derived as described in Appendix~\ref{app:nodal-continuity-equation} as
\begin{equation}
    \mathbf{C}_\mathrm{h} \ddt{\mathbf{p}} - \mathbf{A} \dot{\mathbf{m}} = \dot{\mathbf{m}}_\mathrm{src} + \dot{\mathbf{m}}_\mathrm{exp},
    \label{eq:hydraulic-nodal-continuity-equation}
\end{equation}
where $\dot{\mathbf{m}}_\mathrm{src}$ gathers mass flow sources in the network, $\dot{\mathbf{m}}_\mathrm{exp}$ represents the mass source from thermal expansion of the node content defined by~\eqref{eq:advective-heating}, and $\mathbf{C}_\mathrm{h} \in \mathbb{R}^{\mathcal{N}\times\mathcal{N}}$ is the hydraulic capacitance matrix, which is a diagonal matrix whose entries represent the hydraulic capacitance in \si{\kilo\gram\per\pascal} per node $n$, 
\begin{equation}
    C_{\mathrm{h},n} = V_n \varrho_n \kappa_{s,n} = V_n \varrho_n \left( \kappa_{T,n} - \frac{\beta^2_n T_n}{\varrho_n c_{p,n}} \right) = \frac{V_n}{c_n^2}
    \label{eq:hydraulic-capacitance}
\end{equation}
with $V_n$ being the volume of node $n$, $\kappa_{s,n}$ the isentropic compressibility and $c_n$ the speed of sound of its content. In this sense, the hydraulic capacitance represents a volume in which fluid can gather and be stored, much like a capacitor for electric charges. 

In electromagnetism, the second circuit equation is derived straightforwardly from Faraday's law; the derivation of an analogous relation in the hydraulic regime is slightly more involved. The starting point is the momentum conservation law,
\begin{equation}
    \varrho \left( \partial_t \vec{v} + (\vec{v}\cdot\nabla) \vec{v} \right) = -\nabla p + \nabla \cdot \vec{\tau} + \varrho \vec{g} \label{eq:momentum-law}
\end{equation}
with the Cauchy stress tensor $\vec{\tau}$ and the gravitational acceleration $\vec{g}$.
From this, the branch momentum equation can be derived as elaborated in Appendix~\ref{app:branch-momentum-equation}, 
\begin{equation}
    \mathbf{L}_\mathrm{h} \ddt{\dot{\mathbf{m}}} + \mathbf{R}_\mathrm{h} \dot{\mathbf{m}} + \mathbf{A}^\top \mathbf{p} = \mathbf{p}_\mathrm{src}.
    \label{eq:branch-momentum-equation}
\end{equation}
Herein, $\mathbf{p}_\mathrm{src}$ represents pressure sources from pumps and elevation, $\mathbf{L}_\mathrm{h}$ is the inertance matrix gathering the hydraulic inductances (called inertances) $L_{\mathrm{h},b}$ per branch $b$ in \si{\per\meter}, which are defined as
\begin{equation}
    L_{\mathrm{h},b} = \frac{\ell_b}{A_b}
    \label{eq:hydraulic-inductance}
\end{equation}
with $\ell_b$ being the $b$-th branch's length and $A_b$ the $b$-th branch's area;
$\mathbf{R}_\mathrm{h}$ is the resistance matrix gathering the hydraulic resistances $R_{\mathrm{h},b}$ per branch $b$ in \si{\per\meter\per\second}, which are defined as 
\begin{equation}
    R_{\mathrm{h},b} = \frac{2f\ell_b}{D_\mathrm{h}\varrho A^2} |\dot{m}_b| .
    \label{eq:hydraulic-resistance}
\end{equation}
Lastly, \eqref{eq:hydraulic-nodal-continuity-equation} and~\eqref{eq:branch-momentum-equation} can be expressed together as a coupled system of equations reading 
\begin{equation}
    \begin{bmatrix}
        \mathbf{C}_\mathrm{h} & \mathbf{0} \\
        \mathbf{0} & \mathbf{L}_\mathrm{h}
    \end{bmatrix}
    \ddt{}
    \begin{bmatrix}
        \mathbf{p} \\ \dot{\mathbf{m}}
    \end{bmatrix} + 
    \begin{bmatrix}
        \mathbf{0} & -\mathbf{A} \\
        \mathbf{A}^\top & \mathbf{R}_\mathrm{h}
    \end{bmatrix}
    \begin{bmatrix}
        \mathbf{p} \\ \dot{\mathbf{m}}
    \end{bmatrix} =
    \begin{bmatrix}
        \dot{\mathbf{m}}_\mathrm{src} + \dot{\mathbf{m}}_\mathrm{exp} \\ \mathbf{p}_\mathrm{src}
    \end{bmatrix}.
    \label{eq:hydraulic-network-system}
\end{equation}

\subsection{Comparison to Electric Circuit Equations}\label{subsec:comparison-to-circuits}
The hydraulic network equations~\eqref{eq:hydraulic-network-system} can be identified as an instance of modified nodal analysis (MNA)~\cite{ho_modified_1975}. Writing the MNA of an electric circuit containing capacitors, resistors, inductors and voltage sources in the notation of~\cite{schops_winding_2013}, 
\begin{align}
    \mathbf{A}_\mathrm{C} \mathbf{C}_\mathrm{q} \mathbf{A}_\mathrm{C}^\top \ddt{\mathbf{u}} + \mathbf{A}_\mathrm{R} \mathbf{G}_\mathrm{R} \mathbf{A}_\mathrm{R}^\top \mathbf{u} + \mathbf{A}_\mathrm{L} \mathbf{i}_\mathrm{L} + \mathbf{A}_\mathrm{V} \mathbf{i}_\mathrm{V} &= -\mathbf{A}_\mathrm{I} \mathbf{i}_\mathrm{s}, \\
    \mathbf{L}_\phi \ddt{\mathbf{i}_\mathrm{L}} - \mathbf{A}_\mathrm{L}^\top \mathbf{u} &= 0, \label{eq:inductive-branch-law} \\
    \mathbf{A}_\mathrm{V}^\top \mathbf{u} &= \mathbf{v}_\mathrm{s},
\end{align}
with lumped-element-wise incidence matrices $\mathbf{A}_*$, the correspondence is $\mathbf{u} \leftrightarrow \mathbf{p}$, $\mathbf{i}_\mathrm{L} \leftrightarrow \dot{\mathbf{m}}$ and $\mathbf{A}_\mathrm{L} \leftrightarrow \mathbf{A}$ up to orientation convention. However, two structural differences can be observed.

First, every capacitance is grounded in a hydraulic network. An electric capacitor stores charge in proportion to the voltage across its two terminals, whereas a hydraulic capacitance stores mass in proportion to the absolute pressure at its node (see Appendix~\ref{app:nodal-continuity-equation}). The incidence matrix of a set of grounded capacitors is the identity matrix, $\mathbf{A}_\mathrm{C} = \mathbf{I}$, such that $\mathbf{A}_\mathrm{C} \mathbf{C}_\mathrm{q} \mathbf{A}_\mathrm{C}^\top$ reduces to the diagonal matrix $\mathbf{C}_\mathrm{h}$ in~\eqref{eq:hydraulic-nodal-continuity-equation}. Consequently and in contrast to an electric circuit of floating capacitors, whose capacitance matrix has a constant vector in its null space, the hydraulic network possesses no floating-potential state, i.e.\ a purely capacitive hydraulic network has a unique solution.

Second, the hydraulic resistance enters the system in a different place than in the electric system. MNA eliminates the current of a purely resistive branch $\mathbf{i}_\mathrm{R}$ via Ohm's law, transforming it into the nodal conductance term $\mathbf{A}_\mathrm{R} \mathbf{G}_\mathrm{R} \mathbf{A}_\mathrm{R}^\top $, but keeps the current of an inductive branch $\mathbf{i}_\mathrm{L}$ as an unknown, resulting in Eq.~\eqref{eq:inductive-branch-law}. However, in hydraulics, the resistance appears in the branch equation~\eqref{eq:branch-momentum-equation} instead of the nodal equation. A consequence is that branches with $L_{\mathrm{h},b} = 0$ turn the corresponding rows of~\eqref{eq:hydraulic-network-system} into algebraic equations. The same is true for $C_{\mathrm{h},n} = 0$ anywhere in the hydraulic network. In such a case, \eqref{eq:hydraulic-network-system} becomes a system of differential-algebraic equations, similar to the electric case.


\subsection{Hydraulic Mass-Flow Formulation}\label{subsec:mass-flow-formulation}
For the hydraulic field model, \opensc\ originally solves~\eqref{eq:original-quasilinear-system} in terms of $(v, p, T)$.
With regard to the previously introduced hydraulic field-network coupling, however, it is more convenient to choose a slightly different set of primary variables, $\widetilde{\mathbf{U}} = [\dot{m}, p, T]^\top$, i.e.\ to replace the velocity with the mass flow $\dot{m} = \varrho(p, T) Av$ in \si{\kilo\gram\per\second}. Therefore, we perform a change of variables~\cite{warming_diagonalization_1975, toro_riemann_2009}, 
\begin{equation}
    \mathbf{U} = g(\widetilde{\mathbf{U}}) : \begin{bmatrix}
        v \\ p \\ T
    \end{bmatrix} \rightarrow \begin{bmatrix}
        \frac{\dot{m}}{\varrho(p, T) A} \\ p \\ T
    \end{bmatrix}.
\end{equation}
To that end, we determine the Jacobian and its inverse,
\begin{equation}
    \mathbf{J} = \frac{\partial \mathbf{U}}{\partial \widetilde{\mathbf{U}}} = \begin{bmatrix}
        (\varrho A)^{-1} & -v \kappa_T & v \beta \\
        0 & 1 & 0 \\
        0 & 0 & 1
    \end{bmatrix}, \quad 
    \mathbf{J}^{-1} = \begin{bmatrix}
        \varrho A & \varrho A v \kappa_T & -\varrho A v \beta \\
        0 & 1 & 0 \\
        0 & 0 & 1
    \end{bmatrix}
\end{equation}
with $\kappa_T = \varrho^{-1} (\partial \varrho / \partial p)_T$ being the isothermal compressibility and $\beta = -\varrho^{-1} (\partial \varrho / \partial T)_p$ the isobaric thermal expansion in \si{\per\kelvin}. By chain rule, the derivative terms transform to $\partial_t \mathbf{U} = \mathbf{J} \partial_t \widetilde{\mathbf{U}}$ and $\partial_x \mathbf{U} = \mathbf{J} \partial_x \widetilde{\mathbf{U}}$. Left-multiplying by $\mathbf{J}^{-1}$ to normalize the mass matrix finally yields 
\begin{equation}
    \partial_t \widetilde{\mathbf{U}} + \widetilde{\mathbf{K}}(\widetilde{\mathbf{U}}) \partial_x \widetilde{\mathbf{U}} + \widetilde{\mathbf{S}}(\widetilde{\mathbf{U}}) \widetilde{\mathbf{U}} = \widetilde{\mathbf{q}}, 
\end{equation}
where 
\begin{equation}
    \widetilde{\mathbf{K}} = \mathbf{J}^{-1} \mathbf{K} \mathbf{J} = \frac{1}{\varrho A}
    \begin{bmatrix}
        2\dot{m} & \varrho A^2 -\kappa_T \dot{m}^2 & \beta \dot{m}^2 \\
        \varrho c^2 & -(\gamma-1)\dot{m} & \varrho \beta c^2 \dot{m}\\
        \phi T & -\phi T \kappa_T \dot{m} & \gamma \dot{m}
    \end{bmatrix}
    \label{eq:transformed-flux-jacobian}
\end{equation}
with $\gamma = \frac{c_p}{c_v} = 1+\phi \beta T$ being the heat capacity ratio, 
\begin{equation}
    \widetilde{\mathbf{S}} = \mathbf{J}^{-1} \mathbf{S} \mathbf{P} = \frac{2f|\dot{m}|}{\varrho A D_\mathrm{h}}
    \begin{bmatrix}
        1 & 0 & 0 \\
        -\frac{\phi \dot{m}}{\varrho A^2} & 0 & 0 \\
        -\frac{\dot{m}}{\varrho^2 A^2 c_v} & 0 & 0 
    \end{bmatrix}
    \label{eq:transformed-source-jacobian}
\end{equation}
and $\widetilde{\mathbf{q}} = \mathbf{J}^{-1} \mathbf{q} = [\dot{m}(\kappa_T \dot{q}_p - \beta \dot{q}_T), \dot{q}_p, \dot{q}_T]^\top = [0, \dot{q}_p, \dot{q}_T]^\top$. Herein, $\mathbf{P} = \mathrm{diag((\varrho A)^{-1}, 1, 1)}$ is the secant satisfying $\mathbf{P}\widetilde{\mathbf U} = g(\widetilde{\mathbf{U}})$ at the linearization point.
Substituting the remaining velocity occurrences yields the transformed hydraulic system in terms of the new primary variable set $(\dot{m}, p, T)$,
\begin{align}
    \partial_t \dot{m} + \frac{2\dot{m}}{\varrho A} \,\partial_x \dot{m}
    + \left(A - \frac{\kappa_T \dot{m}^2}{\varrho A}\right) \partial_x p
    + \frac{\beta \dot{m}^2}{\varrho A} \,\partial_x T
    + \frac{2f|\dot{m}|}{\varrho A D_\mathrm{h}} \dot{m} &= 0, \label{eq:mass-flow-form-1} \\
    \partial_t p + \frac{c^2}{A} \,\partial_x \dot{m}
    - \frac{(\gamma-1)\dot{m}}{\varrho A} \,\partial_x p
    + \frac{\beta c^2 \dot{m}}{A} \,\partial_x T
    - \frac{2f\phi|\dot{m}|\dot{m}^2}{\varrho^2 A^3 D_\mathrm{h}} &= \dot{q}_p, \label{eq:mass-flow-form-2} \\
    \partial_t T + \frac{\phi T}{\varrho A} \,\partial_x \dot{m}
    - \frac{\phi T \kappa_T \dot{m}}{\varrho A} \,\partial_x p
    + \frac{\gamma \dot{m}}{\varrho A} \,\partial_x T
    - \frac{2f|\dot{m}|\dot{m}^2}{\varrho^3 A^3 c_v D_\mathrm{h}} &= \dot{q}_T. \label{eq:mass-flow-form-3} 
\end{align}
After spatial discretization with lowest-order FE basis functions $N_j$ and test functions $N_i$~\cite{braess_finite_2007}, the semi-discrete system of equations reads 
\begin{equation}
    \begin{bmatrix}
        \mathbf{M} & & \\
        & \mathbf{M} & \\
        & & \mathbf{M}
    \end{bmatrix} \ddt{} 
    \begin{bmatrix}
        \dot{\mathbf{m}} \\ \mathbf{p} \\ \mathbf{T}
    \end{bmatrix}
    + 
    \begin{bmatrix}
        \mathbf{K}_{\dot{m}\dot{m}} + \mathbf{S}_{\dot{m}\dot{m}} & \mathbf{K}_{\dot{m}p} & \mathbf{K}_{\dot{m}T} \\
        \mathbf{K}_{p\dot{m}} + \mathbf{S}_{p\dot{m}} & \mathbf{K}_{pp} & \mathbf{K}_{pT} \\
        \mathbf{K}_{T\dot{m}} + \mathbf{S}_{T\dot{m}} & \mathbf{K}_{Tp} & \mathbf{K}_{TT}
    \end{bmatrix}
    \begin{bmatrix}
        \dot{\mathbf{m}} \\ \mathbf{p} \\ \mathbf{T}
    \end{bmatrix}
    = 
    \begin{bmatrix}
        \mathbf{0} \\ \dot{\mathbf{q}}_p \\ \dot{\mathbf{q}}_T
    \end{bmatrix},
    \label{eq:hydraulic-massflow-system}
\end{equation}
with $\dot{\mathbf{m}}, \mathbf{p}, \mathbf{T}$ collecting the nodal unknowns of the mass flow, pressure and temperature in the channel, respectively, and the FE matrices and vectors being~\cite{braess_finite_2007}
\begin{align}
    \left( \mathbf{M} \right)_{ij} = \int\limits_\Omega N_j N_i \dd x , \quad
    \left( \mathbf{K}_{ab} \right)_{ij} = \int\limits_\Omega \widetilde{K}_{ab} \partial_x N_j N_i \dd x , \quad
    \left( \mathbf{S}_{ab} \right)_{ij} = \int\limits_\Omega \widetilde{S}_{ab} N_j N_i \dd x , \quad
    \left( \dot{\mathbf{q}}_* \right)_i = \int\limits_\Omega \dot{q}_* N_i \dd x ,
\end{align}
where $\widetilde{K}_{ab}, \widetilde{S}_{ab}$ represent the respective entries in~\eqref{eq:transformed-flux-jacobian} and~\eqref{eq:transformed-source-jacobian}. A characteristic-speed streamline-upwind diffusion~\cite{hughes_finite_1984} already provided by \opensc\ is applied for stabilizing the system.

\subsection{Verification of the Mass-Flow Formulation}
The $\dot{m}$-formulation of Sec.~\ref{subsec:mass-flow-formulation} replaces the $v$-formulation on which \opensc\ was benchmarked against THEA in Sec.~\ref{subsec:validation-against-thea}. Before it is used in the coupled model, its implementation is verified by the method of manufactured solutions~\cite{oberkampf_verification_2010, roache_code_2001}: a smooth, time-periodic set of fields $\mathbf{W}_\mathrm{M} = (\dot{m}_\mathrm{M}, p_\mathrm{M}, T_\mathrm{M})$ is prescribed, its residual under the continuous equations~\eqref{eq:mass-flow-form-1}-\eqref{eq:mass-flow-form-3} is added to the system as a source term, so that $\mathbf{W}_\mathrm{M}$ is the exact solution of the manufactured problem. The same $\mathbf{W}_\mathrm{M}$ is imposed on the $v$-formulation~\eqref{eq:velocity-system-velocity}-\eqref{eq:velocity-system-temperature} through the derived velocity $v_\mathrm{M} = \dot{m}_\mathrm{M}/(\varrho_\mathrm{M} A)$, so that both formulations solve the same problem and their convergence orders are directly comparable. The manufactured fields and the boundary value problem are given in Appendix~\ref{app:manufactured-solutions}. It also accounts for flow reversal. 

The test problem is a piece of the bundle channel of the W7-X conductor with length $L= 20\,$\si{\meter}, $A=3.75\cdot 10^{-5}\,$\si{\square\meter}, $D_\mathrm{h} = 0.376\,$\si{\milli\meter}, void fraction 0.375, supercritical helium at 5\,\si{\bar} and 4.5\,\si{\kelvin}, an imposed pressure drop of 0.5\,\si{\bar}, and a base mass flow of 2\,\si{\gram\per\second}. Mesh and time step are refined together over four levels, such that the observed convergence order is the combined order of the space-time discretization. The discretization error is computed as the root-mean-square of the point-wise relative difference between the computed and manufactured fields. It is evaluated after one full period $\tau = 0.5$\,\si{\second}.

\begin{figure}[tbp]
    \centering
    \includegraphics[width=1\linewidth]{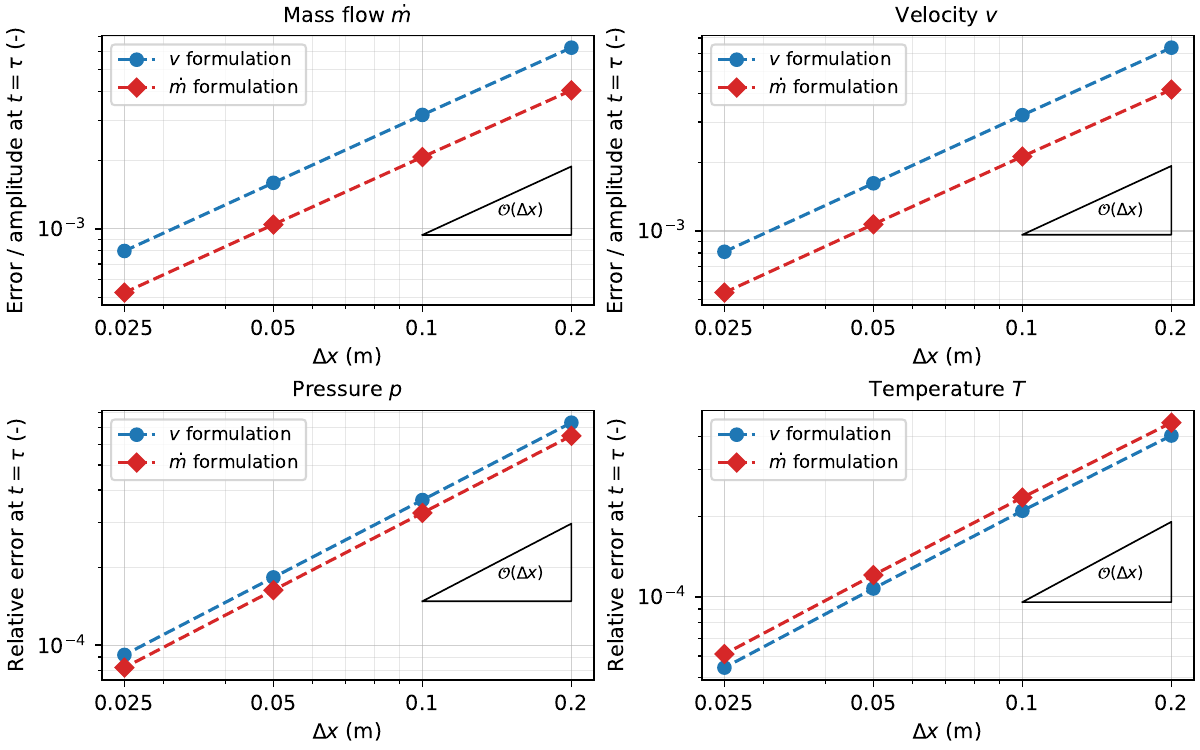}
    \caption{Root-mean-square of the point-wise relative error at $t=\tau$ against the mesh size $\Delta x$ for all fields and both formulations.}
    \label{fig:convergence-study-imposed-m}
\end{figure}

Figure~\ref{fig:convergence-study-imposed-m} shows that both formulations converge with first order in all fields as expected from theory. The $\dot{m}$-formulation is, thus, implemented consistently with~\eqref{eq:mass-flow-form-1}-\eqref{eq:mass-flow-form-3} and is of the same order as the established $v$-formulation. 

\subsection{Interface Conditions and Coupling}\label{subsec:field-network-coupling}
To distinguish between quantities of the field model and the network model, the subscripts f for field and n for network are used in the following. 

Each conductor may be connected through the boundaries of its channels to nodes of the hydraulic network. These connections are called ports. At a port $\ell$ joining the boundary mesh node $i_\ell$ of a channel of conductor $k$ to network node $n_\ell$, two interface conditions hold: pressure continuity, 
\begin{equation}
    p_\mathrm{f}(i_\ell) = p_\mathrm{n}(n_\ell),
\end{equation}
and mass conservation, i.e.\ the mass flow leaving the channel through the ported boundary enters the node exactly. With the channel mass flow counted positive in $x$-direction and the flow sign indicator defined as
\begin{equation}
    \sigma_\ell = \left\lbrace \begin{array}{ll}
        -1 & \text{for a port at $x=0$}, \\
        +1 & \text{for a port at $x=L$},
    \end{array} \right. 
\end{equation}
the mass conservation condition is expressed as
\begin{equation}
    \dot{m}_{\mathrm{port},\ell} = \sigma_\ell \dot{m}_\mathrm{f}(i_\ell),
\end{equation}
and the nodal continuity equation~\eqref{eq:hydraulic-nodal-continuity-equation} is extended by the port mass flows $\dot{m}_{\mathrm{port},\ell}$,
\begin{equation}
    \mathbf{C}_\mathrm{h} \ddt{\mathbf{p}_\mathrm{n}} - \mathbf{A}\dot{\mathbf{m}} - \mathbf{A}_\mathrm{port} \dot{\mathbf{m}}_\mathrm{port} = \dot{\mathbf{m}}_\mathrm{src} + \dot{\mathbf{m}}_\mathrm{exp},
\end{equation}
where $\mathbf{A}_\mathrm{port}$ is the port-to-node incidence matrix, similarly to $\mathbf{A}_\mathrm{sol}$ in~\cite{schops_winding_2013}.
Analogously to the winding matrices in~\cite{schops_winding_2013}, a selector matrix $\mathbf{X} \in \mathbb{R}^{\mathcal{N}_\mathrm{f} \times \mathcal{N}_\mathrm{ports}}$ can be defined with $\mathcal{N}_\mathrm{f}$ being the number of mesh nodes and $\mathcal{N}_\mathrm{ports}$ the number of ports, which incidence-wise associates port nodes of the network with mesh nodes in the field models. Then, the interface conditions can be rewritten in compact form as
\begin{align}
    \mathbf{X}^\top \mathbf{p}_\mathrm{f} &= \mathbf{A}_\mathrm{port}^\top \mathbf{p}_\mathrm{n}, \label{eq:discrete-pressure-condition} \\
    \mathbf{ZX}^\top \dot{\mathbf{m}}_\mathrm{f} &=  \dot{\mathbf{m}}_\mathrm{port}, \label{eq:discrete-massflow-condition}
\end{align}
where $(\mathbf{Z})_{\ell\ell} = \sigma_\ell$ is a diagonal matrix gathering the mass flow signs. Combining the network equations~\eqref{eq:hydraulic-network-system} with the field equations~\eqref{eq:hydraulic-massflow-system} via these interface conditions finally yields the field-network coupled system
\begin{equation}
    \mathcal{M} \ddt{\mathbf{u}} + \mathcal{K} \mathbf{u} = \mathbf{f}
    \label{eq:field-network-coupled-system}
\end{equation}
with the unknowns $\mathbf{u} = [\dot{\mathbf{m}}_\mathrm{f}, \mathbf{p}_\mathrm{f}, \mathbf{T}_\mathrm{f}, \mathbf{p}_\mathrm{n}, \dot{\mathbf{m}}_\mathrm{n}]^\top$, the mass matrix
\begin{equation}
    \mathcal{M} = \begin{bmatrix}
        \mathbf{M} & & & & \\
        & \mathbf{P}_\mathrm{int} \mathbf{M} & & & \\
        & & \mathbf{M} & & \\
        & & & \mathbf{C}_\mathrm{h} & \\
        & & & & \mathbf{L}_\mathrm{h}
    \end{bmatrix}, 
\end{equation}
the stiffness matrix
\begin{equation}
    \mathcal{K} = \begin{bmatrix}
        \mathbf{K}_{\dot{m}\dot{m}} + \mathbf{S}_{\dot{m}\dot{m}} & \mathbf{K}_{\dot{m}p} & \mathbf{K}_{\dot{m}T} & \mathbf{0} & \mathbf{0} \\
        \mathbf{P}_\mathrm{int} (\mathbf{K}_{p\dot{m}} + \mathbf{S}_{p\dot{m}}) & \mathbf{P}_\mathrm{int} \mathbf{K}_{pp} + \mathbf{P}_\Gamma & \mathbf{P}_\mathrm{int} \mathbf{K}_{pT} & - \mathbf{XA}_\mathrm{port}^\top & \mathbf{0} \\
        \mathbf{K}_{T\dot{m}} + \mathbf{S}_{T\dot{m}} & \mathbf{K}_{Tp} & \mathbf{K}_{TT} & \mathbf{0} & \mathbf{0} \\
        -\mathbf{A}_\mathrm{port} \mathbf{ZX}^\top & \mathbf{0} & \mathbf{0} & \mathbf{0} & -\mathbf{A} \\
        \mathbf{0} & \mathbf{0} & \mathbf{0} & \mathbf{A}^\top & \mathbf{R}_\mathrm{h}
    \end{bmatrix}, 
\end{equation}
the load vector $\mathbf{f} = [\mathbf{0}, \mathbf{P}_\mathrm{int} \dot{\mathbf{q}}_p, \dot{\mathbf{q}}_T, \dot{\mathbf{m}}_\mathrm{src} + \dot{\mathbf{m}}_\mathrm{exp}, \mathbf{p}_\mathrm{src}]^\top$, and projector matrices $\mathbf{P}_\Gamma = \mathbf{XX}^\top$ projecting onto port nodes and $\mathbf{P}_\mathrm{int} = \mathbf{I} - \mathbf{P}_\Gamma$ projecting onto interior nodes.

The coupled system is solved monolithically in time using a $\theta$-scheme. The unknowns $\mathbf{u}$ are ordered conductor-wise and network last, which after time discretization results in a time step matrix of the form
\begin{equation}
    \begin{bmatrix}
        \mathbf{A}_1 & \mathbf{0} & \dots & \mathbf{B}_1 \\
        \mathbf{0} & \ddots & & \vdots \\
        \vdots & & \mathbf{A}_K & \mathbf{B}_K \\
        \mathbf{C}_1 & \dots & \mathbf{C}_K & \mathbf D
    \end{bmatrix}
    \begin{bmatrix}
        \mathbf{x}_1 \\ \vdots \\ \mathbf{x}_K \\ \mathbf{x}_\mathrm{net} 
    \end{bmatrix} = 
    \begin{bmatrix}
        \mathbf{b}_1 \\ \vdots \\ \mathbf{b}_K \\ \mathbf{f}_\mathrm{net}
    \end{bmatrix},
\end{equation}
where $\mathbf{A}_k$ is the time step matrix of the $k$-th conductor, $\mathbf{B}_k$ and $\mathbf{C}_k$ are the coupling blocks following from the interface conditions~\eqref{eq:discrete-pressure-condition}-\eqref{eq:discrete-massflow-condition}, and $\mathbf{D}$ is the step matrix of the hydraulic network. To solve this system efficiently, the banded structure of the conductor block matrices is exploited and the system is rewritten into a Schur complement system by eliminating the conductor unknowns exactly by block Gaussian elimination~\cite{golub_matrix_2013}, yielding 
\begin{align}
    \left( \mathbf{D} - \sum\limits_{k=1}^K \mathbf{C}_k \mathbf{A}_k^{-1} \mathbf{B}_k \right) \mathbf{x}_\mathrm{net} &= \mathbf{f}_\mathrm{net} - \sum\limits_{k=1}^K \mathbf{C}_k \mathbf{A}_k^{-1} \mathbf{b}_k, \label{eq:schur-complement} \\
    \mathbf{x}_k &= \mathbf{A}_k^{-1} \left( \mathbf{b}_k - \mathbf{B}_k \mathbf{x}_\mathrm{net} \right).
\end{align}
The inverse $\mathbf{A}_k^{-1}$ is not computed explicitly; instead, for each conductor a single banded factorization is applied which gives the particular solutions. 

Up until now, the pressure and mass flow were treated at the ports. The temperature requires a different treatment. The temperature is imposed at a port only when fluid flows from the node into the channel, $\dot{m}_{\mathrm{port},\ell} < 0$, and is convected out of the channel otherwise. On the network side, the branches are isenthalpic by assumption 3 and carry the upstream enthalpy. The temperature $T_{\mathrm{n},i}$ of a node $i$ with volume $V_i$ obeys the upwinded enthalpy balance
\begin{equation}
    \varrho_i V_i c_{p,i} \ddt{T_{\mathrm{n},i}} = \sum\limits_{\text{inflows}} \dot{m}_\mathrm{in} \left[ \bar{c}_p (T_\mathrm{up}-T_{\mathrm{n},i}) + \Delta h_p \right] + V_i \beta_i T_i \ddt{p_i}, 
    \label{eq:nodal-enthalpy-balance}
\end{equation}
where the sum runs over all branches and ports carrying fluid into the node, $T_\mathrm{up}$ is the upstream node temperature in case of a branch and the channel end temperature $T_\mathrm{f}(i_\ell)$ in case of a port, and the outflows carrying $T_{\mathrm{n},i}$ cancel against mass conservation~\eqref{eq:hydraulic-nodal-continuity-equation}. Zero-volume junctions reduce to a mass-flow-weighted mixing rule. The bracket is the enthalpy difference $h_\mathrm{up} - h_\mathrm{n}$ of~\eqref{eq:enthalpy-difference}, split at the intermediate state $(p_n, T_\mathrm{up})$ into a temperature part and a pressure part, 
\begin{equation}
    h(p_\mathrm{up}, T_\mathrm{up}) - h(p_n, T_n) = \underbrace{ \frac{h(p_n, T_\mathrm{up})-h(p_n,T_n)}{T_\mathrm{up}-T_n}}_{=\bar{c}_p} (T_\mathrm{up}-T_n) + \underbrace{ h(p_\mathrm{up},T_\mathrm{up}) - h(p_n,T_\mathrm{up}) }_{=\Delta h_p},
\end{equation}
where $\bar{c}_p$ is the secant isobaric specific heat of an inflow, which is the mean of $c_p$ along the isobar of the node pressure between the node temperature and the upstream temperature, 
\begin{equation}
   \bar c_p(p_n; T_n, T_{\text{up}})
= \frac{h(p_n, T_{\text{up}}) - h(p_n, T_n)}{T_{\text{up}} - T_n}
= \frac{1}{T_{\text{up}} - T_n} \int_{T_n}^{T_{\text{up}}} c_p(p_n, T)\,\dd T.
\end{equation}

The nodal enthalpy balance~\eqref{eq:nodal-enthalpy-balance} and the monolithic system~\eqref{eq:field-network-coupled-system} are solved in an alternating manner: at $t^{n+1}$, the inflow port temperatures of the channels are imposed from the node temperatures of the previous time step $t^n$, the field-network coupled system~\eqref{eq:field-network-coupled-system} is solved, and the node temperatures are then advanced with~\eqref{eq:nodal-enthalpy-balance} using the new branch and port flows and the new channel end temperatures. 

The coupled solution is thus monolithic in the hydraulic variables $(p, \dot m)$, which are solved simultaneously to solver precision, while the node temperatures are exchanged in a staggered manner. A node temperature changes for two reasons that act on different time scales, see Eq.~\eqref{eq:nodal-enthalpy-balance}: compression of the node content, which follows the pressure and is therefore as fast as the hydraulic transient, and advection of enthalpy into the node, which proceeds at the flow velocity. The fast, pressure-driven part does not appear in the staggered exchange: it was eliminated analytically in Appendix~\ref{app:nodal-continuity-equation} and is contained implicitly in the isentropic hydraulic capacitance~\eqref{eq:hydraulic-capacitance}. Only the advective part lags by one time step, namely the expansion source $\dot{\mathbf{m}}_\mathrm{exp}$ in \eqref{eq:field-network-coupled-system} and the temperatures imposed at the inflow ports, which are both taken from the previous step. The largest lag error occurs when a quench front reaches a port, where it amounts to a delay of one time step. Halving the time step shifts the secondary quenches of the W7-X model by a few milliseconds only (Appendix~\ref{subsec:time-step-sensitivity}).

\begin{table}[p]
    \centering
    \footnotesize
    \caption{Comparison of the extended \opensc\ with the coupled
    field-network models in Flower/SuperMagnet and 4C, as described
    in~\cite{bottura_flower_2003, cryosoft_supermagnet_2021, savoldirichard_4c_2010}.}
    \begin{tabularx}{\linewidth}{@{}p{2.3cm}XXX@{}}
        \toprule
         & Extended \opensc\ (this work)
         & Flower/SuperMagnet
         & 4C \\
        \midrule
        Availability
            & open source (Python)
            & proprietary (CryoSoft)
            & not openly available; built on commercial tools \\ \\
        Conductor model
            & 1D FE, multi-component, multi-conductor
            & THEA: 1D FE, multi-component
            & Mithrandir: 1D FE, multi-component, multi-conductor \\ \\
        Network model
            & lumped network: volume nodes with isentropic capacitance,
              resistive and inertial branches, pumps,
              reservoirs
            & volume nodes with perfect mixing, connected by 1D compressible
              pipes, valves, pumps, compressors, turbines and heat
              exchangers
            & Modelica (Cryogenics library, based on ThermoPower): pipes,
              valves, volumes, circulators, helium bath, heat exchangers,
              controllers \\ \\
        Hydraulic variables
            & $(\dot m, p, T)$ in the conductors, $(p, \dot m)$ in the
              network
            & $(v, p, T)$ in the pipes, $(p, T)$ in the volumes
            & $(p, T, v)$ in the circuit pipes; $(p, \dot m, h)$ at the
              connectors \\ \\
        Field-network coupling
            & monolithic in $(p, \dot m)$ within one system of equations;
              node temperatures staggered by one time step
            & separate processes; explicit exchange at every time step
              without iterations (THEA receives $p$ and $T$, Flower the mass
              and enthalpy fluxes); independent time steps with
              interpolation
            & explicit in time at fixed exchange times; the
              circuit provides $T_\mathrm{in}$, $p_\mathrm{in}$,
              $p_\mathrm{out}$, the winding $\dot m_\mathrm{in}$,
              $\dot m_\mathrm{out}$, $T_\mathrm{out}$ \\ \\
        Thermal coupling between turns
            & direct coupling of adjacent turns through the insulation,
              energy-conserving on non-matching meshes, explicit
            & through a separate heat conduction code (HEATER) coupled to
              the THEA components by temperature and heat flux
            & through 2D FE models of the structures (radial plates, case)
              at selected cross sections, coupled to the jacket by the
              insulation resistance \\ \\
        Electric circuit
            & prescribed current
            & electric network code (POWER): cable current and quench
              resistance
            & no \\ 
        \bottomrule
    \end{tabularx}
    \label{tab:code-comparison}
\end{table}

\begin{figure}[p]
    \centering
    \includegraphics[width=0.8\linewidth]{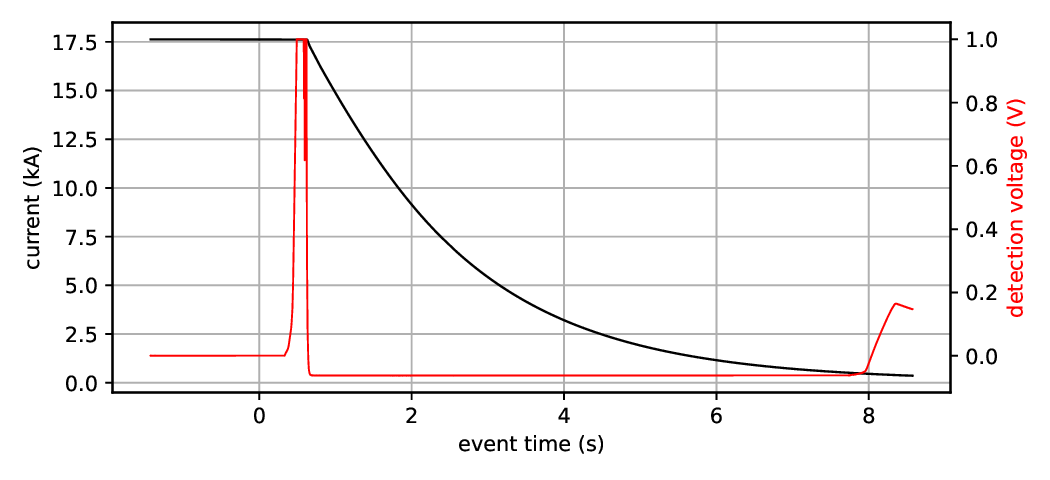}
    \caption{Measured current discharge and detection voltage for the AAB-51 coil.}
    \label{fig:aab51-current-and-voltage}
\end{figure}

\subsection{Comparison to Other Hydraulic Field-Network Coupled Codes}
Table~\ref{tab:code-comparison} places this work in the context of the two existing coupled models Flower/SuperMagnet~\cite{bottura_flower_2003, cryosoft_supermagnet_2021} and 4C~\cite{savoldirichard_4c_2010}. The new elements of the extended \opensc\ code are the open-source implementation, the monolithic coupling of conductors and network in pressure and mass flow instead of an explicit exchange between separate codes, and the direct, energy-conserving thermal coupling between turns without a separate heat conduction solver. In turn, the network of \opensc\ provides lumped volumes, pipes, valves, relief valves and pumps, whereas Flower and 4C additionally offer compressors, turbines, heat exchangers and structural components.

\section{Comparison to W7-X Quench Measurements}\label{sec:validation}
The extended \opensc\ code is applied to the W7-X quench measurement scenario conducted in 2006 at CEA, Saclay~\cite{baldzuhn_coil_2006}, and compared with the measurements of the type 1 non-planar coil AAB-51 therein. 
The W7-X coil tests of 2006 are the only available quench measurements on a full-size stellarator winding pack, but they were not designed for code validation: the external hydraulic circuit and the exact sensor positions are only partly documented, no quantities were measured inside the winding pack, and several operating parameters are uncertain. A quantitative validation in the sense of~\cite{oberkampf_verification_2010} is therefore not possible with these data. Instead, we use them for three purposes: 
\begin{enumerate}
    \item to test whether the winding-pack features move the simulation closer to the measured behavior than the established adiabatic 1D model;
    \item to identify which physical mechanisms (transverse conduction, induced losses, flow redistribution, compression heating) can produce the observed signatures;
    \item to derive from these mechanisms first implications for the design of the coils of future stellarator power plants.
\end{enumerate}
The mechanisms identified in this way are model predictions consistent with the data, not experimentally confirmed effects.

To this end, this section first introduces the measurement setup and the simulation model (Sec.~\ref{subsec:validation-measurement-setup} and Sec.~\ref{subsec:validation-simulation-model}). The quench initiation location is predicted via the current-sharing temperature (Sec.~\ref{subsec:quench-prediction}). The results of a full-feature simulation run are directly compared with the measurement data (Sec.~\ref{subsec:validation-results}), the impact of the single features is discussed (Sec.~\ref{subsec:feature-impacts}), and the compression heating phenomenon is examined in detail (Sec.~\ref{subsec:compression-heating}), before the findings are summarized and first design implications are drawn (Sec.~\ref{subsec:validation-summary} and Sec.~\ref{sec:design-implications}).

\subsection{Measurement Setup}\label{subsec:validation-measurement-setup}
The coil is initially cooled down to a test temperature of 5.7\,\si{\kelvin}. While it is operated at a nominal current of 17.6\,\si{\kilo\ampere}, the inlet helium temperature is increased in steps of 0.1\,\si{\kelvin}, each held until steady state is reached, until a quench occurs. The corresponding inlet temperature is recorded as quench temperature, $T_\mathrm{q} = 6.28\,$\si{\kelvin} in the case of AAB-51. When the voltage threshold of 0.5\,\si{\volt} for the quench detection electronics is surpassed for more than 100\,\si{\milli\second}, the quench protection system triggers a current dump with a time constant of $\tau \approx 2\,$\si{\second} (figure~\ref{fig:aab51-current-and-voltage}); in the test considered here, the dump started about \SI{0.18}{\second} after the threshold was crossed. The recorded quantities are the DL inlet and outlet temperatures, coil casing temperatures, quench detection voltage, coil voltage, helium mass flows and pressures, and stresses. There are no measurements inside the coil, so that the hot-spot temperatures in the windings are unknown. 

\begin{figure}[tp]
    \centering
    \includegraphics[width=1\linewidth]{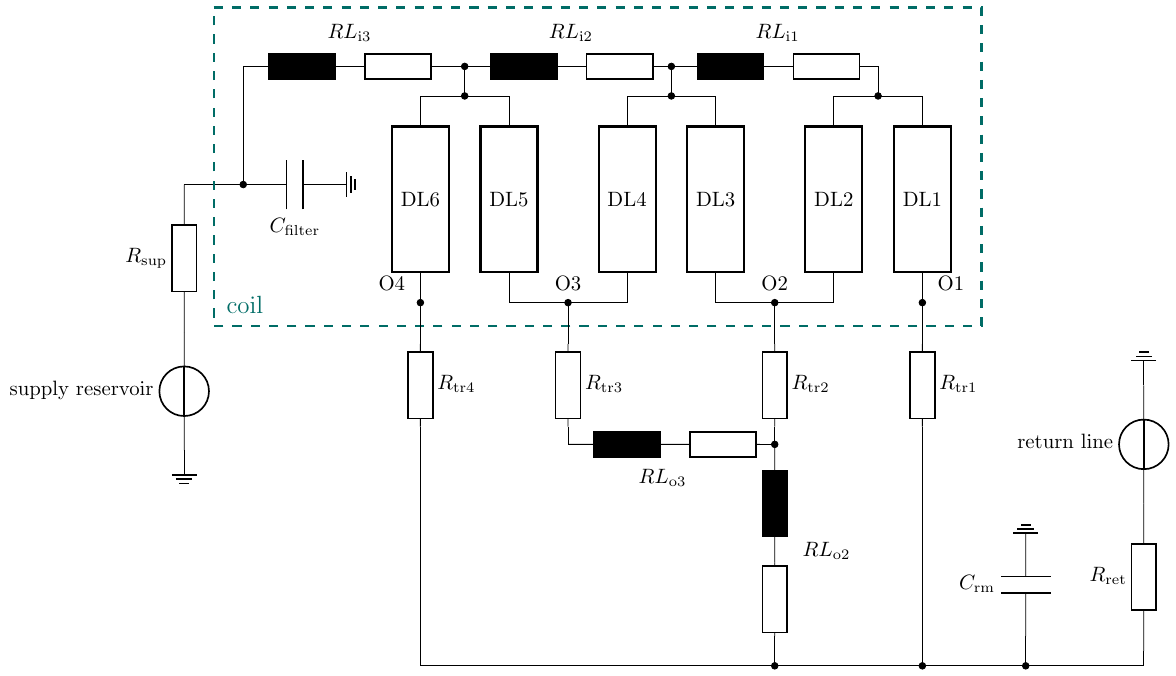}
    \caption{Hydraulic network model for the W7-X coil represented by electric circuit elements following the analogy between electric circuits and hydraulic networks.}
    \label{fig:hydraulic-network}
\end{figure}

\subsection{Simulation Model}\label{subsec:validation-simulation-model}
The coil is modeled in \opensc\ by six conductors representing the six DLs (see figure~\ref{fig:w7x-winding-pack}) with their corresponding winding lengths of 144.5-164.5\,\si{\meter}, discretized by 14 000 FEs each. The DLs are electrically connected in series, which is realized by imposing the same source current $I(t)$ in each DL. Each DL forms one cooling channel: the helium enters a DL at the odd layer, winds through its nine turns to the even layer and returns through its nine turns. 

The DLs are hydraulically connected in parallel as shown in figure~\ref{fig:hydraulic-network}: a common inlet is shared by DL1 \& DL2, DL3 \& DL4, and DL5 \& DL6, while DL2 \& DL3 and DL4 \& DL5 share a common outlet (O2 and O3) and DL1 and DL6 have separate outlets for themselves (O1 and O4), respectively. The resistive-inductive branches represent pipes, while the purely resistive branches represent valves. The hydraulic topology and geometry are known for the coil itself (dashed box), but not exactly for the measurement setup outside of it. For that, we assume a minimalistic external network consisting of a supply and a return reservoir, valves in the supply and return lines ($R_\mathrm{sup}$, $R_\mathrm{ret}$), a capacitance at the return line ($C_\mathrm{rm}$), and a throttle valve in each of the four outlet lines ($R_\mathrm{tr1}$ to $R_\mathrm{tr4}$).

\begin{figure}[p]
    \centering
    \includegraphics[width=0.65\linewidth]{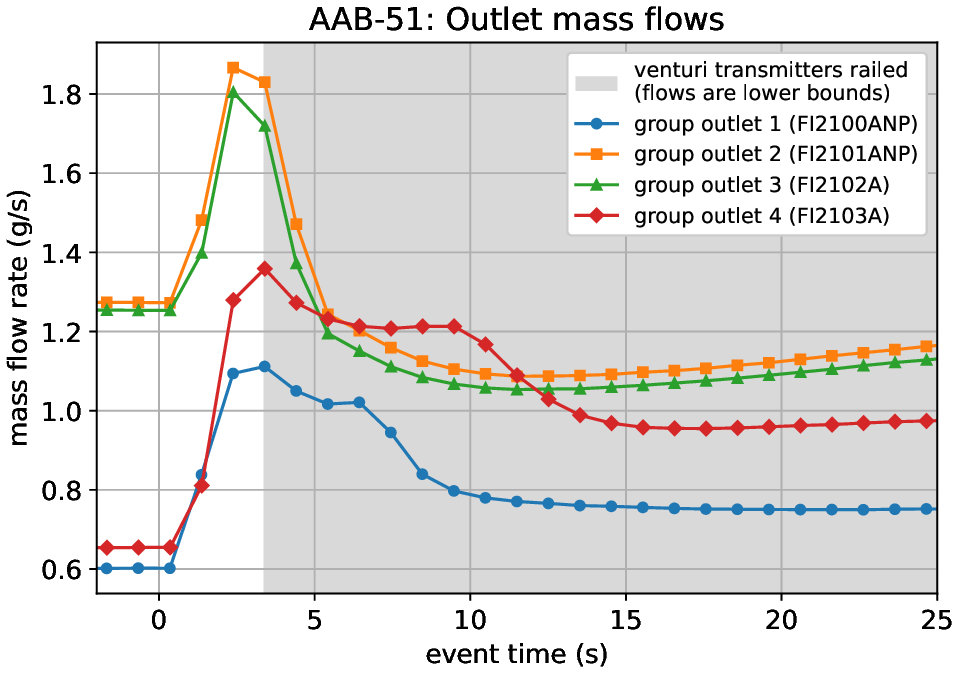}
    \caption{Measured mass flows during the quench event at the outlets of coil AAB-51.}
    \label{fig:massflow-measured}
\end{figure}

The throttle valves are motivated by the mass flows measured at the four outlets (figure~\ref{fig:massflow-measured}). Before the quench, the measured flow is 0.60--\SI{0.65}{\gram\per\second} per DL; afterwards, it rises to about 1.5 to 2 times this value within \SI{3}{\second}. About \SI{3.3}{\second} after the quench (\SI{3.7}{\second} for O4), the venturi transmitters of the outlet lines reach the limit of their range of \SI{1000}{\milli\bar}, so that the mass flows recorded after that time are at best lower bounds of the actual flows. Moreover, a venturi does not register the direction of the flow. The measurement therefore constrains the model in the steady state before the quench and during the first three seconds of the transient.

A model with outlet lines of negligible resistance disagrees with the measurement in both respects: it carries 1.6 to 1.9 times the measured flow before the quench, and its outlet flows are an order of magnitude too high during the transient (Appendix~\ref{sec:network-sensitivity}). This indicates a flow resistance between the DL outlets and the return line that the model of the coil alone does not contain. It is represented by valves with the quadratic characteristic $\Delta p = R_{\mathrm{tr}}\,\dot m^2$, whose resistances are calibrated to the steady state before the quench such that the model carries \SI{0.6}{\gram\per\second} per DL; no transient data enter this calibration. The resulting values and the influence of the throttle valves and of the other assumed elements on the results are given in Appendix~\ref{sec:network-sensitivity}.

The initial temperature is set to $T_0 = 6.1\,\si{\kelvin}$ everywhere. The supply reservoir follows the measured conditions at the winding inlet: its temperature is taken from sensor TT2113C and its pressure from sensor PT2111. If the model has not detected a quench at the time of the measured one, the supply temperature is raised further at \SI{3.6}{\milli\kelvin\per\second}, the slope of the last measured step, until the model detects; the excess reached at that point is kept as a constant offset afterwards. If the computed voltage along a DL surpasses the detection threshold $U_\mathrm{th} = 0.5\,\si{\volt}$, the current is dumped after a delay of 0.184\,\si{\second}, derived from the measured current. The current transient follows the measured one (sensor IAL-C1, see figure~\ref{fig:aab51-current-and-voltage}). The effective coupling loss time constant $n\tau_\mathrm{eff} = 58\,\si{\milli\second}$ is taken from~\cite{takacs_loss_2006}.

\subsection{Quench Initiation Prediction}\label{subsec:quench-prediction}
\begin{figure}[p]
    \centering
    \includegraphics[width=0.75\linewidth]{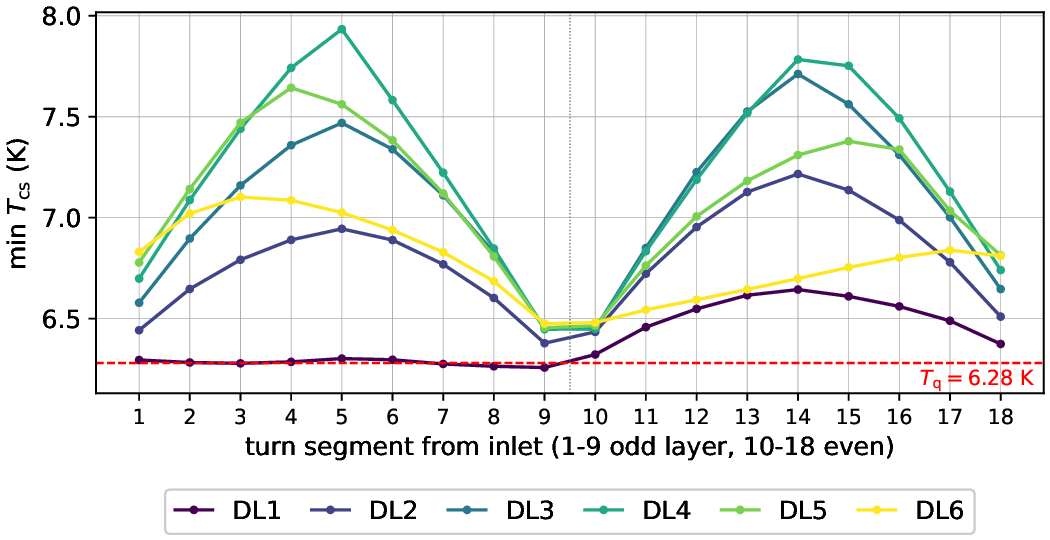}
    \caption{Current-sharing temperature minima for each turn segment and DL in the AAB-51 coil at the test operation point.}
    \label{fig:temperature-margin-minima}
\end{figure}

Since the whole coil is close to current sharing, the local current-sharing temperature $T_\mathrm{cs}$ determines where the quench starts. It is computed from the magnetic flux density of the coil as described in Sec.~\ref{subsec:3d-bfield}, with the critical surface and the conductor parameters given in Appendix~\ref{app:conductor-parameters}. Figure~\ref{fig:temperature-margin-scan} shows that $T_\mathrm{cs}$ is lowest in the first layer of DL1; the minima per turn are shown in figure~\ref{fig:temperature-margin-minima}. The global minimum, $T_\mathrm{cs} = 6.257\,\si{\kelvin}$, lies in the ninth turn of DL1 and is \SI{23}{\milli\kelvin} below the measured quench temperature $T_\mathrm{q}$. The computation thus predicts DL1 as the quench initiation site, in agreement with the measurement, in which the quench was detected by the compensated voltage between DL1 and DL2 (sensor DUP11-3).

\begin{figure}[p]
    \centering
    \includegraphics[width=0.75\linewidth]{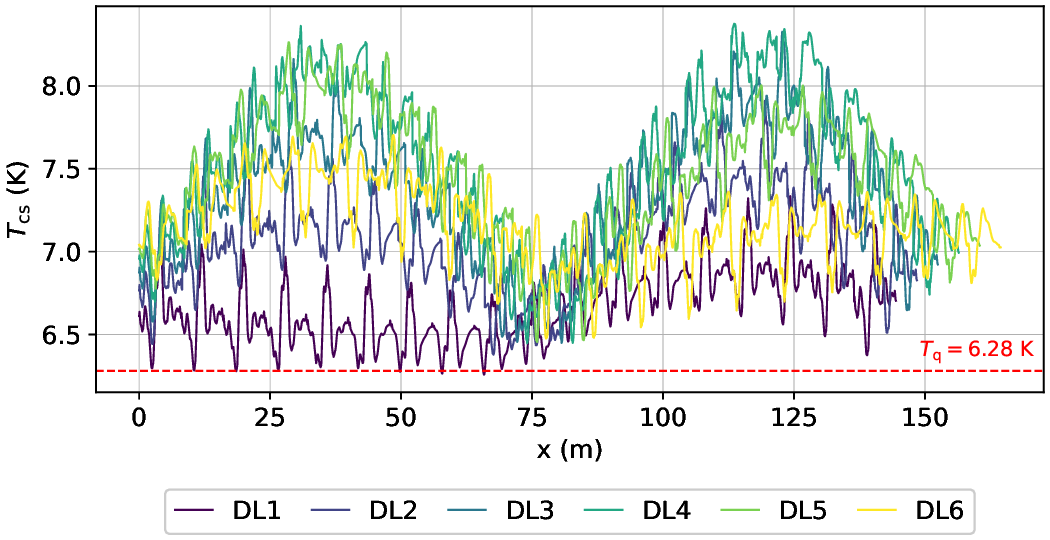}
    \caption{Current-sharing temperature along the windings of the AAB-51 coil at the test operation point.}
    \label{fig:temperature-margin-scan}
\end{figure}

The measurement does not resolve the position within DL1, and neither does the prediction reliably: the minima of $T_\mathrm{cs}$ of the nine turns of the first layer of DL1 differ by less than \SI{0.05}{\kelvin} (figure~\ref{fig:temperature-margin-minima}), and the position of the field map along the winding is uncertain by up to one turn length (Appendix~\ref{subsec:magnetic-field-sensitivity}). The agreement is therefore claimed at the level of the DL only.

\subsection{Full-Feature Simulation Results}\label{subsec:validation-results}
\begin{figure}[tp]
    \centering
    \includegraphics[width=\linewidth]{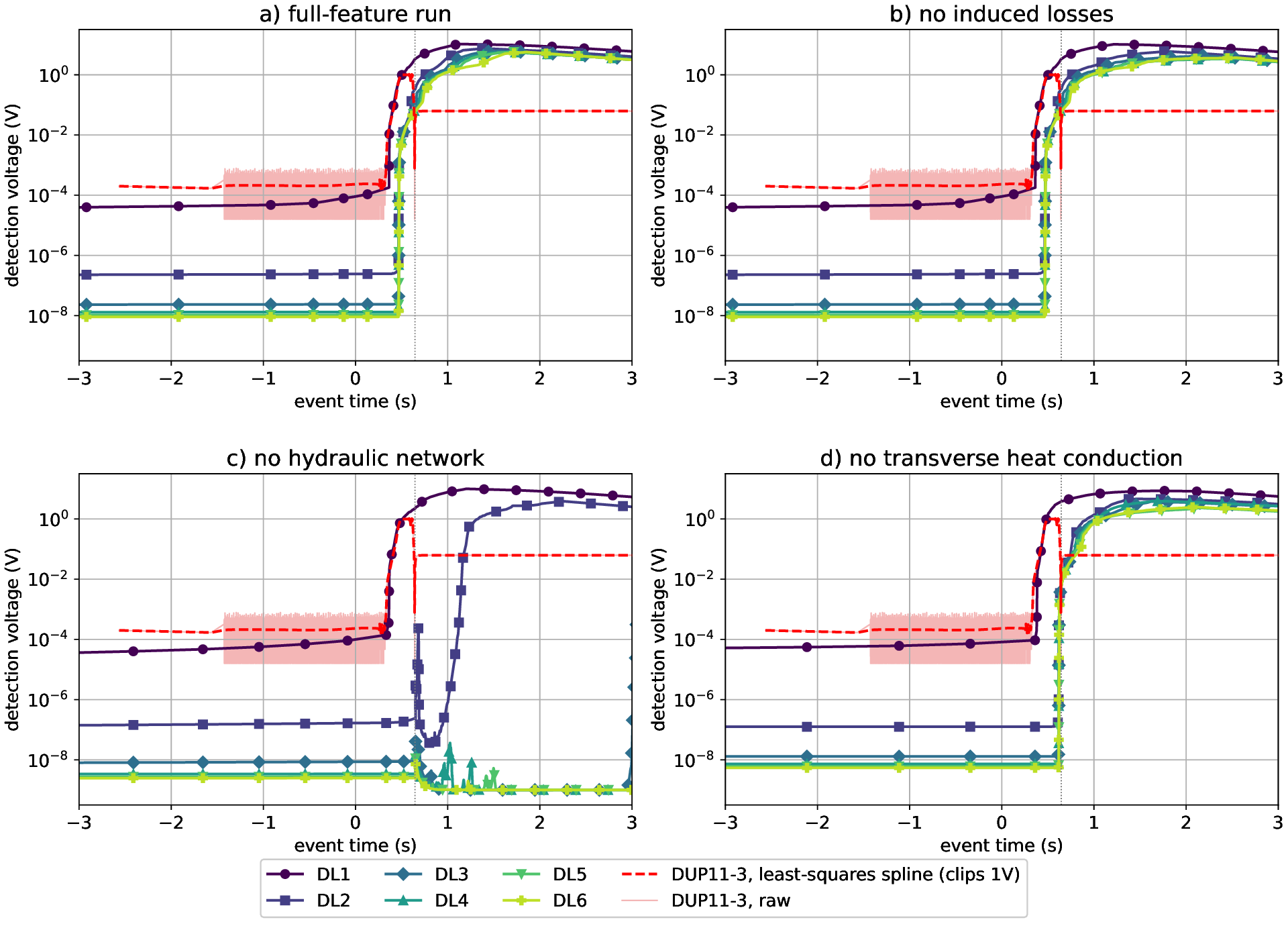}
    \caption{Measured detection voltage vs.~simulated detection voltages.}
    \label{fig:results-detection-voltage}
\end{figure}

\begin{figure}[tp]
    \centering
    \includegraphics[width=0.8\linewidth]{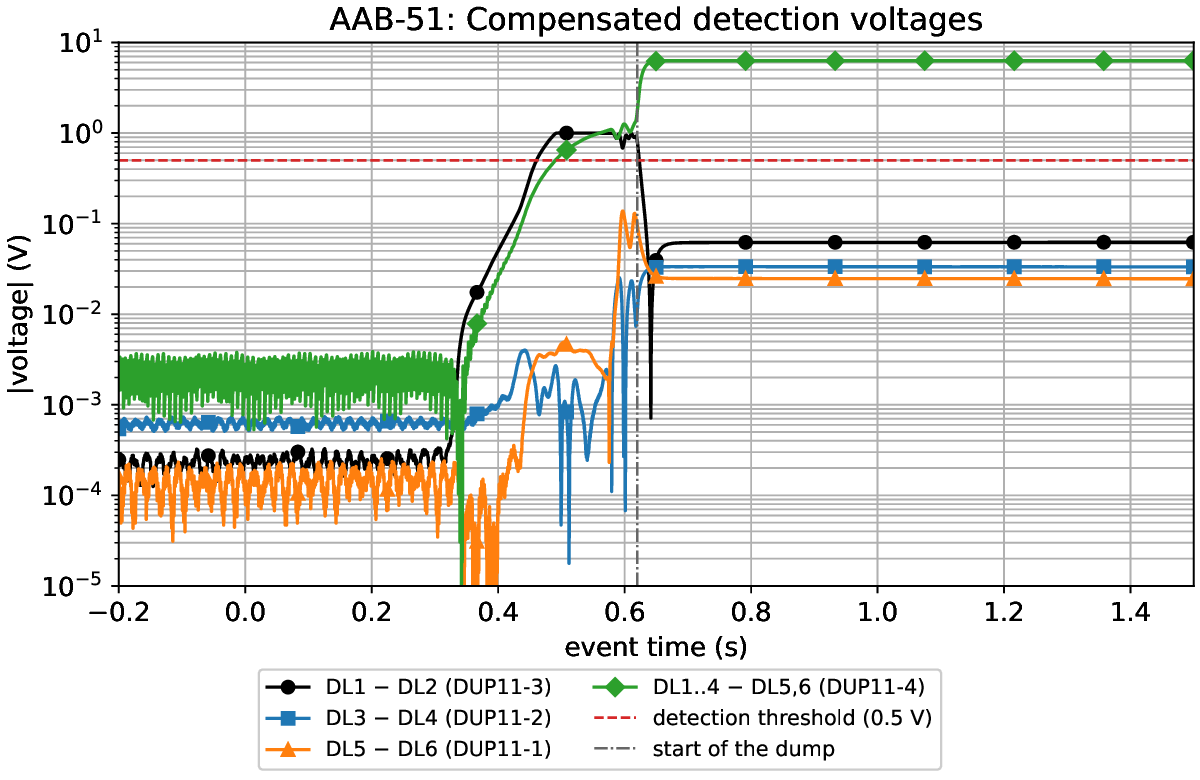}
    \caption{Measured compensated detection voltages of coil AAB-51 (absolute values). Each signal is the difference between the voltages of the indicated DLs or groups of DLs. The signal of DL1 and DL2 saturates at \SI{1}{\volt}; after the start of the current dump, none of the signals is usable.}
    \label{fig:detection-voltages-measured}
\end{figure}

\begin{figure}[tp]
    \centering
    \includegraphics[width=\linewidth]{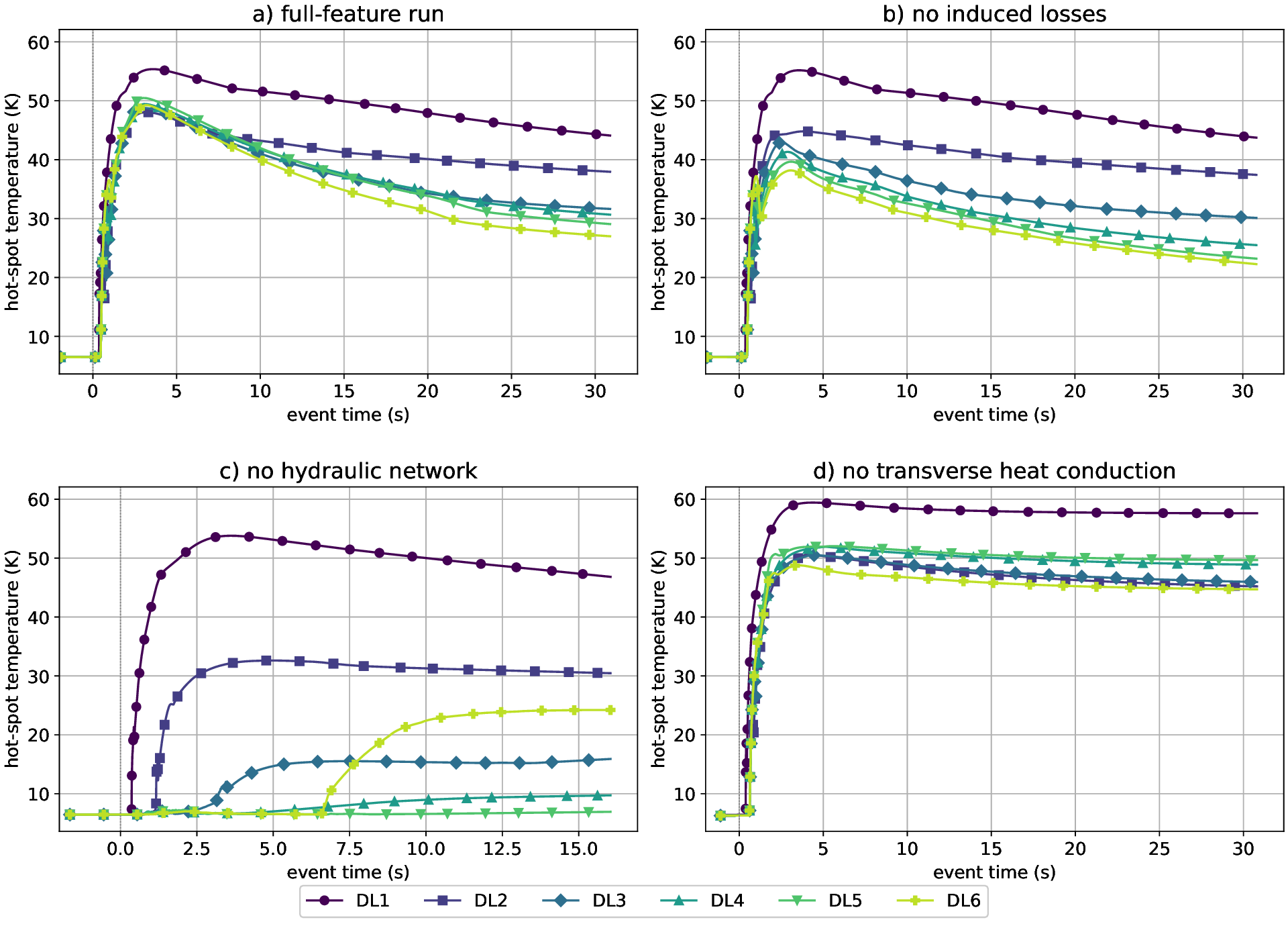}
    \caption{Simulated hot-spot temperatures in the strands of each DL.}
    \label{fig:results-strand-hotspots}
\end{figure}

\begin{figure}[tbp]
    \centering
    \includegraphics[width=0.85\linewidth]{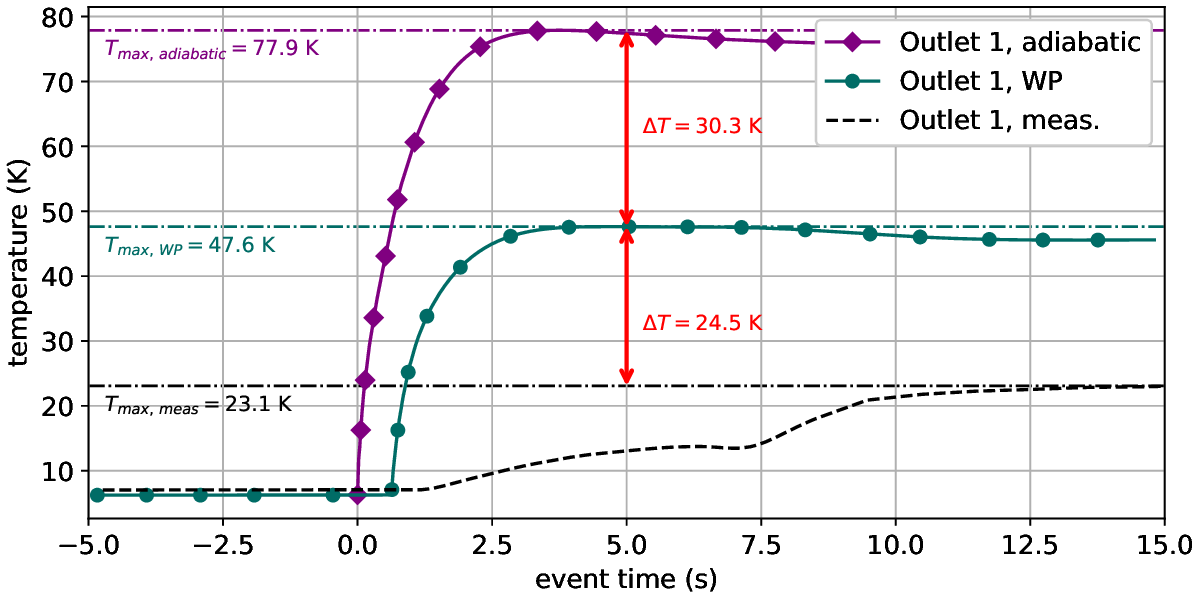}
    \caption{Outlet 1 temperature computed by original \opensc\ (adiabatic model) and extended \opensc\ (WP model, all features) compared to the measurement.}
    \label{fig:results-outlet1-temperature}
\end{figure}

\begin{figure}[tp]
    \centering
    \includegraphics[width=\linewidth]{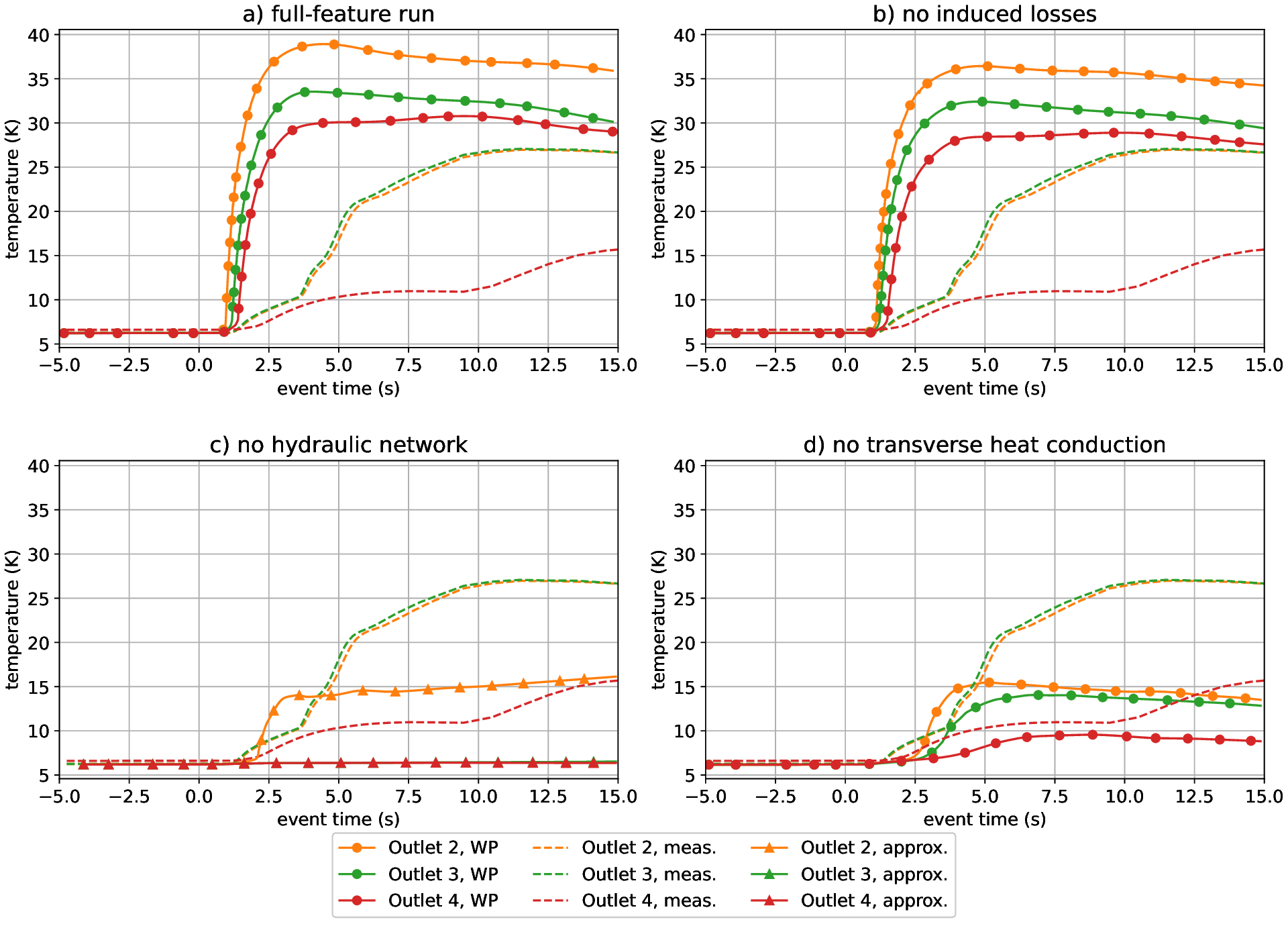}
    \caption{Comparison of simulated and measured outlet temperatures of outlet groups 2-4. In the case of no hydraulic network (c), the outlet temperatures are approximated from the computed boundary temperatures in the individual conductors.}
    \label{fig:results-outlet-temperatures}
\end{figure}

In the simulation, the quench starts in DL1 as predicted in Sec.~\ref{subsec:quench-prediction}. It is detected \SI{96.8}{\second} later than in the measurement. All results are therefore given in an event time $t$ whose origin is the trigger of the fast data acquisition in the experiment. On this axis, the measured detection voltage crosses the threshold $U_\mathrm{th}$ at $t = \SI{0.461}{\second}$, and the measured current dump starts at about \SI{0.65}{\second}. Each simulation is shifted in time such that its own detection, i.e.\ the instant at which the voltage of any DL reaches $U_\mathrm{th}$, coincides with the measured one at $t = \SI{0.461}{\second}$; its current dump then starts \SI{0.184}{\second} later, at $t = \SI{0.645}{\second}$. The origin $t = 0$ thus marks neither the start of the quench nor its detection; where times are given as ``after the quench'' in the following, they refer to this event time. In the following, various computed quantities are compared with the measured ones. The comparison is summarized in the end in table~\ref{tab:comparison-measurement}.

\subsubsection{Detection voltage}
The growth of the detection voltage of DL1 before the current dump, i.e.\ at constant current, agrees with the measured one in shape and slope (see figure~\ref{fig:results-detection-voltage}a). The voltages of DL2 to DL6 start to rise at about \SI{0.5}{\second}, i.e.\ within 0.1 to 0.2\,s after the steep rise in DL1 and before the current dump begins. All DLs thus begin to quench almost simultaneously.

The measurement allows this to be checked to some extent. Figure~\ref{fig:detection-voltages-measured} shows the compensated voltages recorded by the quench detection system, each being the difference between the voltages of two DLs or groups of DLs. The signal of DL1 and DL2 leaves its noise level at \SI{0.33}{\second} and saturates at \SI{0.5}{\second}. The signals of the pairs DL3--DL4 and DL5--DL6 rise at about \SI{0.58}{\second}, i.e.\ while the coil current is still constant, so that these voltages are resistive. The measurement thus indicates that normal zones appeared in DL3 to DL6 about \SI{0.25}{\second} after the quench of DL1, which is of the same order as the 0.2\,s of the model.

This evidence is limited: a compensated signal shows only the difference between the two DLs of a pair, so that the individual onset times and the lengths of the normal zones cannot be derived from it, and the signals are lost with the start of the dump, so that the further spreading of the quench is not recorded.

\subsubsection{Hot-spot temperatures}
\begin{figure}[tp]
    \centering
    \includegraphics[width=\linewidth]{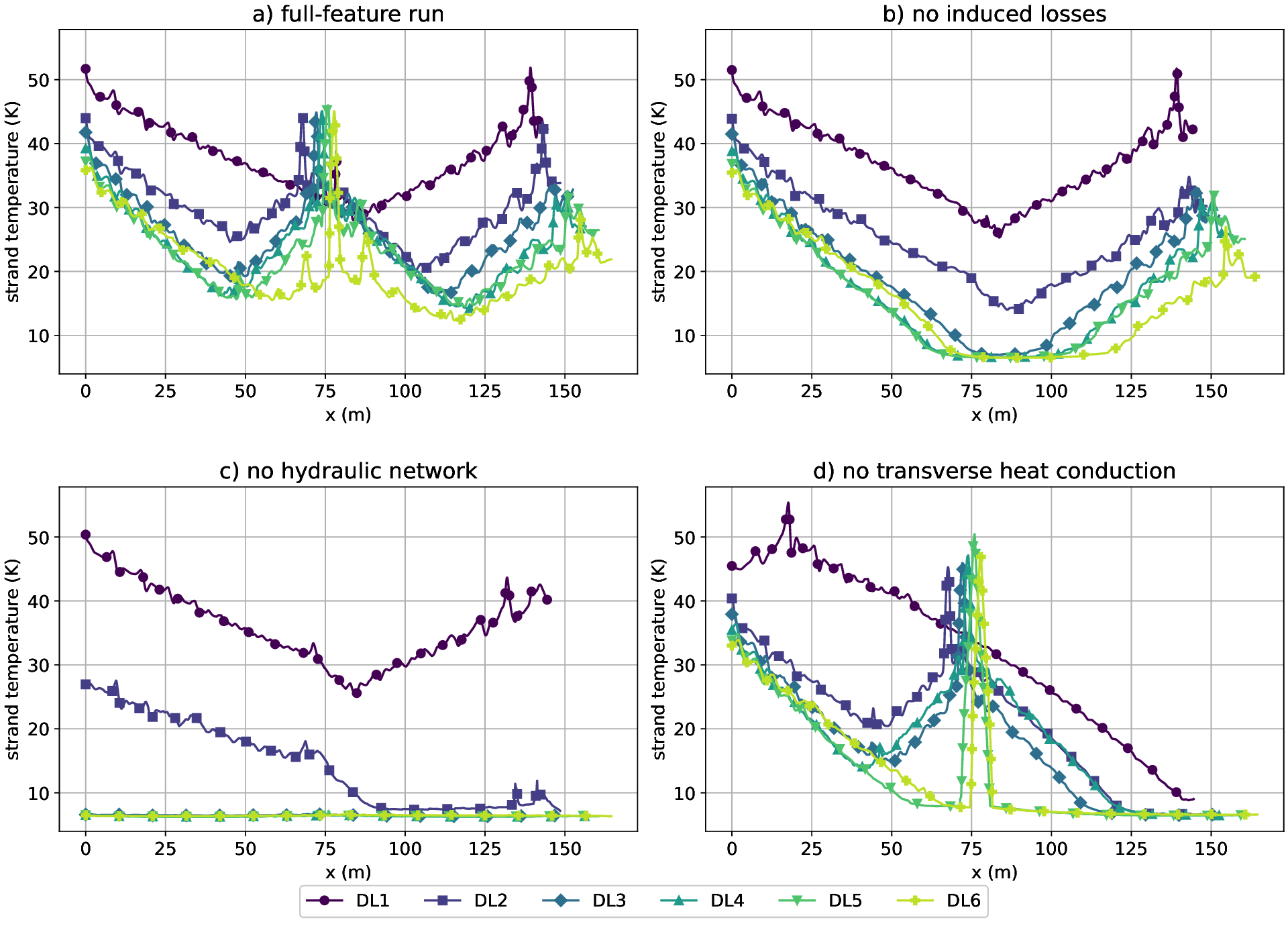}
    \caption{Computed temperatures along the DLs at $t=2\,\si{\second}$.}
    \label{fig:results-temperature-profiles}
\end{figure}

Figure~\ref{fig:results-strand-hotspots}a shows the evolution of the hot-spot temperature in the strands of each DL. The primarily quenched DL1 reaches a maximum of \SI{55.4}{\kelvin} about 3.5\,s after the quench. The other DLs follow closely and reach 48.1 to \SI{50.5}{\kelvin}, i.e.\ only 5 to \SI{7}{\kelvin} less than DL1. At $t = \SI{2}{\second}$, the highest temperatures in DL1 occur at the inlet and close to the outlet (figure~\ref{fig:results-temperature-profiles}a), i.e.\ not at the position of the minimum $T_\mathrm{cs}$, but at the ends of the conductor, where the helium is expelled. These temperatures cannot be compared with measurements, since no sensors were positioned inside the winding pack.

\subsubsection{Outlet temperatures}
Figures~\ref{fig:results-outlet1-temperature} and~\ref{fig:results-outlet-temperatures}a compare the computed outlet temperatures with the measured ones (sensors TT2100A to TT2103A); the values are listed in table~\ref{tab:comparison-measurement}. At the outlet O1 of the primarily quenched DL1 (figure~\ref{fig:results-outlet1-temperature}), the 1D approach of the original \opensc\ code, which models the DL as one straight piece of cable without adjacent turns, overestimates the measured maximum by \SI{54.8}{\kelvin}, and the extended \opensc\ by \SI{24.5}{\kelvin}. In both models, however, the outlet temperature rises within about \SI{2}{\second}, whereas the measured one takes more than \SI{10}{\second}.

At the outlets O2 to O4 (figure~\ref{fig:results-outlet-temperatures}a), which the 1D model cannot consider, the computed and the measured temperatures start to rise about \SI{1}{\second} after the quench. The model overestimates the maxima by 6 to \SI{15}{\kelvin} and, above all, predicts their rise several times too fast. Furthermore, the measured temperatures at O2 and O3 exceed that at O1 by about \SI{4}{\kelvin}, which the model does not reproduce: there, O1 remains the hottest outlet.

\begin{figure}[tbp]
    \centering
    \includegraphics[width=1\linewidth]{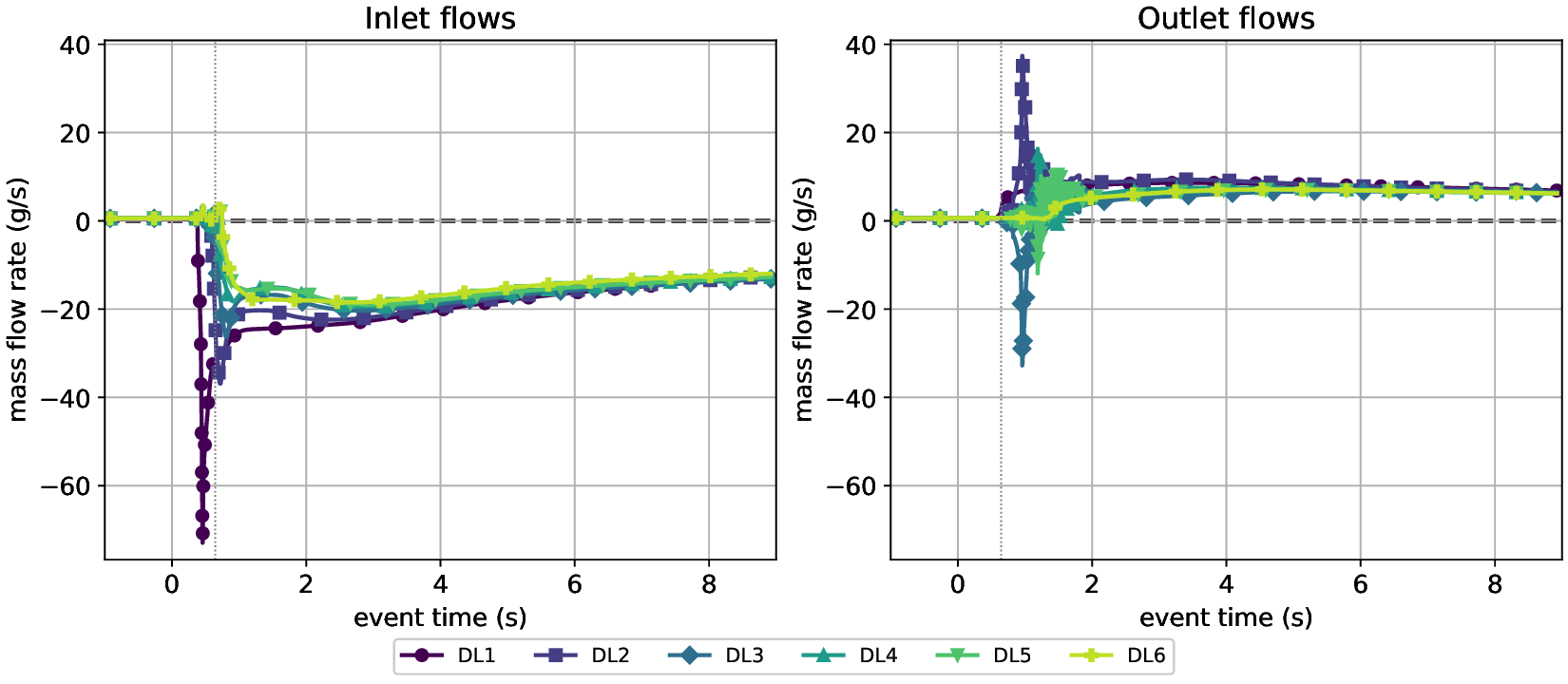}
    \caption{Computed mass flow rates at the DL inlets and outlets in the full-feature run.}
    \label{fig:results-inlet-outlet-flows}
\end{figure}

Figure~\ref{fig:results-inlet-outlet-flows} shows the computed mass flows at the inlets and outlets of the individual DLs. About \SI{0.45}{\second} after the quench, DL1 expels helium through its inlet at up to \SI{73}{\gram\per\second}; within the following half second the inlet flows of the other DLs reverse as well, and from about \SI{1.5}{\second} on all six DLs expel 15 to \SI{25}{\gram\per\second} through their inlets. At the outlets, the two DLs of a common outlet briefly exchange helium at about \SI{1}{\second}, before all DLs settle at 5 to \SI{9}{\gram\per\second} directed out of the winding. With the throttled outlet lines, most of the helium thus leaves the winding through the inlets, which is consistent with the measured pressures (Sec.~\ref{subsec:pressure-and-massflow}); the flow rates themselves are not confirmed by the measurement.

\subsubsection{Spatial temperature distribution}
Figure~\ref{fig:results-temperature-profiles}a shows the strand temperature along the DLs at $t = \SI{2}{\second}$. Along each DL, $x = 0$ is the helium inlet, the first half of the length is the odd layer, the second half the even layer, and the end is the outlet. Three features can be distinguished. First, the inlet regions of all DLs are strongly heated, to \SI{52}{\kelvin} in DL1, where this is the hottest region, and to 36 to \SI{44}{\kelvin} in DL2 to DL6. This is where the helium is expelled (figure~\ref{fig:results-inlet-outlet-flows}). Second, DL2 to DL6 show narrow temperature peaks of 43 to \SI{45}{\kelvin} at mid-length, between $x \approx \SI{68}{\meter}$ and \SI{78}{\meter}. These are the turns at the transition between the two layers, where $T_\mathrm{cs}$ has its minimum in all of these DLs (figure~\ref{fig:temperature-margin-minima}). The peaks are caused by compression heating, which is examined in Sec.~\ref{subsec:compression-heating}. Third, the temperature rises again towards the outlets. The last turns of a DL are adjacent to the first turns of the same DL and of the next one (figure~\ref{fig:transv_thermal}), from which they are heated through the insulation (Sec.~\ref{subsec:no-transverse}). Between these regions the temperatures are lower, down to 13 to \SI{20}{\kelvin} in DL3 to DL6, but everywhere above the critical temperature: \SI{2}{\second} after the quench, all six DLs are normal-conducting over their entire length.

\subsubsection{Pressures}\label{subsec:pressure-and-massflow}
The test facility also recorded the pressure at the winding inlet (sensor PT2111) and at the winding outlet (PT2115Bis). The inlet pressure is prescribed in the model as the pressure of the supply reservoir and is therefore a boundary condition, not a result.

\begin{figure}[tp]
    \centering
    \includegraphics[width=0.6\linewidth]{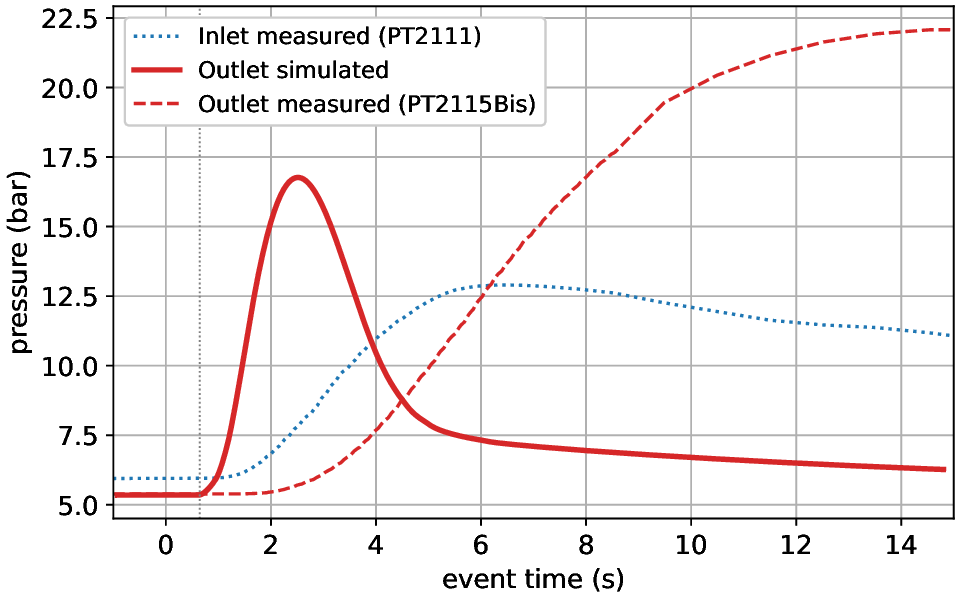}
    \caption{Measured inlet and outlet pressures and simulated outlet pressure. The simulated inlet pressure is not shown since it is prescribed from the measured one.}
    \label{fig:winding_pressures}
\end{figure}

Figure~\ref{fig:winding_pressures} compares the measured outlet pressure with the pressure computed at the return line. The computed maximum of \SI{16.8}{\bar} is comparable to the measured \SI{22.1}{\bar}, but it is reached after \SI{2.5}{\second} instead of about \SI{15}{\second}: the model reproduces the level of the pressure rise, but builds it up about six times faster than measured.

Two observations do not depend on the model. First, the inlet pressure starts to rise \SI{1.0}{\second} after the quench, i.e.\ before the outlet pressure (\SI{1.8}{\second}), and exceeds its initial value by \SI{7}{\bar}, which can only be caused by helium that leaves the winding through the inlets. Second, from \SI{6}{\second} on the outlet pressure exceeds the inlet pressure, by up to \SI{11}{\bar}, and remains higher for about a minute: the outlet side is the more restricted one, while the inlet side relieves the winding. Both observations support the predicted expulsion of helium through the inlets and the restriction assumed in the outlet lines, but not the computed mass flow rates.

\begin{figure}[p]
    \centering
    \includegraphics[width=.85\linewidth]{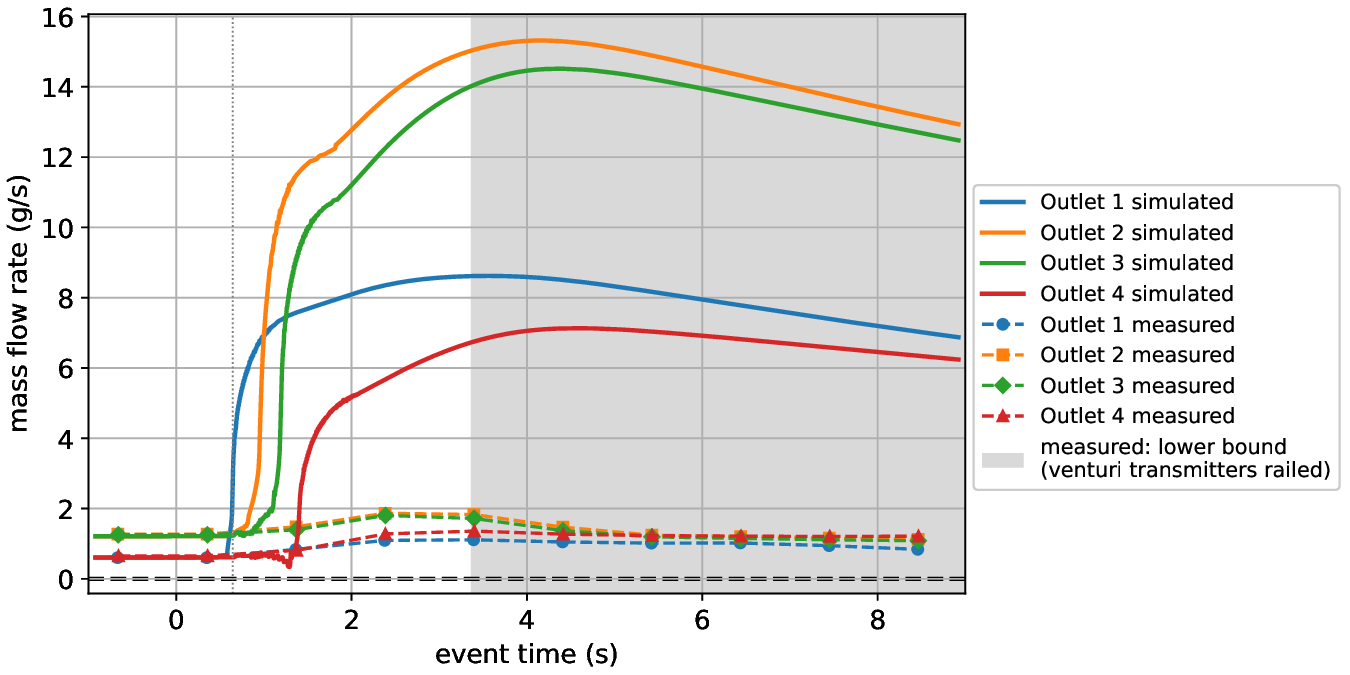}
    \caption{Computed (solid) and measured (dashed, with markers) mass flows at the outlets O1 to O4 during the quench. In the shaded interval, the venturi transmitters are at their limit of \SI{1000}{\milli\bar}, so that the measured values are at best lower bounds of the actual flows.}
    \label{fig:outlet-flows}
\end{figure}

\subsubsection{Mass flows}\label{subsec:massflows}
Figure~\ref{fig:outlet-flows} compares the computed and the measured mass flows at the outlets O1 to O4 (sensors FI2100 to FI2103); the values are listed in table~\ref{tab:comparison-measurement}. Before the quench they agree, which follows from the calibration of the throttle valves. In the model, the outlet flows rise steeply one after the other between \SI{0.65}{\second} (O1) and \SI{1.4}{\second} (O4), whereas the measured flows start to rise in the same time window, but far more gradually. At \SI{2.4}{\second}, the last sample at which all four measurements are valid, the computed outlet flows are four to eight times higher than the measured ones. The computed maxima are reached between \SI{3.5}{\second} and \SI{4.6}{\second}, i.e.\ in the interval in which the measurement provides only lower bounds, and can therefore not be compared. Whether the flows reverse cannot be decided from the measurement either, because a venturi does not register the flow direction.

\begin{table}[p]
    \centering
    \small
    \caption{Computed and measured quantities for the quench of coil AAB-51.
    Mass flows and temperatures are given for the outlets O1 to O4. All times
    are event times as defined in Sec.~\ref{subsec:validation-results}.}
    \begin{tabularx}{\linewidth}{@{}p{3.3cm}p{2.9cm}p{2.9cm}X@{}}
        \toprule
        Quantity (sensor) & Measured & Computed & Assessment \\
        \midrule
        Minimum $T_\mathrm{cs}$ vs.\ quench temperature
            & $T_\mathrm{q} = 6.28$\,K
            & $T_\mathrm{cs,min} = 6.257$\,K
            & static estimate agrees within 23\,mK \\ \\
        Quench onset in the transient
            & reference
            & 96.8\,s later
            & supply temperature 356\,mK above $T_\mathrm{q}$ at detection \\ \\
        Detection voltage of DL1 before the dump 
            & $2\cdot10^{-4}$ to 1\,V
            & --
            & agrees in shape and slope \\
        \midrule
        Outlet mass flow before the quench 
            & 0.60 / 1.27 / 1.25 / 0.65\,g/s
            & 0.61 / 1.21 / 1.21 / 0.60\,g/s
            & agrees by calibration of the throttle valves \\ \\
        Outlet mass flow at 2.4\,s
            & 1.09 / 1.87 / 1.80 / 1.28\,g/s
            & 8.4 / 13.7 / 12.3 / 5.7\,g/s
            & overestimated by a factor of 4 to 8 \\ \\
        Outlet mass flow after 3.3\,s
            & lower bounds only
            & up to 15.3\,g/s
            & not comparable \\
        \midrule
        Inlet pressure 
            & up to 12.9\,bar
            & prescribed
            & boundary condition \\ \\
        Maximum outlet pressure 
            & 22.1\,bar after 15.6\,s
            & 16.8\,bar after 2.5\,s
            & 24\,\% lower, reached six times too early \\
        \midrule
        Maximum outlet temperature O1 
            & 23.1\,K after 15\,s
            & 47.6\,K, risen by 2.2\,s
            & overestimated by 24.5\,K (adiabatic model: 54.8\,K) \\ \\
        Maximum outlet temperatures O2 / O3 / O4 
            & 27.0 / 27.1 / 15.9\,K after 12 to 18\,s
            & 38.9 / 33.5 / 30.8\,K, risen by 2.4 to 3.0\,s
            & overestimated by 6 to 15\,K, rise several times too fast \\ \\
        Hottest outlet
            & O2 and O3
            & O1
            & not reproduced \\
        \bottomrule
    \end{tabularx}
    \label{tab:comparison-measurement}
\end{table}

\subsection{Impact of Single Features}\label{subsec:feature-impacts}
In the following, the three features added to \opensc\ -- induced losses, external hydraulic network and transverse heat conduction -- are deactivated one at a time in order to assess their individual contribution to the full-feature result (Sec.~\ref{subsec:validation-results}).

\subsubsection{No induced losses}
With the induced losses deactivated, the voltages of DL2 to DL6 start to rise at the same time as in the full-feature run (figure~\ref{fig:results-detection-voltage}b). The quench of their inlet regions therefore does not depend on the induced losses. The temperature profiles at \SI{2}{\second} differ, however (figure~\ref{fig:results-temperature-profiles}b): the peaks at mid-length are absent, and the central part of DL3 to DL6 is still at about \SI{7}{\kelvin}. As shown in Sec.~\ref{subsec:compression-heating}, the pressure in the winding pack builds up later without the induced losses, so that compression heating has not yet brought these turns to current sharing. The hot-spot temperature of DL1 is unchanged, whereas those of DL2 to DL6 are lower by about 3 to \SI{11}{\kelvin} (figure~\ref{fig:results-strand-hotspots}b), and the outlet temperatures are lower by up to \SI{2.5}{\kelvin} (figure~\ref{fig:results-outlet-temperatures}b).

\subsubsection{No hydraulic network}
Deactivating the hydraulic network reduces the model to six conductors that are coupled only by transverse heat conduction. Each conductor has its own boundary conditions: the measured inlet temperature is prescribed at the inlets together with the pressure, and pressure at the outlet. The quench is therefore triggered by the rising inlet temperature, as in the full-feature run. What differs is that the helium expelled from one DL cannot reach the others, and that the pressure at the conductor ends is fixed by the boundary conditions instead of following from the network.

The quench then spreads only by conduction through the insulation. DL2 follows DL1 with a delay of about 0.7\,s (figure~\ref{fig:results-detection-voltage}c) and DL3 after about \SI{3}{\second}, whereas DL4 and DL5 stay below \SI{10}{\kelvin} within the simulated \SI{16}{\second} (figure~\ref{fig:results-strand-hotspots}c). The hot spots reach \SI{53.8}{\kelvin} in DL1, \SI{32.6}{\kelvin} in DL2 and \SI{15.9}{\kelvin} in DL3. Neither the heating of the inlet regions nor the peaks at mid-length occur (figure~\ref{fig:results-temperature-profiles}c). The hot spot occurring in DL6 is an artifact of the prescribed boundary conditions following the measurement and is restricted to the inlet alone. 

There are no outlet groups in this configuration, so that figure~\ref{fig:results-outlet-temperatures}c shows the temperatures at the outlet boundaries of the conductors instead. Only the one assigned to O2 rises, to about \SI{16}{\kelvin}; those assigned to O3 and O4 remain at their initial value, in contrast to the measurement.

\subsubsection{No transverse heat conduction}\label{subsec:no-transverse}
Without transverse heat conduction, the conductors are thermally insulated from each other and coupled only through the hydraulic network. DL2 loses its heating by conduction from DL1 and now quenches together with DL3 to DL6, about \SI{0.1}{\second} later than in the full-feature run (figure~\ref{fig:results-detection-voltage}d). The hot spots are higher, \SI{59.4}{\kelvin} in DL1 and 49 to \SI{52}{\kelvin} in the other DLs, and they hardly decrease within \SI{30}{\second} (figure~\ref{fig:results-strand-hotspots}d), because the heat can no longer spread to neighboring turns.

The temperature profiles at \SI{2}{\second} (figure~\ref{fig:results-temperature-profiles}d) show the heated inlets and the peaks at mid-length as in the full-feature run, but most of the second layer of each DL is still cold: DL4 to DL6 are at about \SI{7}{\kelvin} beyond $x \approx \SI{85}{\meter}$. In the full-feature run, the last turns are heated through the insulation by the first turns of the same and of the neighboring DL, which are the hottest ones. Consequently, the outlet temperatures rise later and stay much lower without transverse heat conduction: they reach \SI{15.5}{\kelvin} at O2, \SI{14}{\kelvin} at O3 and \SI{9.5}{\kelvin} at O4 (figure~\ref{fig:results-outlet-temperatures}d), compared with 30.8 to \SI{38.9}{\kelvin} in the full-feature run. The measured temperatures at these outlets lie between the two cases. In DL1, the highest temperature at \SI{2}{\second} is found at $x \approx \SI{18}{\meter}$ instead of at the inlet.

\begin{figure}[tp]
    \centering
    \includegraphics[width=0.75\linewidth]{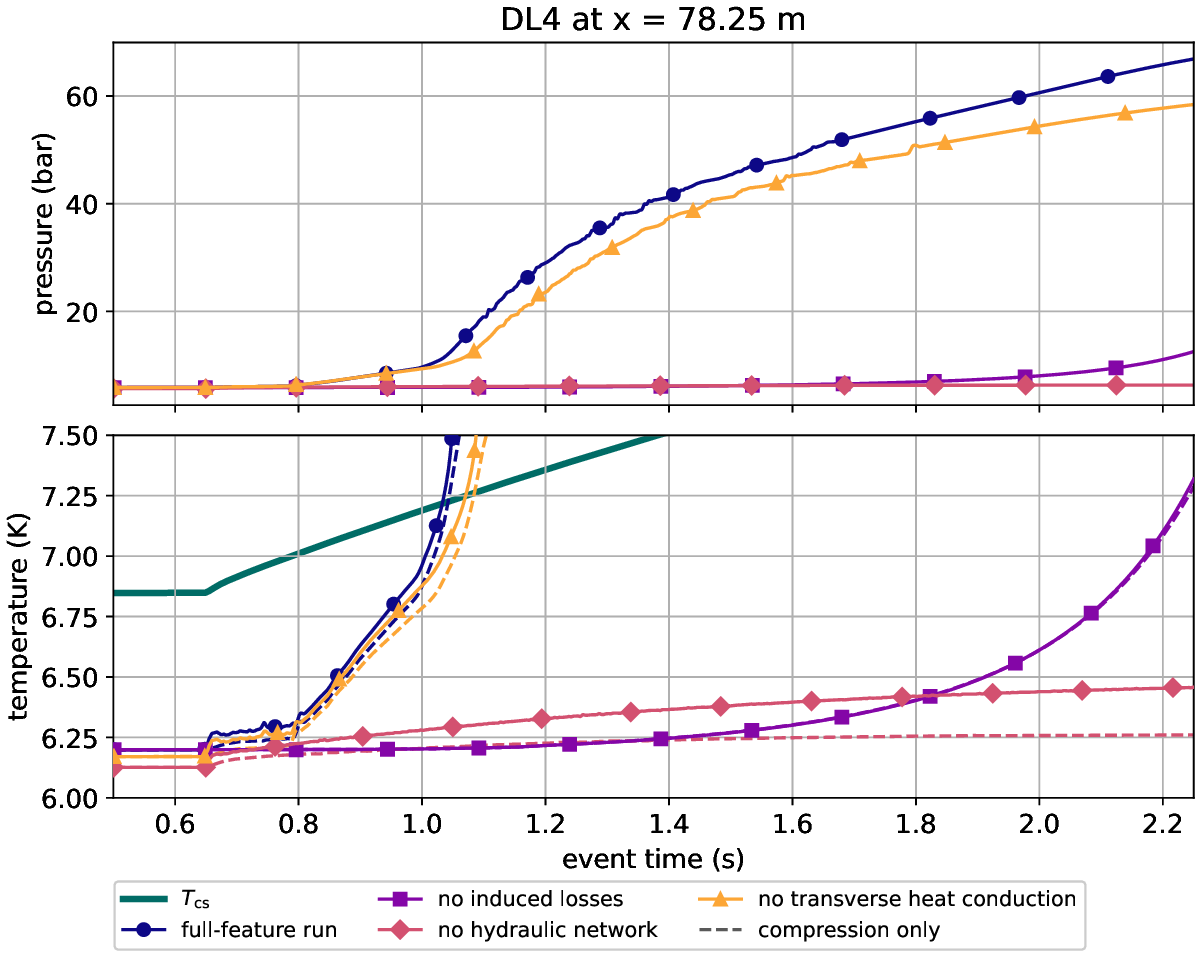}
    \caption{Pressure (top) and strand temperature (bottom, solid lines) in DL4 at $x = 78.25$\,m for the full-feature run and with single features deactivated. Dashed lines: temperature reached by isentropic compression alone, estimated from the pressure history of the respective run.}
    \label{fig:dl4-margin-budget}
\end{figure}

\subsection{Compression Heating}\label{subsec:compression-heating}
The helium expelled from the quenching regions raises the pressure in all DLs, since they are connected through the common inlets and outlets. Compression heating therefore occurs in DL2 to DL6 alike and leads to local quenches where the temperature margin is smallest (figure~\ref{fig:results-temperature-profiles}). DL4 is examined as an example, since it is not in contact with the primarily quenched DL1, so that heat conduction from DL1 can be excluded as the cause. Figure~\ref{fig:dl4-margin-budget} compares the pressure and the strand temperature at $x = \SI{78.25}{\meter}$, about \SI{4}{\meter} from the hot spot of DL4, with the local current-sharing temperature $T_\mathrm{cs}$, evaluated with the local field (\SI{3.16}{\tesla} before the dump). Before the current dump, the strand is at \SI{6.2}{\kelvin} and $T_\mathrm{cs}$ at \SI{6.85}{\kelvin}; with the dump, $T_\mathrm{cs}$ rises as current and field decay, so that the temperature margin would grow if the strand temperature stayed constant.

In the full-feature run, however, the pressure at this position starts to rise about \SI{0.8}{\second} after the quench and reaches about \SI{12}{\bar} at \SI{1.05}{\second}, when the strand temperature crosses $T_\mathrm{cs}$. Up to the crossing, the isentropic estimate (dashed lines) stays within about \SI{0.05}{\kelvin} of the strand temperature: nearly the entire rise of about \SI{1}{\kelvin} is caused by compression, and the small remainder by the induced losses. That the pressure is imposed from outside, and is not the result of local heating, follows from its magnitude: heating the helium locally at constant density to the same temperature would raise the pressure to about \SI{9.5}{\bar} only, whereas isentropic compression requires about \SI{12.4}{\bar} (calculated with CoolProp~\cite{bell_pure_2014}), as observed. Once the strand has crossed $T_\mathrm{cs}$, Joule heating takes over.

The three reduced models identify what this mechanism depends on: 
\begin{itemize}
    \item \textit{Without the hydraulic network}, the pressure stays at its initial value and $T_\mathrm{cs}$ is not reached.
    \item \textit{Without transverse heat conduction}, the evolution is nearly the same as in the full-feature run; heat conduction between the turns is therefore not required for this secondary quench.
    \item \textit{Without induced losses}, the compression still occurs, but the pressure starts to rise only after about \SI{1.8}{\second}. The induced losses thus do not heat this position to current sharing themselves, but they determine how quickly the pressure builds up in the winding pack, presumably by accelerating the quench of the regions that are already close to current sharing when the dump starts.
\end{itemize}
The isentropic temperature is an estimate derived from the coupled solution and not an independent calculation; it shows that the computed state at this position follows the isentrope, not that compression could be switched on separately. Moreover, the timing of this mechanism depends on how quickly the pressure rises in the model, which is faster than in the measurement.

\subsection{Summary}\label{subsec:validation-summary}
Compared with the adiabatic 1D model, the winding-pack model reduces the overestimation of the outlet temperature of the primarily quenched DL from \SI{54.8}{\kelvin} to \SI{24.5}{\kelvin}, at the cost of a runtime of two to three hours instead of roughly \SI{20}{\minute} on a standard workstation. Each of the added features changes the result qualitatively (Sec.~\ref{subsec:feature-impacts} and~\ref{subsec:compression-heating}).

The model agrees with the measurement in the quench initiation in DL1, in the growth of the detection voltage before the dump and in the level of the pressure rise at the outlet, and the measured pressures support the expulsion of helium through the inlets that the model predicts (table~\ref{tab:comparison-measurement}). What the model does not reproduce is the time scale: the outlet pressure and the outlet temperatures rise several times too fast, and the outlet mass flows and temperatures are overestimated. The model therefore overestimates how quickly the quench heats and pressurizes the helium in the winding pack.

The external hydraulic circuit is only partly documented, but it is not the main cause of this discrepancy, which appears with throttled and with open outlet lines and with an increased conductor friction alike (Appendix~\ref{sec:network-sensitivity}). More likely causes are the induced losses, which set the onset of the pressure rise (Sec.~\ref{subsec:compression-heating}), the transverse conductance, since the measured outlet temperatures lie between the runs with and without transverse heat conduction (Sec.~\ref{subsec:no-transverse}), and the temperature margin along the conductors before the quench together with the position of the field map (Appendix~\ref{subsec:magnetic-field-sensitivity}). The discrepancy may also simply originate from the simplifications of the hydraulic model itself: the flow in the cable bundle is described by 1D equations for cross-section averaged quantities, in which the Navier-Stokes equations are reduced to a friction factor correlation that was established for steady flow and is applied here to a fast transient with flow reversal. Further candidates are the lumping of the external network, neglected three-dimensional effects such as the curvature of the windings, and uncertainties in the material data. The identified mechanisms, i.e.\ the expulsion of helium through the inlets, its redistribution among the DLs and the resulting compression heating, are consistent with the measured pressures, whereas their timing remains a model prediction that the measurement does not confirm (table~\ref{tab:claims-evidence}).

\begin{table}[tp]
    \centering
    \small
    \caption{Mechanisms predicted by the model and the experimental evidence
    available for them. Absolute times are event times as defined in
    Sec.~\ref{subsec:validation-results}.}
    \begin{tabularx}{\linewidth}{@{}p{4.0cm}Xp{3.6cm}@{}}
        \toprule
        Model prediction & Experimental evidence & Status \\
        \midrule
        Quench starts in DL1
            & detected by the compensated voltage between DL1 and DL2
              (DUP11-3)
            & confirmed at the level of the DL \\ \\
        Axial position of the quench initiation
            & none; the prediction is uncertain by up to one turn length
            & not testable \\ \\
        Helium is expelled through the inlets
            & the inlet pressure rises before the outlet pressure and by
              7\,bar; from 6\,s on the outlet pressure exceeds the inlet
              pressure
            & supported qualitatively; flow rates not confirmed \\ \\
        Flow resistance in the outlet lines
            & mass flows before the quench; the outlet side stays at higher
              pressure than the inlet side for about a minute
            & supported \\ \\
        Exchange of helium between DLs at the common outlets, flow reversal
            & none; the venturis do not register the flow direction
            & not testable \\ \\
        DL2 to DL6 quench within tenths of a second after DL1
            & the compensated voltages of DL3--DL4 and DL5--DL6 rise about
              0.25\,s after that of DL1--DL2 and before the current dump
            & supported; onset in the model after about 0.2\,s \\ \\
        Compression heating causes local quenches at mid-length
            & none; no sensors inside the winding pack
            & model prediction; its timing depends on the pressure rise,
              which is too fast in the model \\ \\
        All outlets are heated
            & observed at all four outlets
            & confirmed, but several times more slowly than computed \\ \\
        Hot-spot temperatures
            & none
            & not testable \\
        \bottomrule
    \end{tabularx}
    \label{tab:claims-evidence}
\end{table}

\subsection{Implications for the Design of Future Coils}
\label{sec:design-implications}
Although the timing of the mechanisms is not confirmed by the measurement, the mechanisms and their sensitivities point to aspects that deserve attention in the design of coils for stellarator power plants:
\begin{itemize}
    \item \textit{Position of the inlets and outlets relative to the field.} In a non-planar coil, the magnetic flux density varies along every turn, in the W7-X coil by about \SI{1}{\tesla}. Shifting the position at which the windings start and end by $3/8$ of a turn length moved a field maximum into the last turn of every DL, where a pronounced local quench then develops (Appendix~\ref{subsec:magnetic-field-sensitivity}). The end turns are particularly exposed in a quench, since the inlet turns receive the expelled helium and the outlet turns are heated by them through the insulation. Inlets, outlets and layer transitions should therefore be placed in the low-field part of the coil perimeter, and the temperature margin should be assessed with the field resolved along the conductor. Since the smallest margin of DL2 to DL6 lies at the layer transition, where the compression-induced quenches develop, feeding the helium there instead of at one end could be considered.

    \item \textit{Hydraulic interconnection of the DLs.} The helium expelled from a quenching DL is taken up mainly by the other DLs, and the mass flows at the DL ends hardly depend on the valves of the supply and return lines, whereas the resistance of the outlet lines directly at the coil and the friction of the conductor change the hot-spot temperatures of the neighboring DLs by up to \SI{12}{\kelvin}, most strongly in the outermost DL6 (Appendix~\ref{sec:network-sensitivity}). Which DLs share inlets and outlets thus governs how a quench spreads within the winding pack, so that the manifolds and the outlet lines should be treated as part of the quench protection design.

    \item \textit{Turn insulation.} Within a factor of two, the transverse conductance has little influence on the hot-spot temperatures, but a higher conductance delays the quench onset (Appendix~\ref{sec:conductance-sensitivity}); without any transverse conduction, the hot spots are up to \SI{4}{\kelvin} higher and hardly decay (Sec.~\ref{subsec:no-transverse}). A well-conducting turn insulation is thus favorable for stability and for the hot spot, at the price of spreading the heat to the neighboring turns.

    \item \textit{Current dump.} Since the induced losses during the dump determine how quickly the pressure builds up in the winding pack (Sec.~\ref{subsec:compression-heating}), the time constant of the dump and the coupling time constant of the cable influence the pressure and the compression-induced secondary quenches, and not only the hot-spot temperature.
\end{itemize}

\section{Conclusion}\label{sec:conclusion}
This work extended the open-source code \opensc~\cite{savoldi_thermalhydraulic_2022} for CICC quench simulation in winding packs, so that it now resolves secondary quenches caused by transverse heat conduction, induced effects and hydraulic flow redistribution in stellarator coils. This was achieved by modeling the transverse heat conduction between neighboring turns as a nonlocal, energy-conserving coupling, by computing the local current-sharing temperature from a three-dimensional magnetic flux density distribution within the coil, and by coupling the conductor models to an external hydraulic network. For the latter, a dedicated hydraulic formulation was derived, implemented and verified, which uses the mass flow rate rather than the velocity as primary variable in order to obtain a consistent coupling between the field models and the network. To the best of our knowledge, this is the first open-source implementation of a hydraulic field-network coupling for quench simulation; comparable couplings exist in the proprietary Flower/SuperMagnet and 4C codes~\cite{bottura_flower_2003, savoldirichard_4c_2010}.

The extended model was applied to the quench test of the W7-X coil AAB-51 carried out at CEA Saclay~\cite{baldzuhn_coil_2006}. These data were not recorded for code validation, and the test stand is only partly documented, so that the comparison is not a quantitative validation. The only quantities adjusted to the measurement are the resistances of the outlet lines, which were calibrated to the mass flows before the quench; no transient data were used. In agreement with the measurement, the model predicts the quench initiation in the innermost double layer from the computed current-sharing temperature, the growth of the detection voltage before the current dump, and the appearance of normal zones in the other double layers within tenths of a second. It reduces the overestimation of the outlet temperature of the primarily quenched double layer from \SI{54.8}{\kelvin} for the adiabatic 1D model to \SI{24.5}{\kelvin}, and it reaches a level of the outlet pressure comparable to the measured one. The measured pressures furthermore support the expulsion of helium through the coil inlets, by which the quench spreads to the neighboring double layers in the model.

The model does not reproduce the time scale of the transient. The outlet pressure and the outlet temperatures rise within two to three seconds instead of ten to twenty, the outlet mass flows are several times too high during the first three seconds, and the quench sets in at an inlet temperature about \SI{0.36}{\kelvin} above the measured one. The model thus overestimates how quickly the quench heats and pressurizes the helium in the winding pack. The sensitivity studies indicate that the partly unknown external hydraulic circuit is not the main cause. More likely candidates are the induced losses, the transverse thermal conductance and the temperature margin along the conductors before the quench. The discrepancy may also originate from the 1D hydraulic model itself, which reduces the Navier-Stokes equations to cross-section averaged equations with a steady-state friction factor correlation. Accordingly, the compression heating and the timing of the secondary quenches are predictions of the model that the available data cannot confirm. A repetition of such a quench test with a documented hydraulic circuit, with pressure, temperature and bidirectional mass-flow measurements of sufficient range at the inlets and outlets of the individual double layers, and with voltage signals recorded throughout the current dump would allow these predictions to be tested.

Although their timing is not confirmed, the identified mechanisms point to aspects that deserve attention in the design of coils for stellarator power plants (Sec.~\ref{sec:design-implications}). The helium inlets, outlets and layer transitions should be placed in the low-field part of the coil perimeter, and the temperature margin should be assessed with the field resolved along the conductor. The manifolds and outlet lines, which govern how a quench spreads between the double layers, should be treated as part of the quench protection design. A well-conducting turn insulation favors stability and lowers the hot-spot temperature at the price of spreading the heat to the neighboring turns, and the time constant of the current dump and the coupling time constant of the cable influence the compression-induced secondary quenches and not only the hot-spot temperature.

The model is not specific to W7-X: any forced-flow magnet whose windings share a hydraulic circuit, for instance the toroidal-field and central-solenoid coils of tokamaks, exhibits the same coupling between quench-driven expulsion and the external network.

Future work should first address the discrepancy in the time scale, in particular the model of the induced losses and the state of the winding pack before the quench. Furthermore, the solver should be accelerated, for example with a parallel direct solver such as MUMPS~\cite{amestoy_multifrontal_2000}. This will be essential for power-plant-scale stellarator coils, whose windings are substantially longer than those of W7-X: the present winding-pack simulation already required nearly \SI{3}{\hour} for six double layers of 14\,000 finite elements each. Beyond that, the \opensc\ approach of modeling conductors as mutually coupled 1D components could be combined with a quasi-3D hybrid discretization~\cite{dangelo_quasi3d_2022} using a rotation-minimizing frame such as the Bishop frame~\cite{bishop_there_1975} along the winding direction, in order to account for curvature effects.

%
%

\ack{The authors would like to thank Konrad Ri\ss e for his kind support in setting up and running the W7-X THEA model. This work has been supported by the German BMFTR Project ``Bearbeitung grundlegender Themen der Magnetfusion auf dem Weg zum Fusionskraftwerk (FPP-MC) -- Teilvorhaben: Erstellung und Analyse von integrierten Konzepten für ein Fusionskraftwerk vom Typ Tokamak und Stellarator'' (13F1001A).}

\funding{German Federal Ministry of Research, Technology and Space (BMFTR), Grant Nr.\ 13F1001A.}

\roles{\textbf{Laura A.~M.~D'Angelo:} Conceptualization, Methodology, Software, Validation, Formal analysis, Investigation, Data Curation, Writing -- Original Draft, Visualization. \textbf{Felix Warmer:} Conceptualization, Resources, Writing -- Review \& Editing, Supervision, Project Administration, Funding Acquisition.}

\data{The code is available at \url{https://github.com/lamdangelo/OPENSC2} and will be possibly merged into the original repository at \url{https://github.com/MAHTEP/OPENSC2} in the future. W7-X model and measurement data is available upon request.}


\appendix 

\section{Refactoring and Efficiency Improvement of \opensc}\label{app:refactoring}
The observation of \opensc\ demanding 22\,\si{\hour} to run a quench simulation of one W7-X double layer motivated a refactoring of \opensc, since this was an unacceptable runtime considering that the extended model presented in Sec.~\ref{sec:2d} requires the simulation of multiple double layers. 
This section presents highlights of this refactoring work. Throughout the refactoring process, the functionality and correctness of the code was repeatedly tested against the three test cases provided by the creators of \opensc~\cite{savoldi_thermalhydraulic_2022} and finally benchmarked against a W7-X reference simulation (Sec.~\ref{subsec:validation-against-thea}). 
The refactored \opensc\ version is found on the fork 
\url{https://github.com/lamdangelo/OPENSC2}.

\subsection{Structure}
The refactoring branched off the main branch of the original repository \url{https://github.com/MAHTEP/OPENSC2} in June 2026, pursuing three goals: modularization into one package per physical or infrastructural concern with orchestration separated from physics; modernization of the internal data model towards typed structures and explicit enumerations; and a profiling-driven reduction of computational cost. 

At that point, \opensc\ consisted of a flat source code layout dominated by a few very large modules mixing different responsibilities. It was reorganized into the mainly physics-oriented packages listed in table~\ref{tab:opensc_modules}, with each module carrying a single responsibility. For example, a conductor class of over \num{6000} lines with more than 90 methods, mixing mesh geometry, topology, electric pre-processing, heat-transfer physics and diagnostics, was thereby reduced to approximately \num{1300} lines by extracting each of these concerns into its own module. 

\begin{table}[tbp]
    \centering
    \caption{Package layout of the modularized \opensc\ source code.}
    \begin{tabular}{ll}
        \toprule
        Package & Responsibility \\
        \midrule
        \texttt{adapters/} & Input sources, output writers, plotting, CoolProp provider \\
        \texttt{application/} & Simulation build, transient loop, post-processing use cases \\
        \texttt{components/} & Fluid, strand, stack and jacket components and factories \\
        \texttt{conductor/} & Conductor definition, state, mesh, topology, coupling \\
        \texttt{drivers/}, \texttt{simulation.py} & Composition root, command-line and GUI entry points \\
        \texttt{electromagnetics/} & Electric model, topology, resistance, inductance, solver \\
        \texttt{environment/} & External environment (cryostat medium) \\
        \texttt{hydraulics/} & Hydraulic model, coolant state, friction models, network \\
        \texttt{numerics/} & FE assembly, banded solve, time stepping, interpolation \\
        \texttt{physical\_fields/} & Typed field containers (nodal/Gauss fields, time evolutions) \\
        \texttt{ports/} & Protocols implemented by the outer layers \\
        \texttt{properties\_of\_materials/} & Material property correlations and material registry \\
        \texttt{runtime/} & Simulation context (shared run state) and time-step control \\
        \texttt{thermal/} & Thermal model, heat sources, heat transfer correlations \\
        \bottomrule
    \end{tabular}
    \label{tab:opensc_modules}
\end{table}

Selectable physical models and numerical methods replaced the previous selection by raw integer flags read from the Excel input workbooks: the flag is now translated once at load time into a model class instance, so that adding a new model or method is now a local change confined to one model class and one factory branch. Likewise, the string-keyed dictionaries carrying the physical state were replaced by data classes, decreasing both runtime and error proneness by attribute access.

\subsection{Computational Efficiency}\label{subsec:efficiency}
The assembly of FE matrices was enhanced by a SciPy~\cite{virtanen_scipy_2020} sparse block matrix composition, where the indices and calculated entries of the local stiffness matrices per element are first stored in arrays, which are used after the per-element loop to assemble the global stiffness matrix very efficiently with a single vectorized addition.
Numerical routines transcribed from an older Fortran code, including a hand-written LU factorization, were replaced by dedicated SciPy solutions such as LAPACK~\cite{anderson_lapack_1999}.
The repeated re-evaluation of the helium properties by CoolProp was replaced by bicubic interpolation on a grid in $(T, p)$ which is built once at the start of a simulation run.

The adaptive time control offered by \opensc\ struggled with near-stagnant states: its error estimate would diverge and, as a consequence, permanently keep the minimum time step, burdening the computational runtime without necessity. It was therefore replaced by a predictor-corrector local-error estimate~\cite{hairer_solving_1993}, where the error indicator is the deviation of the solution of the current time iteration, $\mathbf{U}^{n+1}$ at $t_{n+1}$, from the predicted solution by a linear extrapolation of the time history, 
\begin{equation}
    \Delta \mathbf{U}^{n+1} = \mathbf{U}^{n+1} - \underbrace{\left[ \mathbf{U}^n + \frac{\Delta t_n}{\Delta t_{n-1}} (\mathbf{U}^n  - \mathbf{U}^{n-1} )\right]}_{=\mathbf{U}^{n+1}_\mathrm{predicted}}, 
\end{equation}
with $\mathbf{U}^n$ and $\mathbf{U}^{n-1}$ being the two stored solutions at the previous time iterations $t_n$ and $t_{n-1}$ with time steps $\Delta t_{n-1}$ and $\Delta t_n$, respectively. The relative indicator is computed by a per-field $w$ normalization using a weighted root-mean-square norm with absolute floors based on the SUNDIALS approach~\cite{hindmarsh_sundials_2005},
\begin{equation}
    e_w = \frac{\|\Delta \mathbf{U}^{n+1}_w\|_2 }{\sqrt{N_w} \, \max(\|\mathbf{U}^{n+1}_w\|_\infty, w_\mathrm{min})},
\end{equation}
where the subscript $w$ denotes the field component of the solution vector and its deviation, respectively, $N_w$ is the number of nodal values of this field, and $w_\mathrm{min}$ are magnitude floor values for each field, preventing any field stagnating near zero from dominating the indicator. The error is chosen as the maximum field indicator, 
\begin{equation}
    \mathrm{err} = \max_w e_w.
\end{equation}
The next time step results from
\begin{equation}
    \Delta t_{n+1} = \Delta t_n \min \left[ f_{\max}, \max\left( f_{\min}, \sigma \sqrt{\frac{\epsilon}{\mathrm{err}}} \right) \right],
\end{equation}
with $\sigma = 0.9$, $f_{\min} = 0.25$, $f_{\max} = 1.5$ and a user-defined tolerance $\epsilon$. The new time step always remains between the user-defined minimum and maximum time steps. A step exceeding $\epsilon$ is rejected if it is still above the minimum time step; the associated time iteration is recalculated again with the new time step. The introduction of this adaptive time control scheme reduced the number of time steps for the W7-X case from 25000 to 2257 time steps; time steps were rejected 16 times. 

All in all, the computational efficiency of \opensc\ has been substantially improved and the runtime for the W7-X conductor model eventually reduced from 22\,h to 19\,min.

\section{Manufactured Solutions for the One-Dimensional Hydraulic Equations}\label{app:manufactured-solutions}

\subsection{Manufactured Fields}
\begin{figure}[tp]
    \centering
    \includegraphics[width=.95\linewidth]{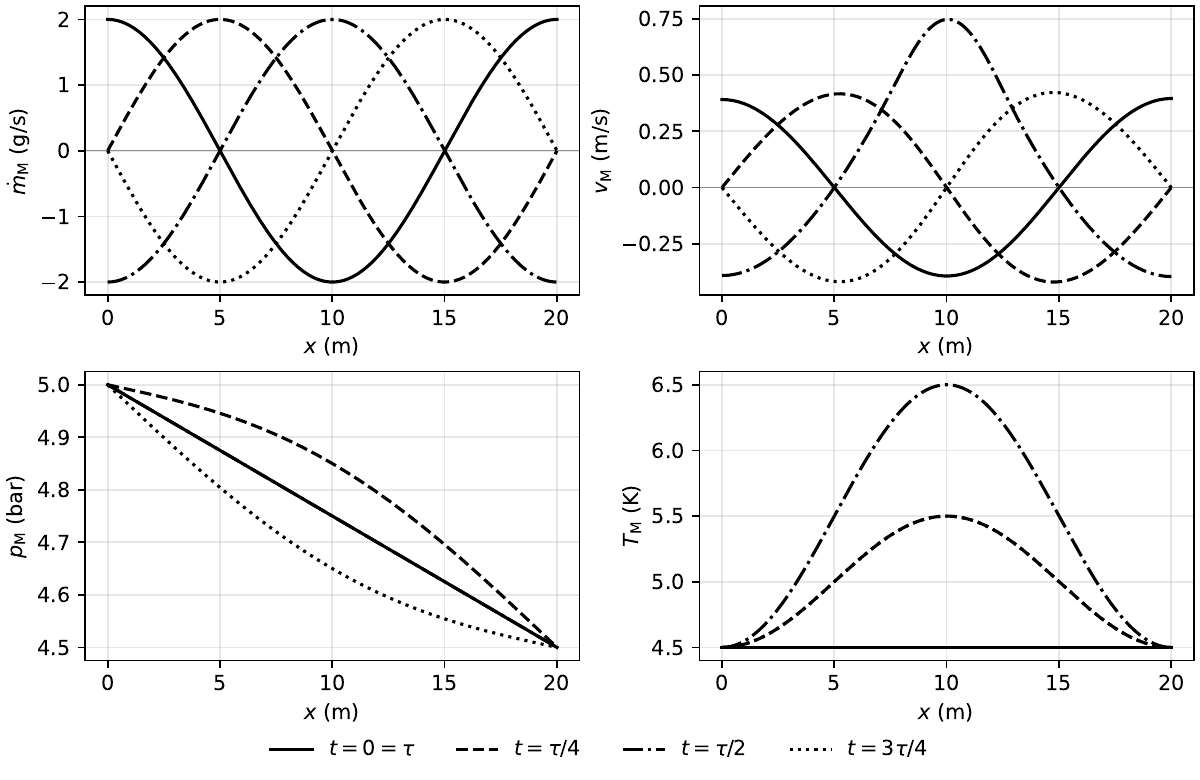}
    \caption{Manufactured fields \eqref{eq:manufactured-mass-flow}-\eqref{eq:manufactured-temperature} and the derived velocity at different points in time.}
    \label{fig:manufactured_fields}
\end{figure}

On $x \in [0,L], t \in [0, \tau]$, the fields are 
\begin{align}
    \dot m_\mathrm{M}(x,t) &= \dot m_0 \cos\left[2\pi \left( \frac{x}{L} - \frac{t}{\tau}\right)\right], \label{eq:manufactured-mass-flow} \\
    p_\mathrm{M} (x,t) &= p_0-\Delta p\,\frac{x}{L}+a_p\,p_0\sin\left(\frac{\pi x}{L}\right)\sin\left(\frac{2\pi t}{\tau}\right), \label{eq:manufactured-pressure} \\
    T_\mathrm{M}(x,t) &= T_0 + \Delta T \sin^2 \left( \frac{\pi x}{L} \right) \frac{1-\cos(2\pi \frac{t}{\tau})}{2}, \label{eq:manufactured-temperature}  
\end{align}
with the derived fields $\varrho_\mathrm{M}=\varrho(p_\mathrm{M},T_\mathrm{M})$ and $v_\mathrm{M}=\dot m_\mathrm{M}/(\varrho_\mathrm{M} A)$. They are shown in figure~\ref{fig:manufactured_fields} for different points in time. At the temperature maximum the density drops from 136.5 to 72.8\,\si{\kilo\gram\per\cubed\meter}, and the state stays supercritical throughout ($p_M\ge4.5$\,\si{\bar}). The parameters are listed in table~\ref{tab:manufactured-parameters}. 
The spatial shapes fix the boundary values in time, $\dot m_\mathrm{M}(0,t)=\dot m_\mathrm{M}(L,t)=\dot m_0 \cos(2\pi t/\tau)$, $T_\mathrm{M}(0,t)=T_\mathrm{M}(L,t)=T_0$, $p_\mathrm{M}(0,t)=p_0$, $p_\mathrm{M}(L,t)=p_0-\Delta p$. The solution is periodic with period $\tau$, and flow reversal occurs: both end flows reverse at $t=\tau/4$ and $t=3\tau/4$. Pressure and temperature at the ends remain constant in time. The channel is the bundle channel of the W7-X conductor with the Darcy–Forchheimer porous-medium friction correlation~\cite{bottura_porous_2008} used as the total Fanning friction factor at every Reynolds number, which for this channel reads $f(\mathrm{Re})=10.04/\mathrm{Re}+0.01893$.

\begin{table}[tp]
  \centering
  \caption{Parameters of the manufactured solution and the base state at
  $(p_0, T_0)$.}
  \label{tab:manufactured-parameters}
  \begin{tabular}{llc}
    \toprule
    quantity & symbol & {value} \\
    \midrule
    base mass flow            & $\dot{m}_0$   & $2.0\cdot10^{-3}$\,\si{\kilogram\per\second} \\
    inlet pressure            & $p_0$       & 5.0\,\si{\bar} \\
    imposed pressure drop     & $\Delta p$  & 0.5\,\si{\bar} \\
    inlet temperature         & $T_0$       & 4.5\,\si{\kelvin} \\
    temperature amplitude     & $\Delta T$  & 2.0\,\si{\kelvin} \\
    pressure amplitude        & $a_p$       & 0.02 \\
    period                    & $\tau$      & 0.5\,\si{\second} \\
    \midrule
    density                   & $\rho(p_0,T_0)$ & 136.5\,\si{\kilo\gram\per\cubed\meter} \\
    speed of sound            & $c(p_0,T_0)$    & 232.9\,\si{\meter\per\second} \\
    velocity                  & $v_0 = \dot{m}_0/(\rho A)$ & 0.391\,\si{\metre\per\second} \\
    Mach number               & $v_0/c$     & $1.68\cdot10^{-3}$ \\
    Reynolds number           & $\mathrm{Re}_0$ & 5239 \\
    friction factor           & $f(\mathrm{Re}_0)$ & 0.0208 \\
    friction coefficient      & $F_0$       & 43.3\,\si{\per\second} \\
    \bottomrule
  \end{tabular}
\end{table}

\subsection{Boundary Value Problem}
The problem solved is: find $\mathbf{W}$ on $\Omega = [0, L]\times(0,\tau]$ such that
\begin{align}
    \partial_t \mathbf{W} + \mathbf{K}(\mathbf{W}) \partial_x \mathbf{W} + \mathbf{S}(\mathbf{W}) \mathbf{W} &= \mathbf{Q}_\mathrm{M}(x,t), \\
    \mathbf{W}(x, 0) &= \mathbf{W}_\mathrm{M}(x, 0), \\
    \dot{m}(0, t) &= \dot{m}_0 \cos\left( \frac{2\pi t}{\tau} \right), \\
    p(L,t) &= p_0 - \Delta p, \\
    T(0,t) &= T_0,
\end{align}
with the source term defined as the residual of the manufactured fields, 
\begin{equation}
    \mathbf{Q}_\mathrm{M}(x,t) := \partial_t \mathbf{W}_\mathrm{M} + \mathbf{K}(\mathbf{W}_\mathrm{M}) \partial_x \mathbf{W}_\mathrm{M} + \mathbf{S}(\mathbf{W}_\mathrm{M}) \mathbf{W}_\mathrm{M},
\end{equation}
such that $\mathbf{W}_\mathrm{M}$ satisfies the problem exactly by construction. For the velocity formulation, the residual is evaluated on $\mathbf{W} = (v_\mathrm{M}, p_\mathrm{M}, T_\mathrm{M})$, for the mass-flow formulation on $(\dot{m}_\mathrm{M}, p_\mathrm{M}, T_\mathrm{M})$.

\section{Derivation of Hydraulic Network Equations}
This section provides the detailed derivation of the hydraulic network equations presented in Sec.~\ref{sec:hydraulic-network}.

\subsection{Nodal Continuity Equation}\label{app:nodal-continuity-equation}
Let us consider a node in the hydraulic network, which in reality represents a junction with volume $V$. Integrating~\eqref{eq:hydraulic-continuity-equation} over $V$, assuming that $V$ is decomposable into a finite number of surfaces $A_k, k=1,2,\dots$, and using Gauss' theorem yields 
\begin{equation}
    \frac{\dd M}{\dd t} + \sum_k \dot{m}_k = 0,
\end{equation}
with the mass flow 
\begin{equation}
    \dot{m} = \int\limits_A \varrho \vec{v} \cdot \dd \vec{A},
\end{equation}
and the mass stored in $V$,
\begin{equation}
    M = \int\limits_V \varrho \dd V.
\end{equation}
Requiring that hydraulic network nodes (junctions) are storage-free analogously to the charge-free property of electric circuit nodes, leads to a hydraulic form of Kirchhoff's current law,
\begin{equation}
    \sum\limits_k \dot{m}_k = 0
\end{equation}
at a network junction, demanding that the sum of incoming and outgoing mass flows cancels out, i.e.\ no mass flow is generated or destroyed at a junction. The mass storage can be represented outside of the node by a lumped element analogously to the capacitor in electric circuits which acts as a charge storage. To derive the hydraulic capacitance, we start from $M=\varrho V$ (following the first network assumption that fluids are homogeneous in the network branches) and perform a total differentiation in time, 
\begin{equation}
    \ddt{M} = V \left[ \left( \frac{\partial \varrho}{\partial p}\right)_T \ddt{p} + \left( \frac{\partial \varrho}{\partial T} \right)_p \ddt{T} \right],
\end{equation}
where the wall compliance $\varrho \dd V/ \dd p$ is neglected since metallic pipe walls are not elastic. The temperature rate follows from the internal energy balance of the adiabatic node, 
\begin{equation}
    \ddt{(M u)} = \sum \dot{m}_\mathrm{in} h_\mathrm{up} - \sum \dot{m}_\mathrm{out} h_n
\end{equation}
where $u$ is the specific internal energy, $h$ is the specific enthalpy, and $h_\mathrm{up} = h(p_\mathrm{up}, T_\mathrm{up})$ and $h_n = h(p_n,T_n)$ are the specific enthalpies of the upstream state and the node $n$, respectively. With $u = h - p/\varrho$, $\dd h = c_p \dd T + (1-\beta T)/\varrho \dd p$ and the mass balance, the outflows cancel and 
\begin{equation}
    \varrho V c_p \frac{\mathrm{d}T_n}{\mathrm{d}t}
	= \underbrace{\sum_\mathrm{inflows} \dot m_\mathrm{in}
		(h_\mathrm{up} - h_n)}_{=H_\mathrm{adv}}
	+ V\beta T \frac{\mathrm{d}p}{\mathrm{d}t} 
    \label{eq:enthalpy-difference}
\end{equation}
with the advective heating source $H_\mathrm{adv}$ in \si{\watt}. Eliminating the temperature rate yields 
\begin{equation}
    \frac{\mathrm{d}M}{\mathrm{d}t}
	= \underbrace{V\varrho\kappa_s}_{=C_\mathrm{h}}
	\frac{\mathrm{d}p}{\mathrm{d}t}
	- \frac{\beta}{c_p} H_\mathrm{adv},
    \label{eq:Ch-definition}
\end{equation}
with the isentropic compressibility
\begin{equation}
    \kappa_s = \kappa_T - \frac{\beta^2 T}{\varrho c_p}
	= \frac{1}{\varrho c^2},
\end{equation}
and the speed of sound $c$, leading to the hydraulic capacitance definition~\eqref{eq:hydraulic-capacitance} used in this work. The hydraulic capacitance of an adiabatic node is thus the isentropic one, $C_\mathrm{h} = V/c^2$, in agreement with the transmission-line model~\cite{wylie_fluid_1978}.
With $\dot{m} = \ddt{M}$, the nodal continuity equation in the hydraulic regime is obtained as~\eqref{eq:hydraulic-nodal-continuity-equation}. The second term in~\eqref{eq:Ch-definition} yields the mass source by thermal expansion of the $n$-th node content,
\begin{equation}
    \dot{m}_{\mathrm{exp},n} = \frac{\beta}{c_p} H_{\mathrm{adv},n}.
    \label{eq:advective-heating}
\end{equation}

\subsection{Branch Momentum Equation}\label{app:branch-momentum-equation}
Following the fourth network assumption, i.e.\ assuming a purely axial flow in the branches, $\vec{v} = u \vec{e}_s$ with $\vec{e}_s$ as unit vector in axial direction, results in the 1D momentum law,
\begin{equation}
    \varrho \partial_t u + \partial_s (\varrho u^2) + u \partial_t \varrho = -\partial_s p + [\nabla \cdot \vec{\tau}]_s + \varrho \vec{g} \cdot \vec{e}_s, 
    \label{eq:momentum-law-axial}
\end{equation}
where the subscript $s$ denotes the axial component of the respective quantity. 
To arrive at a conservative form (integral form) similarly to Faraday's law, an area-averaging is performed. To that end, the bulk area-averaged axial velocity is defined as 
\begin{equation}
    v \equiv \bar{u} = \frac{1}{A} \int\limits_A u \dd A,
    \label{eq:velocity-area-averaging}
\end{equation}
where $u$ is the local axial velocity. Eventually, this gives rise to 
\begin{equation}
    \partial_t (\varrho A v) + \partial_s ( \widetilde{\beta} \varrho A v^2) = -A\partial_s p - \tau_\mathrm{w} P_\mathrm{w} - \varrho g A \partial_s z, 
\end{equation}
where $\widetilde{\beta}$ is the Boussinesq correction factor, $\tau_\mathrm{w}$ the wall shear stress, $P_\mathrm{w}$ the wetted perimeter, and $z(s)$ the elevation. 

The next step is the branch-lumping, i.e.\ we integrate over the branch $b=[n_1, n_2]$ in the sense of a lumped parameter, with $n_1$ and $n_2$ being the branch nodes, and divide by $A$ to produce pressure differences instead of forces (unit-wise) to close the analogy between electric voltages as potential differences. This yields 
\begin{equation}
    \int\limits_{n_1}^{n_2} \frac{1}{A} \partial_t (\varrho A v) \dd s + \int\limits_{n_1}^{n_2} \frac{1}{A} \partial_s (\widetilde{\beta} \varrho A v^2) \dd s = -\int\limits_{n_1}^{n_2} \partial_s p \dd s - \int\limits_{n_1}^{n_2} \frac{\tau_\mathrm{w} P_\mathrm{w}}{A} \dd s - \int\limits_{n_1}^{n_2} \varrho g \partial_s z \dd s.
    \label{eq:branch-lumped-conservation-law}
\end{equation}
With $\dot{m} = \varrho A v$ being constant over the branch, the first term becomes 
\begin{equation}
    \int\limits_{n_1}^{n_2} \frac{1}{A} \partial_t (\varrho A v) \dd s = \ddt{\dot{m}} \int\limits_{n_1}^{n_2} \frac{\dd s}{A} = L_\mathrm{h} \ddt{\dot{m}},
\end{equation}
which, assuming a constant branch area $A$, gives rise to the hydraulic inductance~\eqref{eq:hydraulic-inductance}. 
The second term in~\eqref{eq:branch-lumped-conservation-law}, 
\begin{equation}
    \int\limits_{n_1}^{n_2} \frac{1}{A} \partial_s (\widetilde{\beta} \varrho A v^2) \dd s = \int\limits_{n_1}^{n_2} \frac{1}{A} \partial_s \left( \frac{\widetilde{\beta} \dot{m}^2}{\varrho A} \right) \dd s = \frac{1}{A} \left[ \frac{\widetilde{\beta} \dot{m}^2}{\varrho A} \right]_{n_1}^{n_2} = \frac{\widetilde{\beta} \dot{m}^2}{A^2} \left( \frac{1}{\varrho_2} - \frac{1}{\varrho_1} \right), 
\end{equation}
represents convective effects, which are usually very low and vanish for a quasi-incompressible case ($\varrho_1 \approx \varrho_2$ being the fluid densities at $n_1$ and $n_2$, respectively), which is why this term can be neglected in the following~\cite{matko_pipeline_2000}.
The third term in~\eqref{eq:branch-lumped-conservation-law},
\begin{equation}
    -\int\limits_{n_1}^{n_2} \partial_s p \dd s = p_1 - p_2 = \Delta p,
\end{equation}
becomes a pressure difference and driver, which is analogous to a voltage supply in an electric circuit.
For the wall shear term in~\eqref{eq:branch-lumped-conservation-law}, we use the Fanning friction law
\begin{equation}
    \tau_\mathrm{w} = \frac{f}{2} \varrho v |v|
\end{equation}
with the friction factor $f$, and introduce the hydraulic diameter which is defined by convention as 
\begin{equation}
    D_\mathrm{h} = \frac{4A}{P_\mathrm{w}},
\end{equation}
to obtain 
\begin{equation}
    \int\limits_{n_1}^{n_2} \frac{\tau_\mathrm{w} P_\mathrm{w}}{A} \dd s = \frac{2f \ell}{D_\mathrm{h}} \varrho v|v| = \frac{2f\ell}{D_\mathrm{h} \varrho A^2} |\dot{m}| \dot{m} 
\end{equation}
with $\ell$ as the branch length. The factor of $\dot{m}$ gives rise to the hydraulic resistance~\eqref{eq:hydraulic-resistance}.
The last term in~\eqref{eq:branch-lumped-conservation-law},
\begin{equation}
    \int\limits_{n_1}^{n_2} \varrho g \partial_s z \dd s = \varrho g \Delta z_{21},
\end{equation}
acts as a fixed pressure offset on the branch due to an elevation $\Delta z_{21} = z_2 - z_1$ between the nodes $n_1$ and $n_2$ and corresponds to a constant EMF/battery in series in an electric circuit. 
All terms together lead to 
\begin{equation}
    L_{\mathrm{h}} \ddt{\dot{m}} + R_{\mathrm{h}} \dot{m} + (p_2 - p_1) = \Delta p_\mathrm{src} - \varrho g \Delta z_{21}.
    \label{eq:branch-pressure-differences}
\end{equation}
Taking the whole network into account and summarizing the pressure differences due to pressure drivers and elevation into one single pressure source, results in the branch momentum equation~\eqref{eq:branch-momentum-equation}. Furthermore, Eq.~\eqref{eq:branch-pressure-differences} can be interpreted as a hydraulic form of Kirchhoff's voltage law as it essentially states that all pressure differences $\Delta p_b$ in a hydraulic network loop cancel out, 
\begin{equation}
    \sum\limits_b \Delta p_b = 0.
\end{equation}

\section{Sensitivity Studies}
This section studies the sensitivity of the simulation results with respect to the transverse thermal conductance, the hydraulic network parameters and the external magnetic field map.

\subsection{Transverse Thermal Conductance}\label{sec:conductance-sensitivity}
The W7-X conductor (see figure~\ref{fig:w7x-conductor}) is wrapped in a glass-epoxy insulation, such that adjacent jackets are separated by a total insulation thickness of $t_\mathrm{ins} = \SI{0.6}{\milli\meter}$. The contact width is taken as the side length of the quasi-square jacket, $w_\mathrm{ins} = \SI{16}{\milli\meter}$, which gives $w_\mathrm{ins}/t_\mathrm{ins} = 26.7$. The temperature-dependent thermal conductivity is taken from the glass-epoxy material data provided with \opensc. The winding pack is vacuum-impregnated, so that a contact resistance between the turns is neglected.

\begin{table}[tp]
    \centering
    \caption{Comparison between varying insulation conductances.}
    \begin{tabular}{ccc}
        \toprule 
        Insulation conductance factor & Quench onset & Hot-spot peak in DL1 \\ 
        \midrule 
        $\times 0.5$ &  $-20.97$\,s & 55.2\,K \\
        $\times 1$ (reference) & 0 & 55.4\,K \\
        $\times 2$ & $+26.95$\,s & 55.1\,K \\
        \bottomrule
    \end{tabular}
    \label{tab:insulation-conductance-comparison}
\end{table}

\begin{figure}[tp]
    \centering
    \includegraphics[width=1\linewidth]{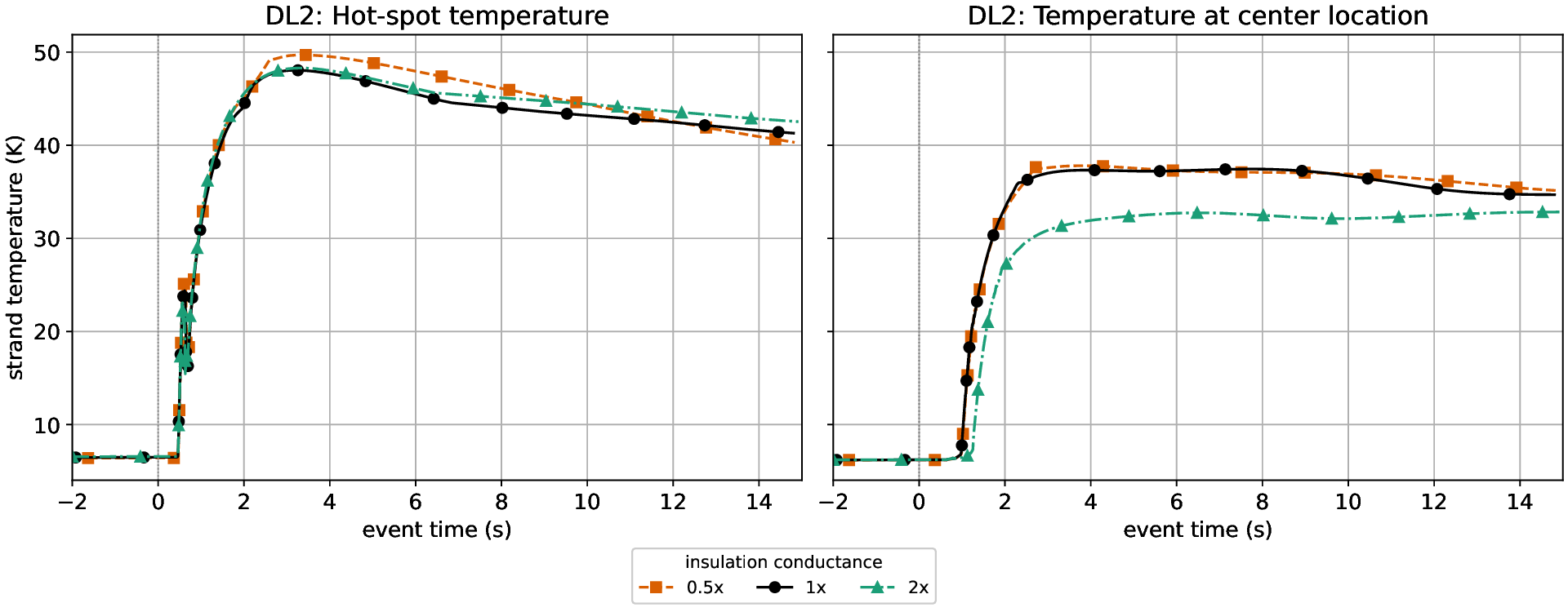}
    \caption{Hot-spot temperature (left) and temperature at the center location at $x=74.25\,$m (right) of DL2 for varying insulation conductances.}
    \label{fig:insulation_conductance_DL2}
\end{figure}

To assess the sensitivity to the transverse conductance (Sec.~\ref{subsec:transverse-thermal}), the full-feature run is repeated with half and with double the nominal conductance (table~\ref{tab:insulation-conductance-comparison}). Doubling the conductance delays the quench onset by \SI{27}{\second}, whereas halving it advances the onset by \SI{21}{\second}: the more heat is conducted from the developing normal zone into the neighboring turns, the longer it takes to reach the detection threshold. The peak hot-spot temperature of DL1, in contrast, stays within \SI{0.3}{\kelvin} of the reference.

Figure~\ref{fig:insulation_conductance_DL2} shows the hot-spot temperature of DL2 and its temperature at the center location ($x = \SI{74.25}{\meter}$). The hot-spot temperature changes by less than \SI{2}{\kelvin}, because the hot spot of the secondarily quenched DLs is driven by the flow redistribution at the inlets and not by the transverse heat conduction. At the center location, halving the conductance has almost no effect, whereas doubling it delays the temperature rise by about \SI{0.3}{\second} and lowers the subsequent plateau by about \SI{5}{\kelvin}.

In summary, an uncertainty of a factor of two in the transverse conductance changes the peak hot-spot temperatures of all DLs by less than \SI{3}{\kelvin} and the temperatures away from the hot spots by up to about \SI{5}{\kelvin}. The most sensitive quantity is the quench onset, since the whole coil is close to current sharing while the inlet temperature rises slowly.

\subsection{Hydraulic Network Parameters}\label{sec:network-sensitivity}
The hydraulic layout of the coil itself is known, whereas the network of the test stand outside the coil is only partly documented (Sec.~\ref{subsec:validation-simulation-model}). The model therefore contains assumed elements: the valves in the supply and return lines and the lines connecting the four outlet groups to the return line. Their influence on the full-feature run is examined in the following.

\subsubsection{Supply and Return Valves}
\begin{table}[tp]
    \centering
    \caption{Quench onset, peak strand hot-spot temperatures and peak mass flows for increased resistances of the supply and return valves. The pre-quench
    flow is the steady-state mass flow per DL. Differences to the reference are given in parentheses.}
    \begin{tabular}{lcccc}
        \toprule
        & Reference & $\times 2$ & $\times 10$ & $\times 100$  \\
        \midrule
        Resistance (\si{\pascal\square\second\per\square\kilo\gram}) & $1 \cdot 10^8$ & $2\cdot10^8$ & $1\cdot10^9$ & $1\cdot10^{10}$ \\
        Quench onset (s) & 1196.85 & 1199.75 ($+2.9$) & 1213.14 ($+16.3$) & 1293.78 ($+96.9$) \\
        Pre-quench flow per DL (g/s) & 0.60 & 0.58 & 0.51 & 0.26 \\
        \midrule
        \multicolumn{5}{l}{Strand hot spot (K)} \\
        \quad DL1 & 55.4 & 55.5 ($+0.1$) & 55.9 ($+0.5$) & 57.3 ($+1.9$) \\
        \quad DL2 & 48.1 & 48.2 ($+0.1$) & 50.6 ($+2.5$) & 52.0 ($+3.9$) \\
        \quad DL3 & 48.8 & 48.9 ($+0.1$) & 50.1 ($+1.3$) & 51.0 ($+2.2$) \\
        \quad DL4 & 49.4 & 49.9 ($+0.5$) & 51.2 ($+1.8$) & 52.3 ($+2.9$) \\
        \quad DL5 & 50.5 & 50.5 ($+0.0$) & 51.4 ($+0.9$) & 54.5 ($+4.0$) \\
        \quad DL6 & 49.3 & 48.8 ($-0.5$) & 48.8 ($-0.5$) & 54.0 ($+4.7$) \\
        \midrule
        \multicolumn{5}{l}{Peak mass flow (g/s)} \\
        \quad DL1 inlet (back-flow)      & 72.9  & 72.0  & 71.9 & 67.6 \\
        \quad DL1 outlet                 & 8.6   & 8.3   & 7.5  & 7.6  \\
        \quad DL2 outlet                 & 37.4  & 37.8  & 39.8 & 44.3 \\
        \quad Supply valve (back-flow)   & 182.9 & 144.0 & 77.0 & 27.6 \\
        \quad Return valve               & 106.9 & 91.9  & 60.1 & 27.0 \\
        \bottomrule
    \end{tabular}
    \label{tab:comparison-resistance-variants}
\end{table}

The quadratic resistance of the supply and return valves is increased by factors of 2, 10 and 100, see table~\ref{tab:comparison-resistance-variants}. The peak mass flows at the DL inlets and outlets hardly change, whereas the peak flows through the line valves drop by a factor of four to seven. The reason is that the helium expelled from a quenching DL is taken up mainly by the winding pack itself: the six DLs contain \SI{34.7}{\liter} of cold, compressible helium, compared with $V_\mathrm{rm} = \SI{1.8}{\liter}$ of the return volume. The large mass flows at the DL ends are therefore exchanged between the DLs through the common inlets and outlets and do not depend on the flow through the line valves.

The quench onset, in contrast, is delayed by up to \SI{97}{\second}, because the valves contribute to the flow resistance before the quench: a higher resistance reduces the steady-state mass flow, which carries the rise of the inlet temperature more slowly into the winding. The hot-spot temperatures change by at most \SI{2.5}{\kelvin} as long as the resistance is increased by up to one order of magnitude. Only for a hundredfold resistance do they rise in all DLs, by 1.9 to \SI{4.7}{\kelvin}, because the valves then let less helium leave the coil during the quench: the pressure and with it the compression heating are higher, and less heat is carried away.

The measured mass flow before the quench, 0.60 to \SI{0.65}{\gram\per\second} per DL, limits the plausible range of the valve resistances: a tenfold resistance already lowers the flow by \SI{15}{\percent}, and a hundredfold resistance by more than half, which is incompatible with the measurement. Within the plausible range, the uncertainty of the valve resistances thus changes the onset by at most \SI{16}{\second} and the hot spots by at most \SI{2.5}{\kelvin}.

\subsubsection{Outlet Lines}
In the reference model, each of the four outlet groups is connected to the return line through a valve with the quadratic characteristic $\Delta p = R\,\dot m^2$. The resistances, $R = 1.17\cdot10^{11}$, $2.86\cdot10^{10}$, $2.80\cdot10^{10}$ and $1.10\cdot10^{11}\,\si{\pascal\second\squared\per\kilogram\squared}$ for outlet groups 1 to 4, are chosen such that the pre-quench flow is \SI{0.6}{\gram\per\second} per DL, in accordance with the measured 0.60--\SI{0.65}{\gram\per\second} (figure~\ref{fig:massflow-measured}), which corresponds to a pressure drop of about \SI{0.4}{\bar} across each valve; no transient data enter this choice. To assess its influence, the run is repeated with outlet lines of negligible resistance (``open outlets'').

Table~\ref{tab:network-sensitivity} compares the two models. Without the throttles, the model carries 1.7 to 1.9 times the reference flow before the quench, the peak outlet flows rise from 7--\SI{15}{\gram\per\second} to 20--\SI{81}{\gram\per\second}, and the flow reverses at the outlets of groups 2 to 4, which the measurement cannot decide, since a venturi does not register the flow direction. At \SI{2.4}{\second}, the last sample at which all four measurements are valid, the outlet flows are 14 to 20 times higher than the measured ones with open outlets, and four to eight times higher in the reference model (Sec.~\ref{subsec:massflows}). The throttle valves thus bring the outlet flows considerably closer to the measurement, without reproducing it. Since more helium leaves through open outlets, all DLs stay colder, the hot spot of DL1 by \SI{1.5}{\kelvin} and those of DL2 to DL6 by 3.0 to \SI{10.0}{\kelvin}, and the quench onset occurs \SI{21}{\second} earlier, because the higher pre-quench flow carries the inlet temperature rise faster into the winding.

\begin{table}[tp]
    \centering
    \caption{Full-feature run with throttled and with open outlets, compared
    with the measurement where available. Differences to the reference are
    given in parentheses.}
    \begin{tabular}{lccc}
        \toprule
         & Throttled outlets (reference) & Open outlets & Measured \\
        \midrule
        Quench onset (s) & 1196.85 & 1175.74 ($-21.1$) & -- \\
        Peak outlet pressure & 16.8\,bar at 2.5\,s & 34.5\,bar at 2.0\,s & 22.1\,bar at 15\,s \\
        \midrule
        \multicolumn{4}{l}{Pre-quench outlet flow (g/s)} \\
        \quad Outlet 1 (DL1)       & 0.61 & 1.12 & 0.60 \\
        \quad Outlet 2 (DL2, DL3)  & 1.21 & 2.19 & 1.27 \\
        \quad Outlet 3 (DL4, DL5)  & 1.21 & 2.13 & 1.25 \\
        \quad Outlet 4 (DL6)       & 0.60 & 1.04 & 0.65 \\
        \midrule
        \multicolumn{4}{l}{Peak outlet flow (g/s)} \\
        \quad Outlet 1 (DL1)       & 8.6  & 79 & \multirow{4}{*}{above range} \\
        \quad Outlet 2 (DL2, DL3)  & 15.3 & 81 & \\
        \quad Outlet 3 (DL4, DL5)  & 14.5 & 57 & \\
        \quad Outlet 4 (DL6)       & 7.1  & 20 & \\
        \midrule
        \multicolumn{4}{l}{Outlet flow at 2.4\,s (g/s)} \\
        \quad Outlet 1 (DL1)       & 8.4  & 22.1 & 1.09 \\
        \quad Outlet 2 (DL2, DL3)  & 13.7 & 32.4 & 1.87 \\
        \quad Outlet 3 (DL4, DL5)  & 12.3 & 29.9 & 1.80 \\
        \quad Outlet 4 (DL6)       & 5.7  & 18.1 & 1.28 \\
        \midrule
        \multicolumn{4}{l}{Strand hot spot (K)} \\
        \quad DL1 & 55.4 & 53.9 ($-1.5$)  & -- \\
        \quad DL2 & 48.1 & 45.1 ($-3.0$)  & -- \\
        \quad DL3 & 48.8 & 43.6 ($-5.2$)  & -- \\
        \quad DL4 & 49.4 & 42.2 ($-7.2$)  & -- \\
        \quad DL5 & 50.5 & 40.6 ($-9.9$)  & -- \\
        \quad DL6 & 49.3 & 39.3 ($-10.0$) & -- \\
        \bottomrule
    \end{tabular}
    \label{tab:network-sensitivity}
\end{table}

\begin{table}[tp]
    \centering
    \caption{Full-feature run with the calibrated and with the unscaled friction correlation of the bundle. Differences to the reference are given in parentheses.}
    \begin{tabular}{lcc}
        \toprule
         & Multiplier 0.376 (reference) & Multiplier 1 \\
        \midrule
        Quench onset (s) & 1196.8 & 1213.1 ($+16.3$) \\
        Secondary-quench onset, DL2 to DL6 (s) & 0.511--0.518 & 0.507--0.520 \\
        Peak pressure at the inlet filter (bar) & 39.7 & 33.2 ($-6.5$) \\
        Peak pressure in the return manifold (bar) & 16.8 & 20.4 ($+3.6$) \\
        Peak back-flow at the inlet of DL1 (g/s) & 72.9 & 54.2 ($-18.7$) \\
        Peak back-flow through the supply valve (g/s) & 183 & 164 ($-19$) \\
        \midrule
        \multicolumn{3}{l}{Peak outlet flow (g/s)} \\
        \quad Outlet 1 (DL1)       & 8.6  & 11.8 ($+3.2$) \\
        \quad Outlet 2 (DL2, DL3)  & 15.3 & 20.4 ($+5.1$) \\
        \quad Outlet 3 (DL4, DL5)  & 14.5 & 19.3 ($+4.8$) \\
        \quad Outlet 4 (DL6)       & 7.1  & 9.8 ($+2.7$) \\
        \midrule
        \multicolumn{3}{l}{Outlet temperature at 9\,s (K)} \\
        \quad Outlet 1 (DL1)       & 46.7 & 46.8 ($+0.1$) \\
        \quad Outlet 2 (DL2, DL3)  & 37.1 & 36.8 ($-0.3$) \\
        \quad Outlet 3 (DL4, DL5)  & 32.5 & 31.5 ($-1.0$) \\
        \quad Outlet 4 (DL6)       & 30.8 & 28.6 ($-2.2$) \\
        \midrule
        \multicolumn{3}{l}{Strand hot spot (K)} \\
        \quad DL1 & 55.4 & 57.6 ($+2.2$) \\
        \quad DL2 & 48.1 & 46.6 ($-1.5$) \\
        \quad DL3 & 48.8 & 49.9 ($+1.1$) \\
        \quad DL4 & 49.4 & 48.2 ($-1.2$) \\
        \quad DL5 & 50.5 & 48.1 ($-2.4$) \\
        \quad DL6 & 49.3 & 37.4 ($-11.9$) \\
        \bottomrule
    \end{tabular}
    \label{tab:friction-sensitivity}
\end{table}

\subsubsection{Conductor Friction}\label{subsec:friction-sensitivity}
The friction factor of the bundle follows a Darcy-Forchheimer porous-medium correlation~\cite{bagnasco_friction_2010} scaled by a factor of 0.376 (Appendix~\ref{app:conductor-parameters}). This factor and the four throttle resistances were calibrated together at the operating point before the quench, i.e.\ against the measured pressure difference of \SI{61.8}{\kilo\pascal} between supply and return and the measured mass flows of the outlet groups. These data fix only the sum of the pressure drops of conductor and throttle in each group, not their split, and no independent measurement of the conductor friction is available. The run is therefore repeated with the unscaled correlation, i.e.\ a 2.7-fold friction, and with the throttle resistances reduced such that the pre-quench flows are unchanged.

Table~\ref{tab:friction-sensitivity} compares the two runs. With the unscaled correlation, less helium is expelled through the inlets and more through the outlets, so that the pressure at the inlets drops and that in the return manifold rises. The secondary quenches of DL2 to DL6, in contrast, start within \SI{4}{\milli\second} of the reference run, and the outlet temperatures and the hot spots change by at most \SI{2.4}{\kelvin}, except in DL6, where the secondary quench is weaker. The undetermined split between conductor friction and throttling thus changes the pressures and the flows by up to \SI{40}{\percent}, but neither the sequence of the quenches nor the outlet temperatures, and it does not explain the discrepancy in the time scale.

\subsection{External Magnetic Field Map}\label{subsec:magnetic-field-sensitivity}
\begin{figure}[tp]
    \centering
    \includegraphics[width=1\linewidth]{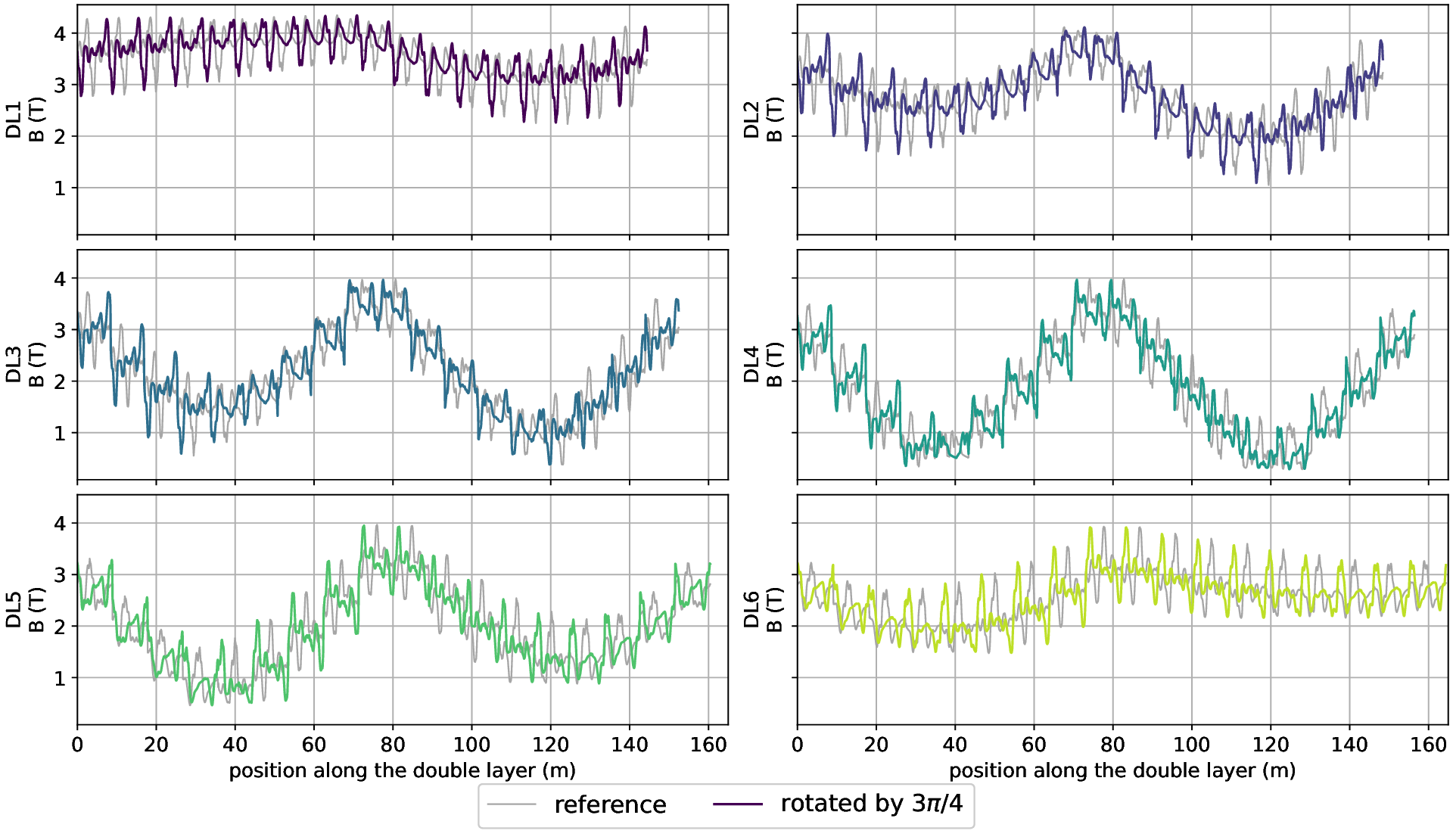}
    \caption{Magnetic field of the rotated map (colored) compared to the reference map (gray) for each DL.}
    \label{fig:rotated_field}
\end{figure}

The position of every turn within the WP cross-section is known, but the position along the coil perimeter at which the windings start and end is not documented for AAB-51 and was chosen arbitrarily. This choice shifts the periodic variation of the magnetic flux density (Sec.~\ref{subsec:3d-bfield}) within each turn along the conductor coordinate $x$. To assess its influence, the full-feature run is repeated with the field map rotated by $3\pi/4$ along the coil perimeter, i.e.\ shifted by $3/8$ of a turn length or about \SI{3}{\meter} (figure~\ref{fig:rotated_field}).

\begin{table}[tp]
    \centering
    \caption{Quench onset and peak strand hot-spot temperatures for the reference and the rotated field map.}
    \begin{tabular}{lccc}
        \toprule 
         & Reference & Rotated & Difference to reference \\
        \midrule  
        Quench onset & 1196.85\,s & 1265.85\,s & $+69.01$\,s \\
        Strand hot spot DL1 & 55.4\,K & 55.6\,K & $+0.2$\,K \\
        Strand hot spot DL2 & 48.1\,K & 52.7\,K & $+4.6$\,K \\
        Strand hot spot DL3 & 48.8\,K & 51.0\,K & $+2.2$\,K \\ 
        Strand hot spot DL4 & 49.4\,K & 50.8\,K & $+1.4$\,K \\
        Strand hot spot DL5 & 50.5\,K & 51.2\,K & $+0.7$\,K \\
        Strand hot spot DL6 & 49.3\,K & 50.4\,K & $+1.1$\,K \\
        \bottomrule
    \end{tabular}
    \label{tab:rotated-field-comparison}
\end{table}

The primary quench still starts in DL1, but its onset shifts by \SI{69}{\second} (table~\ref{tab:rotated-field-comparison}): the rotation changes where the field maxima lie along the conductor, and thus at which inlet temperature the margin is first exhausted.

\begin{figure}[tp]
    \centering
    \includegraphics[width=0.75\linewidth]{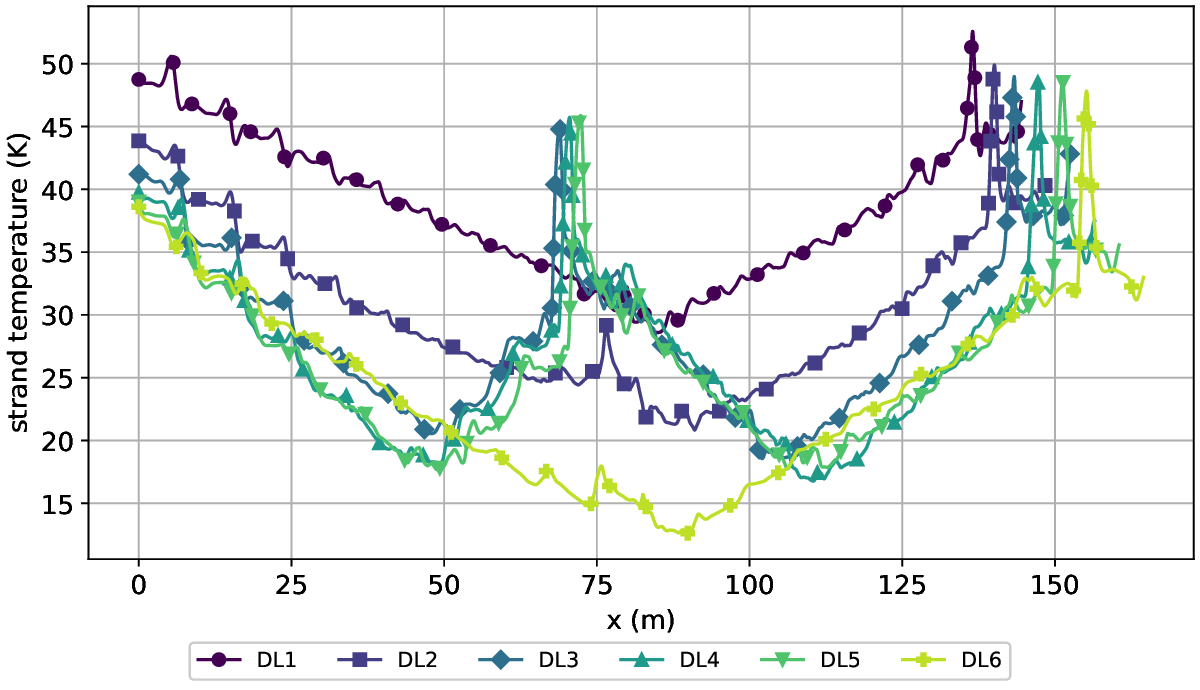}
    \caption{Computed temperatures along the DLs at $t=2\,$s for the rotated field map.}
    \label{fig:rotated-field-spatial-T}
\end{figure}

The peak hot-spot temperatures are affected only moderately: that of the primarily quenched DL1 is practically unchanged, and those of the other DLs rise by 0.7 to \SI{4.6}{\kelvin}.

What changes is where the local quenches develop within the DLs (figure~\ref{fig:rotated-field-spatial-T}). The narrow temperature peaks that compression heating produces at mid-length of DL2 to DL6 in the reference run (figure~\ref{fig:results-temperature-profiles}a) remain in DL3 to DL5 only, whereas DL2 shows only a weak one and DL6 none. Instead, every DL develops a pronounced peak of 47 to \SI{52}{\kelvin} in its last turn before the outlet, because the rotation moves a field maximum, and with it a minimum of the current-sharing temperature, into this turn.

The primary quench in DL1, the secondary quenches in all neighboring DLs and the hot-spot temperatures (to within about \SI{5}{\kelvin}) are therefore robust against the unknown position of the windings along the coil perimeter. The onset time, the positions of the compression-induced local quenches and the axial position of the quench initiation within a turn are not: they depend on local field maxima and cannot be predicted reliably without knowing the terminal positions.

\subsection{Time Step}\label{subsec:time-step-sensitivity}

The transverse heat exchange (Sec.~\ref{subsec:transverse-thermal}) and the node temperatures (Sec.~\ref{subsec:field-network-coupling}) lag the solution by one time step. To quantify the resulting error, the reference run was repeated with the minimum and maximum time steps halved to \SI{0.5}{\milli\second} and \SI{0.5}{\second} and the tolerance $\epsilon$ reduced from $10^{-3}$ to $2.5\cdot 10^{-4}$, which corresponds to half the time step since the error indicator of Appendix~\ref{subsec:efficiency} scales with its square. The number of time steps rises from 7159 to 12808.

The results are compared in table~\ref{tab:time-step-comparison}. The secondary quenches of DL2 to DL6 start in the same sequence and within \SI{4}{\milli\second} of the reference run, and the sustained flow reversal at the inlets shifts by at most \SI{5}{\milli\second}, except in DL2 (\SI{26}{\milli\second}). The peak temperatures and mass flows change by about \SI{1}{\percent}, and the peak pressures rise from 39.7 to \SI{39.9}{\bar} at the inlet filter and from 16.8 to \SI{16.9}{\bar} in the return manifold. On the steep fronts of the first \SI{1.5}{\second}, the traces differ point by point by up to \SI{2.8}{\kelvin} and \SI{2.1}{\gram\per\second}. The quench detection occurs \SI{1.7}{\second} earlier, which corresponds to \SI{6}{\milli\kelvin} of supply temperature at the ramp rate of \SI{3.6}{\milli\kelvin\per\second}.

\begin{table}[tp]
    \centering
    \caption{Reference run (ref.) and run with half the time step (half). Onset and reversal times are event times; the quench onset is the time at which the strand temperature first exceeds \SI{8}{\kelvin}, the flow reversal the time after which the inlet flow stays reversed.}
    \begin{tabular}{lcccccc}
        \toprule
         & \multicolumn{2}{c}{Quench onset (s)} & \multicolumn{2}{c}{Inlet flow reversal (s)} & \multicolumn{2}{c}{Peak hot spot (K)} \\
        \cmidrule(lr){2-3} \cmidrule(lr){4-5} \cmidrule(lr){6-7}
         & ref. & half & ref. & half & ref. & half \\
        \midrule
        DL1 & 0.366 & 0.369 & 0.379 & 0.383 & 55.4 & 55.3 \\
        DL2 & 0.471 & 0.470 & 0.556 & 0.582 & 48.1 & 48.3 \\
        DL3 & 0.472 & 0.471 & 0.652 & 0.647 & 48.8 & 48.8 \\
        DL4 & 0.473 & 0.472 & 0.693 & 0.695 & 49.4 & 49.8 \\
        DL5 & 0.473 & 0.472 & 0.731 & 0.728 & 50.5 & 50.9 \\
        DL6 & 0.474 & 0.472 & 0.752 & 0.749 & 49.3 & 48.9 \\
        \bottomrule
    \end{tabular}

    \medskip
    \begin{tabular}{lcccc}
        \toprule
         & \multicolumn{2}{c}{Peak temperature (K)} & \multicolumn{2}{c}{Peak mass flow (g/s)} \\
        \cmidrule(lr){2-3} \cmidrule(lr){4-5}
         & ref. & half & ref. & half \\
        \midrule
        Outlet 1 & 47.6 & 47.7 & 8.6 & 8.7 \\
        Outlet 2 & 38.9 & 39.4 & 15.3 & 15.4 \\
        Outlet 3 & 33.5 & 33.8 & 14.5 & 14.5 \\
        Outlet 4 & 30.8 & 30.9 & 7.1 & 7.2 \\
        \bottomrule
    \end{tabular}
    \label{tab:time-step-comparison}
\end{table}

\begin{table}[tp]
    \centering
    \caption{Critical-surface and conductor parameters of the W7-X conductor model.}
    \begin{tabular}{llc}
        \toprule
        Quantity & Symbol & Value \\
        \midrule
        \multicolumn{3}{l}{Critical surface of NbTi} \\
        \quad Critical temperature at zero field & $T_\mathrm{c0}$ & \SI{9.03}{\kelvin} \\
        \quad Upper critical field at zero temperature & $B_\mathrm{c20}$ & \SI{14.61}{\tesla} \\
        \quad Scaling constant & $C_0$ & $1.643\cdot10^{11}$\,\si{\ampere\tesla\per\square\meter} \\
        \quad Fit exponents & $\alpha_\mathrm{c}$, $\beta_\mathrm{c}$, $\gamma_\mathrm{c}$, $\nu$ & 1.0, 1.54, 2.1, 1.7 \\
        \quad Power-law exponent & & 25 \\
        \quad Electric field criterion & & \SI{10}{\micro\volt\per\meter} \\
        \midrule
        \multicolumn{3}{l}{Conductor} \\
        \quad Number of strands & & 243 \\
        \quad Strand diameter & & \SI{0.57}{\milli\meter} \\
        \quad Copper to non-copper ratio & & 2.6 \\
        \quad Residual resistivity ratio of the copper & & 160 \\
        \quad Filament diameter & $d_\mathrm{f}$ & \SI{26}{\micro\meter} \\
        \quad Twist pitch factor of the strands & $\cos(\theta)$ & 0.985 \\
        \quad Strand cross-section & & \SI{62.0}{\square\milli\meter} \\
        \quad Helium cross-section & & \SI{37.5}{\square\milli\meter} \\
        \quad Void fraction & & 0.375 \\
        \quad Hydraulic diameter & $D_\mathrm{h}$ & \SI{0.376}{\milli\meter} \\
        \quad Friction correlation of the bundle & & Darcy--Forchheimer~\cite{bagnasco_friction_2010} \\
        \quad Friction multiplier (Appendix~\ref{subsec:friction-sensitivity}) & & 0.376 \\
        \quad Jacket material & & Al 6063 \\
        \quad Jacket cross-section & & \SI{151.1}{\square\milli\meter} \\
        \quad Jacket side length & & \SI{16}{\milli\meter} \\
        \bottomrule
    \end{tabular}
    \label{tab:conductor-parameters}
\end{table}

\section{Conductor Parameters}\label{app:conductor-parameters}
The critical current density of the NbTi strands is described by the practical fit of~\cite{bottura_practical_2000},
\begin{equation}
    J_\mathrm{c}(T,B) = \frac{C_0}{B} \, b^{\alpha_\mathrm{c}} (1-b)^{\beta_\mathrm{c}} \left( 1-t^{\nu} \right)^{\gamma_\mathrm{c}},
\end{equation}
with the reduced temperature $t = T/T_\mathrm{c0}$, the reduced field $b = B/B_\mathrm{c2}(T)$ and the upper critical field $B_\mathrm{c2}(T) = B_\mathrm{c20} \left( 1-t^{\nu} \right)$. It is referred to the non-copper cross-section of the strands. A strain dependence is not considered, as is common for NbTi. The parameters of the fit and of the conductor are those of the W7-X THEA model~\cite{risse_thea_2023, risse_thermohydraulic_2024} and are listed in table~\ref{tab:conductor-parameters}.

\printbibliography

@article{amestoy_multifrontal_2000,
  title = {Multifrontal Parallel Distributed Symmetric and Unsymmetric Solvers},
  author = {Amestoy, P. R. and Duff, I. S. and L'Excellent, J.-Y.},
  date = {2000-04-14},
  journaltitle = {Comput. Methods Appl. Mech. Eng.},
  volume = {184},
  number = {2},
  pages = {501--520},
  doi = {10.1016/S0045-7825(99)00242-X},
}

@book{anderson_lapack_1999,
  title = {{{LAPACK Users}}' {{Guide}}},
  author = {Anderson, E. and Bai, Z. and Bischof, C. and Blackford, L. S. and Demmel, J. and Dongarra, J. and Du Croz, J. and Greenbaum, A. and Hammarling, S. and McKenney, A. and Sorensen, D.},
  date = {1999-01},
  series = {Software, {{Environments}}, and {{Tools}}},
  publisher = {{Society for Industrial and Applied Mathematics}},
  doi = {10.1137/1.9780898719604},
  isbn = {978-0-89871-447-0},
}

@article{bagnasco_friction_2010,
  title = {Friction Factor Correlation for {{CICC}}'s Based on a Porous Media Analogy},
  author = {Bagnasco, M. and Bottura, L. and Lewandowska, M.},
  year = 2010,
  month = nov,
  journal = {Cryogenics},
  volume = {50},
  number = {11},
  pages = {711--719},
  issn = {0011-2275},
  doi = {10.1016/j.cryogenics.2010.06.016},
}

@article{bagni_analysis_2017,
  title = {Analysis of {{ITER NbTi}} and {{Nb3Sn CICCs}} Experimental Minimum Quench Energy with {{JackPot}}, {{MCM}} and {{THEA}} Models},
  author = {Bagni, T and Duchateau, J L and Breschi, M and Devred, A and Nijhuis, A},
  date = {2017-07},
  journal = {Supercond. Sci. Technol.},
  volume = {30},
  number = {9},
  pages = {095003},
  publisher = {IOP Publishing},
  doi = {10.1088/1361-6668/aa7a80},
}

@incollection{bald_supercritical_1971,
  title = {Supercritical {{Helium Cooling}} of {{Hollow Superconductors}}},
  booktitle = {Advances in {{Cryogenic Engineering}}: {{Proceeding}} of the 1970 {{Cryogenic Engineering Conference The University}} of {{Colorado Boulder}}, {{Colorado June}} 17–17, 1970},
  author = {Bald, W. B.},
  editor = {Timmerhaus, K. D.},
  date = {1971},
  pages = {368--377},
  publisher = {Springer US},
  location = {Boston, MA},
  doi = {10.1007/978-1-4757-0244-6_46},
}

@article{baldzuhn_coil_2006,
  title = {Coil Tests and Superconductor Code Calculations for the Stellarator {{W7-X}} Coils},
  author = {Baldzuhn, J. and Ehmler, H. and Hoelting, A. and Hertel, K. and Sborchia, C. and Genini, L. and Schild, T.},
  date = {2006-07-01},
  journaltitle = {Cryogenics},
  volume = {46},
  number = {7},
  pages = {507--516},
  doi = {10.1016/j.cryogenics.2006.01.002},
}

@article{bell_pure_2014,
  title = {Pure and {{Pseudo-pure Fluid Thermophysical Property Evaluation}} and the {{Open-Source Thermophysical Property Library CoolProp}}},
  author = {Bell, Ian H. and Wronski, Jorrit and Quoilin, Sylvain and Lemort, Vincent},
  date = {2014-01-13},
  journaltitle = {Ind. Eng. Chem. Res.},
  volume = {53},
  number = {6},
  pages = {2498--2508},
  doi = {10.1021/ie4033999},
}

@article{bessette_simulations_2006,
  title = {Simulations of the {{ITER Toroidal Field Coil Operation With}} the {{VINCENTA Code}}},
  author = {Bessette, D. and Shatil, N. and Zapretilina, E.},
  date = {2006-06},
  journaltitle = {IEEE Trans. Appl. Supercond.},
  volume = {16},
  number = {2},
  pages = {795--798},
  doi = {10.1109/TASC.2006.873258},
}

@article{bishop_there_1975,
  title = {There Is {{More}} than {{One Way}} to {{Frame}} a {{Curve}}},
  author = {Bishop, Richard L.},
  date = {1975},
  journaltitle = {The American Mathematical Monthly},
  volume = {82},
  number = {3},
  eprint = {2319846},
  eprinttype = {jstor},
  pages = {246--251},
  publisher = {[Taylor \& Francis, Ltd., Mathematical Association of America]},
  doi = {10.2307/2319846},
}

@article{bottura_general_2000,
  title = {A General Model for Thermal, Hydraulic and Electric Analysis of Superconducting Cables},
  author = {Bottura, L. and Rosso, C. and Breschi, M.},
  date = {2000-08-01},
  journaltitle = {Cryogenics},
  series = {4th {{Workshop}} on {{Computation}} of {{Thermal Hydraulic Transients}} in {{Superconductors}}},
  volume = {40},
  number = {8},
  pages = {617--626},
  doi = {10.1016/S0011-2275(01)00019-4},
}

@article{bottura_flower_2003,
  title = {Flower, a Model for the Analysis of Hydraulic Networks and Processes},
  author = {Bottura, L. and Rosso, C.},
  date = {2003-03-01},
  journaltitle = {Cryogenics},
  volume = {43},
  number = {3},
  pages = {215--223},
  doi = {10.1016/S0011-2275(03)00038-9},
}

@article{bottura_porous_2008,
  title = {A Porous Medium Analogy for the Helium Flow in {{CICCs}}},
  author = {Bottura, L. and Marinucci, C.},
  date = {2008-05-01},
  journaltitle = {Int. J. Heat Mass Transfer},
  volume = {51},
  number = {9},
  pages = {2494--2505},
  doi = {10.1016/j.ijheatmasstransfer.2007.07.051},
}

@article{bottura_practical_2000,
  title = {A Practical Fit for the Critical Surface of {{NbTi}}},
  author = {Bottura, L.},
  year = 2000,
  month = mar,
  journal = {IEEE Transactions on Applied Superconductivity},
  volume = {10},
  number = {1},
  pages = {1054--1057},
  issn = {1558-2515},
  doi = {10.1109/77.828413},
}

@book{braess_finite_2007,
  author    = {Braess, Dietrich},
  title     = {Finite Elements: Theory, Fast Solvers, and Applications in Solid Mechanics},
  edition   = {3rd},
  publisher = {Cambridge University Press},
  address   = {Cambridge},
  year      = {2007},
  doi       = {10.1017/CBO9780511618635}
}

@article{campbell_general_1982,
  title = {A General Treatment of Losses in Multifilamentary Superconductors},
  author = {Campbell, A. M.},
  date = {1982-01-01},
  journaltitle = {Cryogenics},
  volume = {22},
  number = {1},
  pages = {3--16},
  doi = {10.1016/0011-2275(82)90015-7},
}

@manual{cryosoft_supermagnet_2021,
  title        = {{SuperMagnet}: Multitasking Code Manager. User's Guide},
  organization = {CryoSoft},
  edition      = {Version 2.1a},
  month        = nov,
  year         = {2021}
}

@article{dangelo_quasi3d_2022,
  title = {Quasi-{{3D Magneto-Thermal Quench Simulation Scheme}} for {{Superconducting Accelerator Magnets}}},
  author = {D’Angelo, Laura A. M. and Späck-Leigsnering, Yvonne and De Gersem, Herbert},
  date = {2022-09},
  journaltitle = {IEEE Trans. Appl. Supercond.},
  volume = {32},
  number = {6},
  pages = {1--5},
  doi = {10.1109/TASC.2022.3159302},
}

@article{dular_general_1998,
  title = {A General Environment for the Treatment of Discrete Problems and Its Application to the Finite Element Method},
  author = {Dular, P. and Geuzaine, C. and Henrotte, F. and Legros, W.},
  date = {1998-09},
  journaltitle = {IEEE Trans. Mag.},
  volume = {34},
  number = {5},
  pages = {3395--3398},
  doi = {10.1109/20.717799},
}

@book{golub_matrix_2013, 
    title = {Matrix Computations},
    author = {Golub, Gene H. and Van Loan, Charles F.},
    data = {2013},
    publisher = {John Hopkins University Press},
    location = {Baltimore},
    edition = {4th ed.},
    isbn = {978-1-4214-0794-4},
}

@book{hairer_solving_1993,
  title = {Solving {{Ordinary Differential Equations I}}},
  author = {Hairer, E. and N\o rsett, S. P. and Wanner, G.},
  date = {1993},
  series = {Springer {{Series}} in {{Computational Mathematics}}},
  volume = {8},
  publisher = {Springer},
  location = {Berlin, Heidelberg},
  doi = {10.1007/978-3-540-78862-1},
  isbn = {978-3-540-78862-1},
}

@article{hindmarsh_sundials_2005,
  title = {{{SUNDIALS}}: {{Suite}} of Nonlinear and Differential/Algebraic Equation Solvers},
  author = {Hindmarsh, Alan C. and Brown, Peter N. and Grant, Keith E. and Lee, Steven L. and Serban, Radu and Shumaker, Dan E. and Woodward, Carol S.},
  date = {2005-09-01},
  journaltitle = {ACM Trans. Math. Softw.},
  volume = {31},
  number = {3},
  pages = {363--396},
  doi = {10.1145/1089014.1089020},
}

@article{ho_modified_1975,
  title = {The Modified Nodal Approach to Network Analysis},
  author = {Ho, Chung-Wen and Ruehli, A. and Brennan, P.},
  date = {1975-06},
  journaltitle = {IEEE Trans. Circuits Syst.},
  volume = {22},
  number = {6},
  pages = {504--509},
  doi = {10.1109/TCS.1975.1084079},
}

@article{hughes_finite_1984,
  title = {Finite Element Methods for First-Order Hyperbolic Systems with Particular Emphasis on the Compressible Euler Equations},
  author = {Hughes, T. J. R. and Tezduyar, T. E.},
  date = {1984-09-01},
  journaltitle = {Comput. Methods Appl. Mech. Eng.},
  volume = {45},
  number = {1},
  pages = {217--284},
  doi = {10.1016/0045-7825(84)90157-9},
}

@article{landreman_efficient_2025,
  title = {Efficient Calculation of Self Magnetic Field, Self-Force, and Self-Inductance for Electromagnetic Coils with Rectangular Cross-Section},
  author = {Landreman, Matt and Hurwitz, Siena and Antonsen, Thomas M.},
  date = {2025-02},
  journaltitle = {Nucl. Fusion},
  volume = {65},
  number = {3},
  pages = {036008},
  publisher = {IOP Publishing},
  doi = {10.1088/1741-4326/adb04e},
}

@article{lewandowska_modelling_2024,
  title = {Modelling and Analysis of Quench in the 15-{{kA HTS}} Conductor},
  author = {Lewandowska, Monika and Dembkowska, Aleksandra and Ortwein, Rafał and Bocian, Dariusz and Dicuonzo, Ortensia and Sedlak, Kamil},
  date = {2024-07-01},
  journaltitle = {Cryogenics},
  volume = {141},
  pages = {103889},
  doi = {10.1016/j.cryogenics.2024.103889},
}

@article{matko_pipeline_2000,
  title = {Pipeline Simulation Techniques},
  author = {Matko, Drago and Geiger, Gerhard and Gregoritza, Withold},
  date = {2000-06-30},
  journaltitle = {Math. Comput. Simul.},
  volume = {52},
  number = {3},
  pages = {211--230},
  doi = {10.1016/S0378-4754(00)00152-X},
}

@article{muzzi_cableinconduit_2015,
  title = {Cable-in-Conduit Conductors: Lessons from the Recent Past for Future Developments with Low and High Temperature Superconductors},
  author = {Muzzi, L and De Marzi, G and Di Zenobio, A and family=Corte, given=A, prefix=della, useprefix=true},
  date = {2015-03},
  journal = {Supercond. Sci. Technol.},
  volume = {28},
  number = {5},
  pages = {053001},
  publisher = {IOP Publishing},
  doi = {10.1088/0953-2048/28/5/053001},
}

@book{oberkampf_verification_2010,
  title = {Verification and {{Validation}} in {{Scientific Computing}}},
  author = {Oberkampf, W. L. and Roy, C. J.},
  date = {2010},
  publisher = {Cambridge University Press}
}

@article{prost_economically_2024,
  title = {Economically Optimized Design Point of High-Field Stellarator Power-Plant},
  author = {Prost, Victor and Volpe, Francesco A.},
  date = {2024-01},
  journaltitle = {Nucl. Fusion},
  volume = {64},
  number = {2},
  pages = {026007},
  publisher = {IOP Publishing},
  doi = {10.1088/1741-4326/ad142e},
}

@article{quarteroni_coupling_2001,
  title = {Coupling between Lumped and Distributed Models for Blood Flow Problems},
  author = {Quarteroni, Alfio and Ragni, Stefania and Veneziani, Alessandro},
  date = {2001-12-01},
  journaltitle = {Comput. Visual. Sci.},
  volume = {4},
  number = {2},
  pages = {111--124},
  doi = {10.1007/s007910100063},
}

@article{ravaioli_lumpedelement_2016,
  title = {Lumped-{{Element Dynamic Electro-Thermal}} Model of a Superconducting Magnet},
  author = {Ravaioli, E. and Auchmann, B. and Maciejewski, M. and family=Kate, given=H. H. J., prefix=ten, useprefix=true and Verweij, A. P.},
  date = {2016-12-01},
  journaltitle = {Cryogenics},
  volume = {80},
  pages = {346--356},
  doi = {10.1016/j.cryogenics.2016.04.004},
}

@techreport{risse_thea_2023,
  type = {Technical Report},
  title = {Thea Model for Quench Analysis on {{W7-X}} Superconductor},
  author = {Ri{\ss}e, Konrad},
  year = 2023,
  month = may,
  number = {1-AAB-T 2F7WLT V1.1},
  address = {Greifswald, Germany},
  institution = {IPP},
}

@article{risse_thermohydraulic_2024,
  title = {Thermo-{{Hydraulic Calculations}} on {{W7-X Coils}} for {{Updated Quench Detection Parameters}}},
  author = {Risse, Konrad and Lewandowska, Monika and Dembkowska, Aleksandra and Rummel, Thomas and Bosch, Hans-Stephan and Team, the W7-X.},
  date = {2024-08},
  journaltitle = {IEEE Trans. Appl. Supercond.},
  volume = {34},
  number = {5},
  pages = {1--5},
  doi = {10.1109/TASC.2024.3376634},
}

@article{roache_code_2001,
  title = {Code {{Verification}} by the {{Method}} of {{Manufactured Solutions}}},
  author = {Roache, Patrick J.},
  date = {2001-11-12},
  journaltitle = {J. Fluids Eng},
  volume = {124},
  number = {1},
  pages = {4--10},
  doi = {10.1115/1.1436090},
}

@article{rummel_superconducting_2012,
  title = {The {{Superconducting Magnet System}} of the {{Stellarator Wendelstein}} 7-{{X}}},
  author = {Rummel, Thomas and Ribe, Konrad and Ehrke, Gunnar and Rummel, Kerstin and John, Andre and Monnich, Thomas and Buscher, Klaus-Peter and Fietz, Walter H. and Heller, Reinhard and Neubauer, Olaf and Panin, Anatoly},
  date = {2012-03},
  journaltitle = {IEEE Trans. Plasma Sci.},
  volume = {40},
  number = {3},
  pages = {769--776},
  doi = {10.1109/TPS.2012.2184774},
}

@article{savoldirichard_4c_2010,
  title = {The {{4C}} Code for the Cryogenic Circuit Conductor and Coil Modeling in {{ITER}}},
  author = {Savoldi Richard, L. and Casella, F. and Fiori, B. and Zanino, R.},
  date = {2010-03-01},
  journaltitle = {Cryogenics},
  volume = {50},
  number = {3},
  pages = {167--176},
  doi = {10.1016/j.cryogenics.2009.07.008},
}

@article{savoldi_thermalhydraulic_2022,
  title = {Thermal-Hydraulic Analysis of Superconducting Cables for Energy Applications with a Novel Open Object-Oriented Software: {{OPENSC2}}},
  author = {Savoldi, Laura and Placido, Daniele and Viarengo, Sofia},
  date = {2022-06-01},
  journaltitle = {Cryogenics},
  volume = {124},
  pages = {103457},
  publisher = {Elsevier},
  doi = {10.1016/j.cryogenics.2022.103457},
}

@article{schops_winding_2013,
  title = {Winding Functions in Transient Magnetoquasistatic Field-Circuit Coupled Simulations},
  author = {Schöps, Sebastian and De Gersem, Herbert and Weiland, Thomas},
  date = {2013-11-11},
  journaltitle = {COMPEL},
  volume = {32},
  number = {6},
  pages = {2063--2083},
  doi = {10.1108/COMPEL-01-2013-0004},
}

@article{schwerg_quench_2008,
  title = {Quench {{Simulation}} in an {{Integrated Design Environment}} for {{Superconducting Magnets}}},
  author = {Schwerg, Nikolai and Auchmann, Bernhard and Russenschuck, Stephan},
  date = {2008-06},
  journaltitle = {IEEE Trans. Mag.},
  volume = {44},
  number = {6},
  pages = {934--937},
  doi = {10.1109/TMAG.2007.916304},
}

@book{simonyi_foundations_1963,
  title = {Foundations of {{Electrical Engineering}}: {{Fields--Networks--Waves}}},
  author = {Simonyi, Károly},
  date = {1963},
  publisher = {Pergamon Press},
  location = {Oxford, England},
  doi = {10.1016/C2013-0-02694-1},
  isbn = {978-0-08-019001-3}
}

@book{smythe_static_1968,
  title = {Static and {{Dynamic Electricity}}},
  author = {Smythe, William R.},
  date = {1968},
  origdate = {1939},
  series = {International {{Series}} in {{Pure}} and {{Applied Physics}}},
  edition = {3rd ed.},
  publisher = {McGraw-Hill International Book Co.},
  location = {USA}
}

@article{takacs_loss_2006,
  title = {Loss Contributions in Superconducting Magnets Caused by Transient Magnetic Fields},
  author = {Takács, S. and Werner, A. and Sochor, M.},
  date = {2006-11-01},
  journaltitle = {Fusion Eng. Des.},
  series = {Proceedings of the {{Fifteenth International}} Toki {{Conference}} on "{{Fusion}} and {{Advanced Technology}}"},
  volume = {81},
  number = {20},
  pages = {2509--2513},
  doi = {10.1016/j.fusengdes.2006.07.008},
}

@book{toro_riemann_2009,
  title = {Riemann {{Solvers}} and {{Numerical Methods}} for {{Fluid Dynamics}}: {{A Practical Introduction}}},
  author = {Toro, Eleuterio F.},
  date = {2009},
  publisher = {Springer},
  location = {Berlin, Heidelberg},
  doi = {10.1007/b79761},
  isbn = {978-3-540-25202-3 978-3-540-49834-6},
}

@article{virtanen_scipy_2020,
  title = {{{SciPy}} 1.0: Fundamental Algorithms for Scientific Computing in {{Python}}},
  shorttitle = {{{SciPy}} 1.0},
  author = {Virtanen, Pauli and Gommers, Ralf and Oliphant, Travis E. and Haberland, Matt and Reddy, Tyler and Cournapeau, David and Burovski, Evgeni and Peterson, Pearu and Weckesser, Warren and Bright, Jonathan and family=Walt, given=Stéfan J., prefix=van der, useprefix=true and Brett, Matthew and Wilson, Joshua and Millman, K. Jarrod and Mayorov, Nikolay and Nelson, Andrew R. J. and Jones, Eric and Kern, Robert and Larson, Eric and Carey, C. J. and Polat, İlhan and Feng, Yu and Moore, Eric W. and VanderPlas, Jake and Laxalde, Denis and Perktold, Josef and Cimrman, Robert and Henriksen, Ian and Quintero, E. A. and Harris, Charles R. and Archibald, Anne M. and Ribeiro, Antônio H. and Pedregosa, Fabian and family=Mulbregt, given=Paul, prefix=van, useprefix=true},
  date = {2020-03},
  journaltitle = {Nat. Methods},
  volume = {17},
  number = {3},
  pages = {261--272},
  publisher = {Nature Publishing Group},
  doi = {10.1038/s41592-019-0686-2},
}

@article{vitrano_opensource_2023,
  title = {An {{Open-Source Finite Element Quench Simulation Tool}} for {{Superconducting Magnets}}},
  author = {Vitrano, Andrea and Wozniak, Mariusz and Schnaubelt, Erik and Mulder, Tim and Ravaioli, Emmanuele and Verweij, Arjan},
  date = {2023-08},
  journaltitle = {IEEE Trans. Appl. Supercond.},
  volume = {33},
  number = {5},
  pages = {1--6},
  doi = {10.1109/TASC.2023.3259332},
}

@article{warmer_system_2016,
  title = {System {{Code Analysis}} of {{HELIAS-Type Fusion Reactor}} and {{Economic Comparison With Tokamaks}}},
  author = {Warmer, F. and Torrisi, S. B. and Beidler, C. D. and Dinklage, A. and Feng, Y. and Geiger, J. and Schauer, F. and Turkin, Y. and Wolf, R. and Xanthopoulos, P. and Kemp, R. and Knight, P. and Lux, H. and Ward, D.},
  date = {2016-09},
  journaltitle = {IEEE Trans. Plasma Sci.},
  volume = {44},
  number = {9},
  pages = {1576--1585},
  doi = {10.1109/TPS.2016.2545868},
}

@article{warmer_overview_2024,
  title = {Overview of {{European}} Efforts and Advances in {{Stellarator}} Power Plant Studies},
  author = {Warmer, Felix and Alguacil, J. and Biek, D. and Bogaarts, T. and Bongiovì, G. and Bykov, V. and Catalán, J. P. and Duligal, R. K. and Fernández-Berceruelo, I. and Giambrone, S. and Hume, C. and Hrecinuc, M. and Kembleton, R. and Lion, J. and Lyytinen, T. and Noguerón Valiente, J. A. and Palermo, I. and Queral, V. and Rapisarda, D. and Rutten, W. J. and Sanchis, L. and Sarasola, X. and Sedlak, K. and Snicker, A. and Sosa, D. and Urgorri, F. R.},
  date = {2024-05-01},
  journaltitle = {Fusion Eng. Des.},
  volume = {202},
  pages = {114386},
  doi = {10.1016/j.fusengdes.2024.114386},
}

@article{warming_diagonalization_1975,
  title = {Diagonalization and {{Simultaneous Symmetrization}} of the {{Gas-Dynamic Matrices}}},
  author = {Warming, R. F. and Beam, Richard M. and Hyett, B. J.},
  date = {1975},
  journaltitle = {Mathematics of Computation},
  volume = {29},
  number = {132},
  eprint = {2005742},
  eprinttype = {jstor},
  pages = {1037--1045},
  publisher = {American Mathematical Society},
  doi = {10.2307/2005742},
}

@book{wilson_superconducting_1987,
  title = {Superconducting {{Magnets}}},
  author = {Wilson, Martin N.},
  date = {1987-03-12},
  series = {Monographs on {{Cryogenics}}},
  publisher = {Oxford University Press},
  location = {Oxford, New York},
  isbn = {978-0-19-854810-2},
}

@book{wylie_fluid_1978,
  title = {Fluid {{Transients}}},
  author = {Wylie, Evan Benjamin and Streeter, Victor L.},
  date = {1978},
  publisher = {McGraw-Hill International Book Co.},
  location = {New York, USA},
}

\end{document}